\documentclass[prd,twocolumn,aps, floatfix,preprintnumbers,showpacs,nofootinbib]{revtex4}
\usepackage{amsthm}   
\usepackage{amsmath}  
\usepackage{amssymb}  
\usepackage{mathrsfs}
\usepackage{color}
\usepackage[all]{xy}
\usepackage[pdftex]{graphicx}
\usepackage{indentfirst}
\usepackage[pdftex]{hyperref}
\usepackage{subfigure}

\newcommand{\be}{\begin{equation}}
\newcommand{\ee}{\end{equation}}
\newcommand{\ba}{\begin{eqnarray}}
\newcommand{\ea}{\end{eqnarray}}

\newcommand{\en}{\nonumber\\}

\newcommand{\de}{\delta}

\newcommand{\dd}[1]{\dot{#1}}

\definecolor{darkred}{RGB}{175,0,0}
\definecolor{darkblue}{RGB}{0,0,175}

\begin{document}
\title{Constraints on Large-Scale Dark Acoustic Oscillations from Cosmology}
\author{Francis-Yan Cyr-Racine\footnote{W.M. Keck Institute for Space Studies Postdoctoral Fellow}}\email{francis-yan.cyr-racine@jpl.nasa.gov}
\affiliation{NASA Jet Propulsion Laboratory, California Institute of Technology, Pasadena, CA 91109, USA}
\affiliation{California Institute of Technology, Pasadena, CA 91125, USA}
\author{Roland de Putter}
\affiliation{NASA Jet Propulsion Laboratory, California Institute of Technology, Pasadena, CA 91109, USA}
\affiliation{California Institute of Technology, Pasadena, CA 91125, USA}
\author{Alvise Raccanelli}
\affiliation{NASA Jet Propulsion Laboratory, California Institute of Technology, Pasadena, CA 91109, USA}
\affiliation{California Institute of Technology, Pasadena, CA 91125, USA}
\author{Kris Sigurdson}%\email{krs@phas.ubc.ca}
\affiliation{Department of Physics and Astronomy, University of British Columbia, Vancouver, BC, V6T 1Z1, Canada}
\date{\today}
\begin{abstract}
If all or a fraction of the dark matter (DM) were coupled to a bath of dark radiation (DR) in the early Universe we expect the combined DM-DR system to give rise to acoustic oscillations of the dark matter until it decouples from the DR.  Much like the standard baryon acoustic oscillations, these dark acoustic oscillations (DAO) imprint a characteristic scale, the sound horizon of dark matter, on the matter power spectrum. We compute in detail how the microphysics of the DM-DR interaction affects the clustering of matter in the Universe and show that the DAO physics also gives rise to unique signatures in the temperature and polarization spectra of the cosmic microwave background (CMB).  We use cosmological data from the CMB, baryon acoustic oscillations (BAO), and large-scale structure to constrain the possible fraction of interacting DM as well as the strength of its interaction with DR. 
  Like nearly all knowledge we have gleaned about DM since inferring its existence this constraint rests on the betrayal by gravity of the location of otherwise invisible DM. 
Although our results can be straightforwardly applied to a broad class of models that couple dark matter particles to various light relativistic species, in order to make quantitative predictions, we model the interacting component as dark atoms coupled to a bath of dark photons.  We find that linear cosmological data and CMB lensing put strong constraints on existence of DAO features in the CMB and the large-scale structure of the Universe. Interestingly, we find that at most $\sim5\%$ of all DM can be very strongly interacting with DR.  We show that our results are surprisingly constraining for the recently proposed Double-disk DM model, a novel example of how large-scale precision cosmological data can be used to constrain galactic physics and sub-galactic structure.
\end{abstract}
\maketitle

%%%%%%
\section{Introduction}
%%%%%%
The fundamental nature of dark matter (DM) has puzzled scientists for decades. While we have long observed the gravitational pull it exerts on regular baryonic matter \cite{1933AcHPh...6..110Z,1937ApJ....86..217Z,1980ApJ...238..471R}, no conclusive hint of the particle physics governing DM has so far shown up in laboratory experiments (see Refs.~\cite{Aalseth:2010vx,Aalseth:2011wp,Angloher:2011uu,Agnese:2013rvf,Bernabei:2013xsa} for tentative signals). This does not necessarily imply that the physics of DM is trivial or uninteresting; it merely tells us that it couples very weakly to the visible sector, thus allowing it to hide its potentially rich physics. To make progress, one can turn to astrophysical observations of DM dominated objects such as dwarf spheroidals \cite{Walker:2011zu,2012MNRAS.420.2034S,BoylanKolchin:2011de,BoylanKolchin:2011dk,Walker:2012td,Zavala:2012us,Laporte:2013fwa,Amorisco:2013uwa}, galaxies \cite{2004MNRAS.351..903G,deNaray:2009xj,deNaray:2011hy,Vogelsberger:2012ku}, and merging galaxy clusters \cite{Markevitch:2003at,Randall:2007ph,Merten:2011wj,Dawson:2011kf,Dawson:2012fx}.  These objects are however highly non-linear and one cannot in general neglect the impact of baryon physics (see e.g.~Refs~\cite{Oh:2010mc,Pontzen:2011ty,Governato:2012fa,Zolotov:2012xd,Brooks:2012vi,Shen:2013wva}) on their evolution. Numerical simulations are therefore necessary to assess the impact of non-minimal DM physics on these objects.

Since DM dominates the matter density on cosmological scales, it is natural to ask whether cosmological data can shed light on the fundamental physics of DM. While the cold DM paradigm \cite{Blumenthal:1984bp,Davis:1985rj} provides a very good fit to data on large cosmological scales, it is possible that a subdominant DM component could display very different properties. For instance, there could be new dark forces \cite{Foot:2004pa,ArkaniHamed:2008qn,Ackerman:2008gi,Feng:2009mn,Baldi:2012ua,Tulin:2012wi,Hooper:2012cw,Aarssen:2012fx,Tulin:2013teo} that couple only to a fraction of the DM particles or a portion of the DM could be warm \cite{Bode:2000gq,Dalcanton:2000hn,Zentner:2003yd,Smith:2011ev}. 

A particularly interesting case is one in which a fraction of the DM can interact with or via a massless (or nearly massless) particle. In this scenario, the interacting DM component is prohibited from forming gravitationally bound structures until it kinematically decouples from the light state. If this decoupling happens at relatively late times (that is, close to or after matter-radiation equality), an imprint similar in many ways to the baryon acoustic oscillation feature should be left on the matter density field on cosmological scales \cite{CyrRacine:2012fz}. This dark acoustic oscillation (DAO) feature generically arises in any model where DM is coupled to relativistic particles until relatively late times. For instance, they occur if DM couples to neutrinos \cite{Mangano:2006mp,Serra:2009uu,Aarssen:2012fx,Diamanti:2012tg} or photons \cite{Boehm:2001hm,McDermott:2010pa,Wilkinson:2013kia}, if DM interacts with a dark $U(1)_D$ gauge boson \cite{Ackerman:2008gi,Feng:2009mn,Kaplan:2009de,Kaplan:2011yj, CyrRacine:2012fz}, if DM couples to a light scalar field \cite{1992ApJ...398..407G}, or for the so-called ``cannibal'' DM models \cite{1992ApJ...398...43C,1994ApJ...431...41M,1995ApJ...452..495D}. We emphasize for the reader unfamiliar with the above body of work that if the DM sector couples purely gravitationally to the visible sector, some of these scenarios are surprisingly unconstrained. For instance, even a model in which the totality of the DM interacts via a massless $U(1)_D$ gauge boson \cite{Ackerman:2008gi,Feng:2009mn} has a large allowed parameter space.

Intriguingly, a subset of these scenarios for which a fraction of the DM couples directly to the light state (as opposed to coupling to the light state via a massive mediator) also has the potential to impact structure formation on galactic scales. Indeed, since a fraction of the DM can lose energy by radiating a light mediator particle, its distribution inside galaxies may differ significantly from that of the standard cold DM component due to the formation of a DM disk \cite{Fan:2013tia,Fan:2013yva,McCullough:2013jma}. As it also generically predicts a DAO feature on large cosmological scales, this Double-disk DM scenario has the unique feature of affecting the DM density field on a large range of scales, hence providing us with a multi-pronged approach to constrain its properties. This contrasts with usually-considered self-interacting DM scenarios where only the smallest astrophysical scales are affected (see e.g.~Refs.~\cite{Rocha:2012jg,Peter:2012jh}).

Moreover, scenarios in which DM couples to a relativistic component naturally incorporates some form of dark radiation (DR), which might be necessary to reconcile the current cosmic microwave background (CMB) data \cite{Hinshaw:2012aka,planckXVI} with recent measurements of the local Hubble expansion rate \cite{Riess:2011yx,Freedman:2012ny}. We emphasize however that the DR in these models is different than that usually parametrized via the quantity $\Delta N_{\rm eff}$, which effectively describes the number of fermionic free-streaming species that have the same temperature as the cosmic neutrino background assuming instant decoupling. Indeed, since DR couples to DM at early times, DR is not free-streaming at all times and therefore does not behave like a neutrino. While this subtle change leaves unchanged the background expansion history of Universe, it \emph{does} affect the way density fluctuations evolve, implying that we cannot straightforwardly interpret the DR in terms of the tacit (and sometimes obscure, see e.g.~Ref.~\cite{Steigman:2013yua}) parameter $\Delta N_{\rm eff}$. As we will see below, the cosmological constraints on the abundance of a DR component that couples to DM until late times can be vastly different than the current limits on $\Delta N_{\rm eff}$, implying that one must exert caution in interpreting bounds on this latter parameter.

In this paper, we use cosmological data from the CMB, baryon acoustic oscillations (BAO), and large-scale structure to constrain the possible fraction of interacting DM as well as the strength of its interaction with light relativistic particles. Since, for the models we consider, interacting DM only affects the cosmological observables through its gravitational interaction, our bounds on interacting DM are very general and apply to any hidden-sector model in which large-scale DAO arises. We will make this correspondence explicit below. As we will see, cosmological data alone place strong limits on both the possible fraction of interacting DM as well as on the strength of its interaction with the light state.

The paper is organized as follow. In section \ref{PIDM}, we present the details of our partially-interacting DM scenario. We then introduce in section \ref{CosmoPert} the physics of DAOs and review the perturbation equations necessary to describe the cosmological evolution of this model. In section \ref{observables}, we describe the consequence of our scenario on the cosmological observables, including galaxy clustering, the CMB, and CMB lensing. The details of the data used for our analysis are given in section \ref{data}. Our main results are presented in section \ref{results} and we discuss their implications for the formation of a dark disk inside galaxies in section \ref{galform}. We discuss the implications of our results in section \ref{sec:conc}.

%%%%%
\section{Partially-Interacting Dark Matter}\label{PIDM}
%%%%%
We consider a hybrid DM sector comprised of particles interacting with a light relativistic state and of a standard non-interacting cold DM component. We shall refer to such a scenario as partially-interacting DM (PIDM).  For definitiveness and simplicity, we take the interacting component to be made of dark atoms \cite{Goldberg:1986nk,Kaplan:2009de,Kaplan:2011yj,Behbahani:2010xa,Petraki:2011mv,Cline:2012is,CyrRacine:2012fz}, a well-studied model that allows us to make exact quantitative predictions. Atomic DM naturally encompasses hidden-charged DM models \cite{Ackerman:2008gi,Feng:2009mn} and can mimic the behavior of CDM in some limits. Importantly, this model readily incorporates a dark radiation (DR) component. Since its impact on DM fluctuations (acoustic oscillations, damping) are very generic, atomic DM can be viewed as a simple toy model to parametrize deviations from a pure CDM scenario. As we discussed in the introduction, the limits we obtain in this work are very general and apply to a variety of models. We briefly review the physics of dark atoms in section \ref{adm} below. We refer the reader to Ref.~\cite{CyrRacine:2012fz} for more details.

Throughout this work, we assume that the relic abundances of both the CDM and the interacting components are set by some unspecified UV physics. Examples of such UV completion are given in Refs.~\cite{Kaplan:2011yj,Petraki:2011mv}. We do not expect the details of the UV completion to affect the low-energy interactions responsible for modifying the growth of DM fluctuations on small scales. To retain generality, we will refer to the fraction of DM made of dark atoms as the ``interacting DM'' component in order to distinguish it from the standard collionsionless CDM. We further assume that the CDM and the interacting component only interact through gravity.

%%%%
\subsection{Atomic Dark Matter}\label{adm}
%%%%
In the atomic DM scenario, two oppositely-charged massive fermions can interact through a new unbroken $U(1)_D$ gauge force to form hydrogen-like bound states. In the early Universe, the dark atoms are all ionized by the hot thermal bath of DR (that is, the $U(1)_D$ gauge boson) and form a plasma similar to that of standard baryons and photons. When the temperature of the DR falls significantly below their binding energy, dark atoms are allowed to recombine into neutral bound states if the recombination rate is larger than the expansion rate of the Universe. The DR eventually decouples from the atomic DM and begin to free-stream across the Universe. We note that the order and the dynamics of the different important transitions of the dark plasma (recombination, onset of DR free-streaming, atomic DM drag epoch, DM thermal decoupling, etc.) can be very different than in the standard baryonic case. We refer the reader to Ref.~\cite{CyrRacine:2012fz} for more details.

To retain generality and emphasize that the PIDM scenario we are considering is quite general, we shall refer to the massless $U(1)_D$ ``dark photons'' simply as DR. For simplicity, we also denote the lightest fermion as  ``dark electron''  (mass $m_{\bf e}$) while the heaviest fermion is referred to as  ``dark proton'' (mass $m_{\bf p}$). We assume that these two oppositely-charged components come in equal number such that the dark sector is overall neutral under the $U(1)_D$ interaction. This model is characterized by five parameters which are the mass of the dark atoms $m_D$, the dark fine-structure constant $\alpha_D$, the binding energy of the dark atoms $B_D$, the present-day ratio of the DR temperature ($T_D$) to the cosmic microwave background temperature $\xi\equiv(T_D/T_{\rm CMB})|_{z=0}$, and the fraction of the overall DM density contained in interacting DM (here, dark atoms), $f_{\rm int}\equiv\rho_{\rm int}/\rho_{\rm DM}$, where $\rho_{\rm DM}=\rho_{\rm int}+\rho_{\rm CDM}$ and where $\rho_{\rm int}$ is the energy density of the interacting DM component. These parameters are subject to the consistency condition $m_D/B_D \geq 8/\alpha_D^2-1$, which ensures that the relationship $m_{\bf e}+m_{\bf p}-B_D=m_D$ is satisfied. We note that if the visible and dark sectors were coupled above the electroweak scale, we naturally expect $\xi\sim0.5$ \cite{Fan:2013yva}. A smaller value would either require new degrees of freedom in the visible sector or that the two sectors were never in thermal equilibrium in the first place.

The evolution of the dark plasma is largely governed by the opacity $\tau_D^{-1}$ of the medium to DR.  For the model we considered, the main contributions\footnote{In this work, we neglect the small contribution to the opacity from photoionization processes.} to this opacity are Compton scatterings of DR off charged dark fermions and Rayleigh scatterings off neutral dark atoms, that is,
\be\label{opacity}
\tau_D^{-1}=\tau_{\rm Compton}^{-1}+\tau_{\rm R}^{-1},
\ee
where
\be
\tau_{\rm Compton}^{-1}=an_{\rm ADM}x_D\sigma_{{\rm T},D}\left[1+\left(\frac{m_{\bf e}}{m_{\bf p}}\right)^2\right],
\ee
and
\ba\label{rayleigh}
\tau_{\rm R}^{-1}&=&an_{\rm ADM}(1-x_D)\langle\sigma_{\rm R}\rangle\en
&\simeq&32\pi^4an_{\rm ADM}(1-x_D)\sigma_{{\rm T},D}\left(\frac{T_D}{B_D}\right)^4.
\ea
Here, $\sigma_{{\rm T},D}\equiv 8\pi\alpha_D^2/(3 m_{\bf e}^2)$ is the dark Thomson cross section, $a$ is the scale factor describing the expansion of the Universe, $x_D$ is the ionized fraction of the dark plasma, $n_{\rm ADM}$ is the number density of dark atoms, $\sigma_{\rm R}$ is the Rayleigh scattering cross section, and where the angular bracket denotes thermal averaging. We note that the second line of Eq.~\ref{rayleigh} is only valid if $T_D<B_D$. It is out of the scope of this paper to discuss in detail the evolution of the ionized fraction and of the DM temperature. We refer the reader to Ref.~\cite{CyrRacine:2012fz} for a thorough investigation of dark atom recombination and thermal history. 
%%%%
\subsection{$\xi$ vs $\Delta N_{\rm eff}$}
%%%%
We note that, as far as the \emph{background cosmological expansion} is concerned, varying the temperature of the DR is equivalent to changing the effective number of relativistic species $N_{\rm eff}$ according to the mapping
\be\label{delta_n_eff}
\Delta N_{\rm eff} \leftrightarrow \frac{8}{7}\left(\frac{11}{4}\right)^{4/3}\xi^4.
\ee
\emph{However}, since $\xi$ affects the evolution of cosmological fluctuations in a different way than $\Delta N_{\rm eff}$ (because the DR couples to DM and is not always free-streaming), we emphasize that one cannot blindly translate the constraints on $\Delta N_{\rm eff}$ from, say, Planck \cite{planckXVI} to a bound on $\xi$. In fact, as we discuss below, the bounds on $\xi$ are generally much more stringent than the equivalent  limit on $\Delta N_{\rm eff}$.

%%%%
\section{Cosmological Evolution}\label{CosmoPert}
%%%%

%%%%
\subsection{Dark Acoustic Oscillation Scale}
%%%%
Since a fraction of the DM forms a tightly-coupled plasma in the early Universe, the evolution of cosmological fluctuations in the PIDM model departs significantly from that of a standard $\Lambda$CDM Universe. Indeed, as Fourier modes enter the causal horizon, the DR pressure provides a restoring force opposing the gravitational growth of over densities, leading to the propagation of dark acoustic oscillations (DAO) in the plasma. These acoustic waves propagate until DR kinematically decouples from the interacting DM component. Similar to the baryon case, the scale corresponding to the sound horizon of the dark plasma at kinetic decoupling remains imprinted on the matter field at late times. This so-called DAO scale is given by
\be\label{rDAO}
r_{\rm DAO}\equiv\int_0^{\eta_{D}}c_D(\eta)d\eta,
\ee
where $c_D$ is the sound speed of the dark plasma, $\eta$ is the conformal time, and $\eta_D$ denotes the conformal time at the epoch at which atomic DM kinematically decouples from the DR bath. The DAO scale is a key quantity of cosmologically-interesting interacting DM models. Indeed, much like the free-streaming length of warm DM models, the DAO scale divides the modes that are strongly affected by the DM interactions (through damping and oscillations) from those that behave mostly like in the CDM paradigm. We note however that, in contrast to warm DM models, the suppression of small-scale fluctuations in the PIDM scenario is mostly due to acoustic (also known as collisional) damping \cite{Boehm:2000gq,CyrRacine:2012fz}, while residual free-streaming after kinematic decoupling can play a minor role.

In the tight-coupling limit of the dark plasma, the sound speed takes the form $c_D=1/\sqrt{3(1+R_D^{-1})}$, where $R_D\equiv4\rho_{\tilde{\gamma}}/3\rho_{\rm int}$. Here, $\rho_{\tilde{\gamma}}$ stands for the the energy density of the DR. In a matter-radiation Universe, the integral of Eq.~\ref{rDAO} can be performed analytically
\ba\label{r_DAO_theory}
r_{\rm DAO} &=& \frac{4 \xi^2\sqrt{\Omega_{\gamma}}}{3 H_0 \sqrt{f_{\rm int}\Omega_{\rm DM}\Omega_{\rm m}}}\times\\
&&\ln{\left[\frac{\sqrt{\gamma_{\rm int}}\sqrt{\Omega_{\rm r}+\Omega_{\rm m} a_D}+\sqrt{\Omega_{\rm m}+\gamma_{\rm int}a_D}}{\sqrt{\gamma_{\rm int}\Omega_{\rm r}}+\sqrt{\Omega_{\rm m}}}\right]},\nonumber
\ea
where we have defined
\be
\gamma_{\rm int} \equiv \frac{3f_{\rm int}\Omega_{\rm DM}}{4\xi^4\Omega_{\gamma}},
\ee
$a_D$ is the scale factor at the epoch of atomic DM kinematic decoupling, and $H_0$ is the present-day Hubble constant. $\Omega_{\gamma}$, $\Omega_{\rm r}$, and $\Omega_{\rm m}$ stand for the energy density in photons, radiation (including neutrinos and DR), and non-relativistic matter, respectively, all in units of the critical density of the Universe. We observe that the DAO scale depends most strongly on the ratio $\xi^2/\sqrt{f_{\rm int}}$ and that the details of the interacting DM microphysics only enter through a logarithmic dependence on $a_D$. The scale factor at the epoch of dark decoupling can be estimated from the criterion $n_{\rm ADM}x_D\sigma_{{\rm T},D}\simeq H$, since Thomson scattering is the dominant mechanism responsible for the opacity of the dark plasma. Here, $H$ is the Hubble parameter. We give an expression for $a_D$ in terms of the dark parameters in Appendix \ref{app:a_D}.

The scale factor at the epoch of dark decoupling (and, consequently, $r_{\rm DAO}$) is largely determined by the following combination of parameters
\be\label{eq:pidm_mod}
\Sigma_{\rm DAO}\equiv \alpha_D \left(\frac{B_D}{\rm eV}\right)^{-1}\left(\frac{m_D}{\rm GeV}\right)^{-1/6}.
\ee
\begin{figure}[t!]
\begin{centering}
\includegraphics[width=0.5\textwidth]{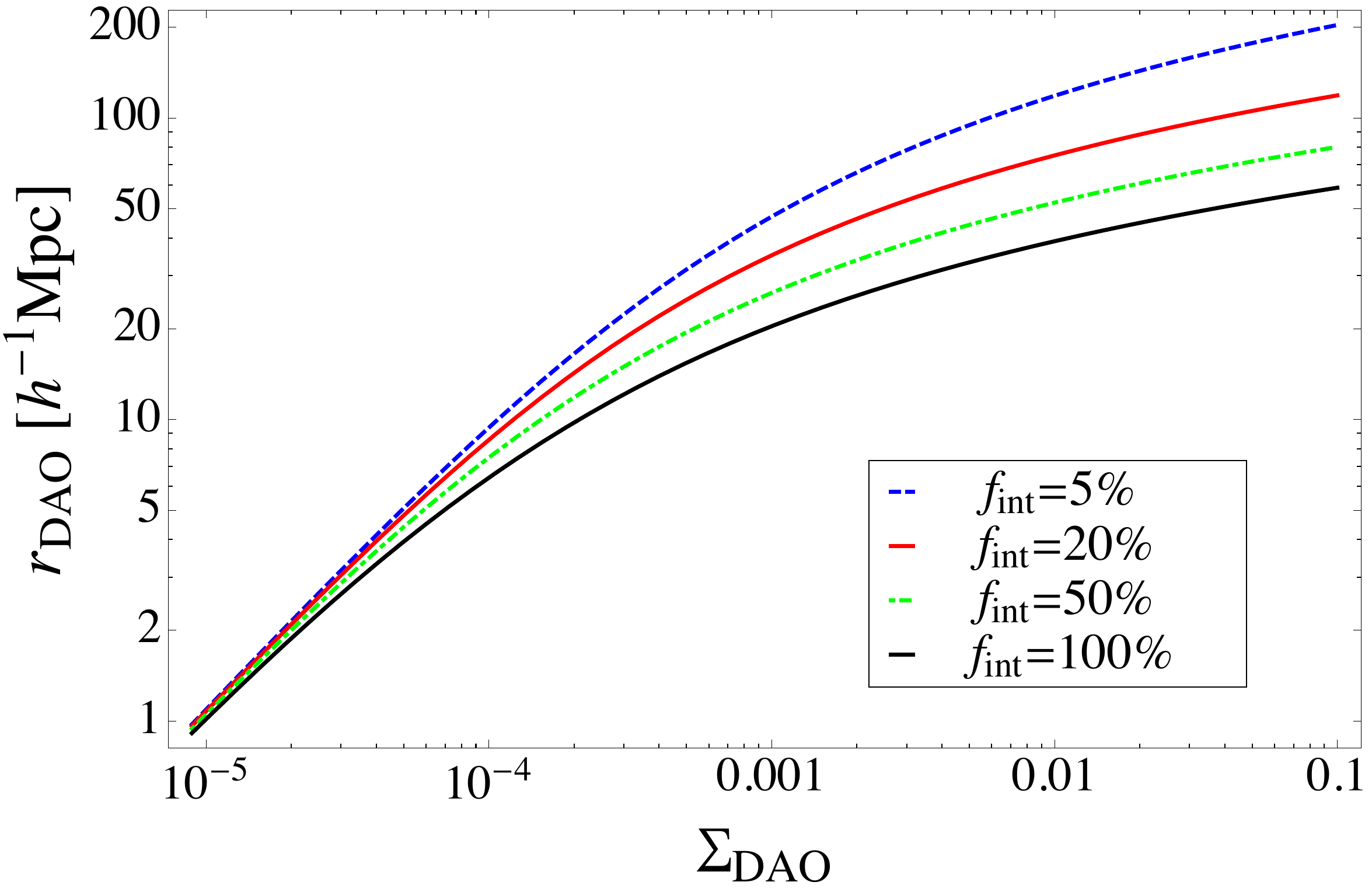}
\includegraphics[width=0.5\textwidth]{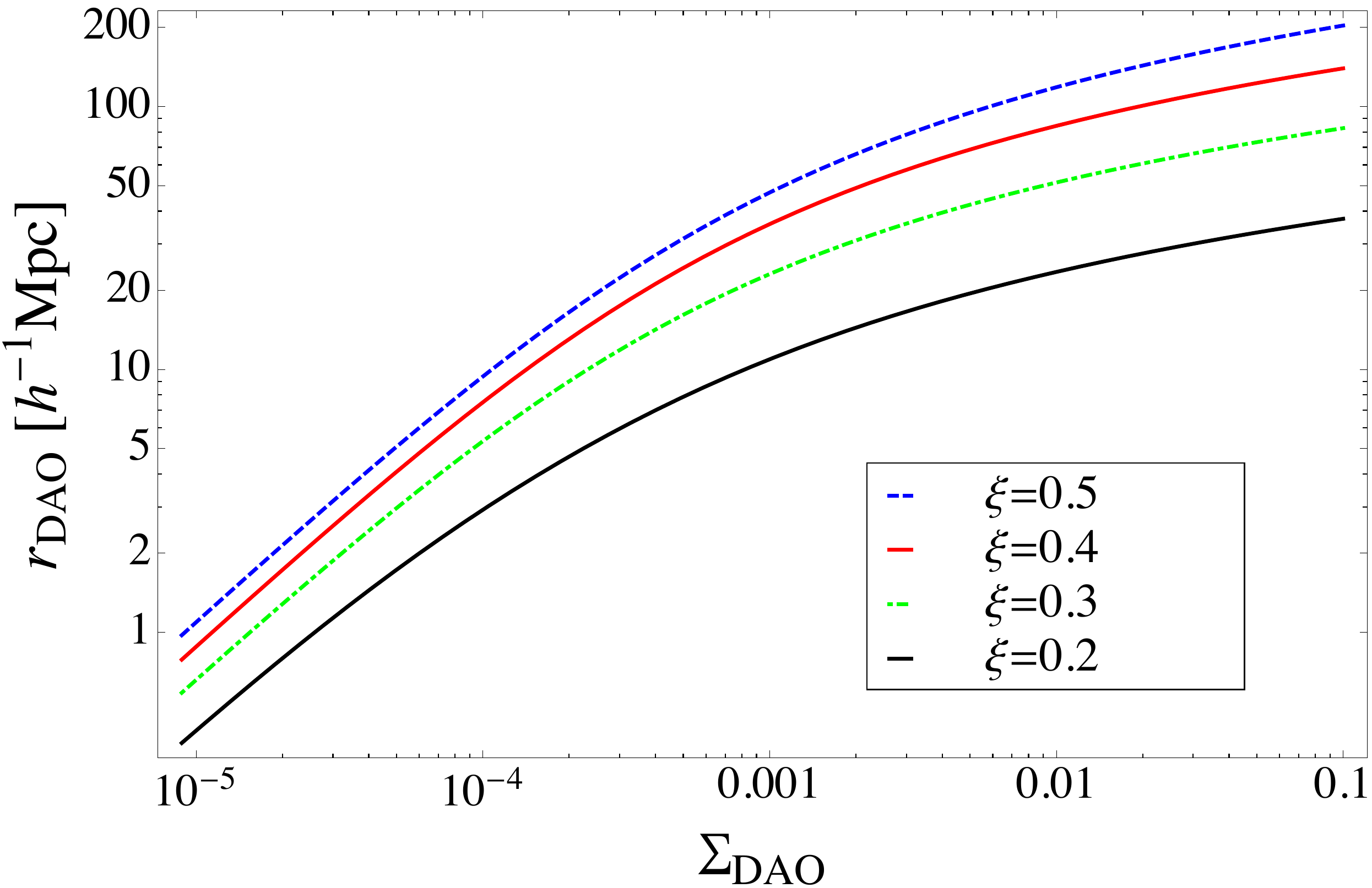}
\caption{Comoving DAO scale as a function of the parameter $\Sigma_{\rm DAO}$ for strongly-coupled atomic DM models ($\alpha_D>0.025$). In the upper panel, we fix $\xi=0.5$ and vary the fraction of interacting DM. In the lower panel, we fix $f_{\rm int}=5\%$ and let $\xi$ vary. Here, take $H_0=69.57$ km/s/Mpc, $\Omega_{\rm m}=0.3048$, $\Omega_{\rm DM}h^2=0.1198$, and $N_{\rm eff}=3.046$.}
\label{fig:rDAO}
\end{centering}
\end{figure}
This quantity is directly proportional to the scattering rate between DR and interacting DM. Its non-trivial dependence on the dark parameters $\alpha_D$, $B_D$, and $m_D$ is caused by $x_D$ which itself depends on these dark parameters (see Appendix \ref{app:a_D}). To give a sense of scale, we note that for regular baryonic hydrogen we have $\Sigma_{\rm BAO}\simeq5.4\times10^{-4}$. We emphasize that, while the definition given in Eq.~\ref{eq:pidm_mod} is very specific to the atomic DM model considered, $\Sigma_{\rm DAO}$ is a simple proxy for the cross section between DM and DR at the epoch of kinematic decoupling ($\sigma_{\rm DM-DR}(a_D)$) over the DM mass. Explicitly, the relation between $\Sigma_{\rm DAO}$ and the DM-DR cross section over the DM mass is
\ba
\left(\frac{\sigma_{\rm DM-DR}(a_D)}{m_D}\right)&= &1.9\times10^{-4}\left(\frac{\xi}{0.5}\right)\left(\frac{\Sigma_{\rm DAO}}{10^{-3}}\right)\en
&&\,\,\,\times\left(\frac{f_{\rm int}\Omega_{\rm DM}h^2}{0.12}\right)^{-1}\frac{\rm cm^2}{\rm g}.
\ea
It should be clear from the above expression that any constraints we put on $\xi$ and $\Sigma_{\rm DAO}$ can be directly translated to model-independent limits on the DM-DR cross section over the DM mass\footnote{We note that this cross section between DM and DR should not be confused with DM self-interaction cross section relevant for small-scale astrophysical objects.} at kinematic decoupling. In the remainder of this work, we shall parametrize the strength of the DM-DR interaction in terms of $\Sigma_{\rm DAO}$ but the reader should keep in mind that our results can be translated to any model in which a fraction of the DM interacts with a bath of DR.

We plot the DAO scale as a function of $\Sigma_{\rm DAO}$  in Fig.~\ref{fig:rDAO} for different values of the interacting DM fraction (top panel) and for different values of $\xi$ (lower panel). We observe that for $\Sigma_{\rm DAO} > 10^{-3}$ and $\xi>0.2$, the $r_{\rm DAO}$ lies in the range of scales currently probed by galaxy surveys and CMB experiments. Looking ahead, we thus expect these datasets to severely constrain PIDM models lying in this region of parameter space. In particular, a typical double-disk DM model \cite{Fan:2013tia,Fan:2013yva} with $\Sigma_{\rm DAO}\sim10^{-2.5}$, $\xi\sim0.5$, and $f_{\rm int}\sim5\%$, has $r_{\rm DAO}\sim80h^{-1}$Mpc, which is ruled out by current data as we will discuss in section \ref{results}.

%%%%
\subsection{Cosmological Fluctuations}
%%%%
We now turn our attention to the evolution of cosmological perturbations in the PIDM scenario. The equations describing the evolution of interacting DM density and velocity fluctuations are 
\be\label{delta_D}
\dd{\de}_{D}+\theta_{D}-3\dd{\phi}=0,
\ee
\be\label{theta_D}
\dd{\theta}_{D}+\frac{\dd{a}}{a}\theta_D-c_D^2k^2\de_{D}-k^2\psi= \frac{R_D}{\tau_D}(\theta_{\tilde{\gamma}}-\theta_D),
\ee
where we closely followed the notation of Ref.~\cite{Ma:1995ey} in conformal Newtonian gauge. Here, a dot atop a quantity denotes a derivative with respect to conformal time, $\de_D$ is the interacting DM density contrast, $\theta_D$ and $\theta_{\tilde{\gamma}}$  are the divergence of the interacting DM and DR velocity, respectively; $\phi$  and $\psi$ are the gravitational scalar potentials, and $k$ is the wavenumber of the mode. Here, the subscript $\tilde{\gamma}$ always refers to the DR. The right-hand side of Eq.~(\ref{theta_D}) represents the collision term between the atomic DM and the DR. At early times, we generally have $R_D\gg1$ and $\tau_D\ll H^{-1}$, implying that the interacting DM component is effectively dragged along by the DR. 
The latter evolves according to the following Boltzmann equations:
\be\label{delta_gD}
\dd{\de}_{\tilde{\gamma}}+\frac{4}{3}\theta_{\tilde{\gamma}}-4\dd{\phi}=0;
\ee
\be\label{theta_gD}
\dd{\theta}_{\tilde{\gamma}}-k^2(\frac{1}{4}\de_{\tilde{\gamma}}-\frac{F_{\tilde{\gamma}2}}{2})-k^2\psi= \frac{1}{\tau_D}(\theta_D-\theta_{\tilde{\gamma}});
\ee
\be\label{FgD2}
\dd{F}_{\tilde{\gamma}2} = \frac{8}{15}\theta_{\tilde{\gamma}}-\frac{3}{5}kF_{\tilde{\gamma}3}-\frac{9}{10\tau_D}F_{\tilde{\gamma}2};
\ee
\be\label{FgDl}
\dd{F}_{\tilde{\gamma}l}=\frac{k}{2l+1}\left[l F_{\tilde{\gamma}(l-1)}-(l+1)F_{\tilde{\gamma}(l+1)}\right]-\frac{1}{\tau_D}F_{\tilde{\gamma}l}.
\ee
Eqs.~(\ref{delta_gD}) and (\ref{theta_gD}) describe the evolution of the DR over-densities ($\de_{\tilde{\gamma}}$) and of the DR velocity, respectively. It is also necessary to solve for the hierarchy of DR multipoles (Eqs.~(\ref{FgD2}) and (\ref{FgDl})) to properly account for DR diffusion and its impact on interacting DM perturbations. We note that the opacity of the dark plasma can be written in terms of $\Sigma_{\rm DAO}$. 

We solve these equations numerically together with those describing the evolution of CDM, baryon, photon, and neutrino fluctuations using a modified version of the code \texttt{CAMB} \cite{Lewis:1999bs}. We first precompute the evolution of the dark plasma opacity as described in Ref.~\cite{CyrRacine:2012fz}. We assume purely adiabatic initial conditions
\be
\de_D(z_{\rm i}) =\de_{\rm c}(z_{\rm i})\qquad \de_{\tilde{\gamma}}(z_{\rm i})=\de_{\gamma}(z_{\rm i}),
\ee
\be
\theta_D(z_{\rm i})=\theta_{\tilde{\gamma}}(z_{\rm i})=\theta_{\gamma}(z_{\rm i}),
\ee
\be
F_{\tilde{\gamma}l} = 0,\quad l\geq2.
\ee
where $z_{\rm i}$ is the initial redshift which is determined such that all modes of interest are superhorizon at early times, $k\tau(z_{\rm i})\ll1$. Here, the subscripts ``c"  and ``$\gamma$" refer to CDM and regular photon, respectively. At early times when $k\tau_D\ll1$ and $\tau_D/\tau\ll1$, Eqs.~(\ref{theta_D}) and (\ref{theta_gD}) are very stiff and we use a second-order tight-coupling scheme similar to that used for the baryon-photon plasma at early times \cite{2011PhRvD..83j3521C,Pitrou:2010ai}. Due to the presence of tightly-coupled DR at early times, the neutrino initial conditions need to be modified to take into account the different free-streaming fraction.
%%%%
\section{Cosmological Observables}\label{observables}
%%%%
Since it modifies the growth of DM fluctuations on a variety of scales, PIDM can imprint its signatures on cosmological observables such as the CMB, CMB lensing, and the matter power spectrum. These observables are mostly sensitive to the momentum transfer rate between the interacting DM and the DR, which itself determines the kinetic decoupling epoch. This rate is largely determined by the parameter $\Sigma_{\rm DAO}$ defined in Eq.~\ref{eq:pidm_mod}. For a relatively strongly-coupled ($\alpha_D > 0.025$) dark sector, any changes to $\alpha_D$, $B_D$, and $m_D$ that leaves $\Sigma_{\rm DAO}$ invariant lead to the same cosmological observables. Indeed, the recombination process for these models is well-described by the Saha approximation until the epoch of dark decoupling, which itself is determined entirely by $\Sigma_{\rm DAO}$ (see Eq.~\ref{eq:sol_a_D}). For smaller values of the dark fine-structure constant however, the details of the dark recombination process becomes important and the observables develop a small explicit dependence on $\alpha_D$. In the following subsections, we focus our attention on strongly-coupled models (unless otherwise stated) which are fully characterized by $\Sigma_{\rm DAO}$, but will address the constraints on weakly-coupled models in section \ref{results}. Unless otherwise noted, the cosmological observables plotted in this section assume $100\Omega_{\rm b}h^2=2.22$, $\Omega_{\rm DM}h^2=0.1253$, $H_0=69.57$ km/s/Mpc, $10^9A_{\rm s}=2.21$, $n_{\rm s}=0.969$, and $\tau=0.092$. In the following, we often compare our PIDM observables with those from a standard $\Lambda$CDM cosmology with a ``equivalent'' number of neutrinos (denoted $\Lambda$CDM$+\nu$) to ensure an identical cosmological expansion history. This equivalent number of neutrinos is given by $N_{\rm eff,equiv}=3.046+(8/7)(11/4)^{4/3}\xi^4$.

%%%%
\subsection{Galaxy Clustering}
%%%%
The matter power spectrum describing the clustering of matter in the Universe depend on a variety of cosmological parameters, and for this reason it has been used (together with its Fourier transform, the correlation function) to set constraints on, among others, dark energy parameters~\cite{Samushia:2013vn}, models of gravity~\cite{Raccanelli:2012fk}, neutrino mass~\cite{Putter:2012kx, Zhao:2012ly}, the growth of structures~\cite{Samushia:2011uq, Reid:2012zr}, and non-Gaussianity~\cite{Ross:2012ys3}. Since PIDM models can generally have a large impact on the clustering of matter in the Universe, we expect that recent measurements of the galaxy power spectrum and correlation function \cite{Anderson:2012ve, Samushia:2013vn, Reid:2012zr, Sanchez:2013qf} can provide useful limits on scales where non-linearities can be neglected\footnote{Since PIDM models generally predict a different shape and amplitude for the small-scale power spectrum as compared to a standard $\Lambda$CDM model, one cannot use tools such as Halofit \cite{Smith:2002dz} to model non-linearities.}.
\begin{figure}[t!]
\begin{centering}
\includegraphics[width=0.5\textwidth]{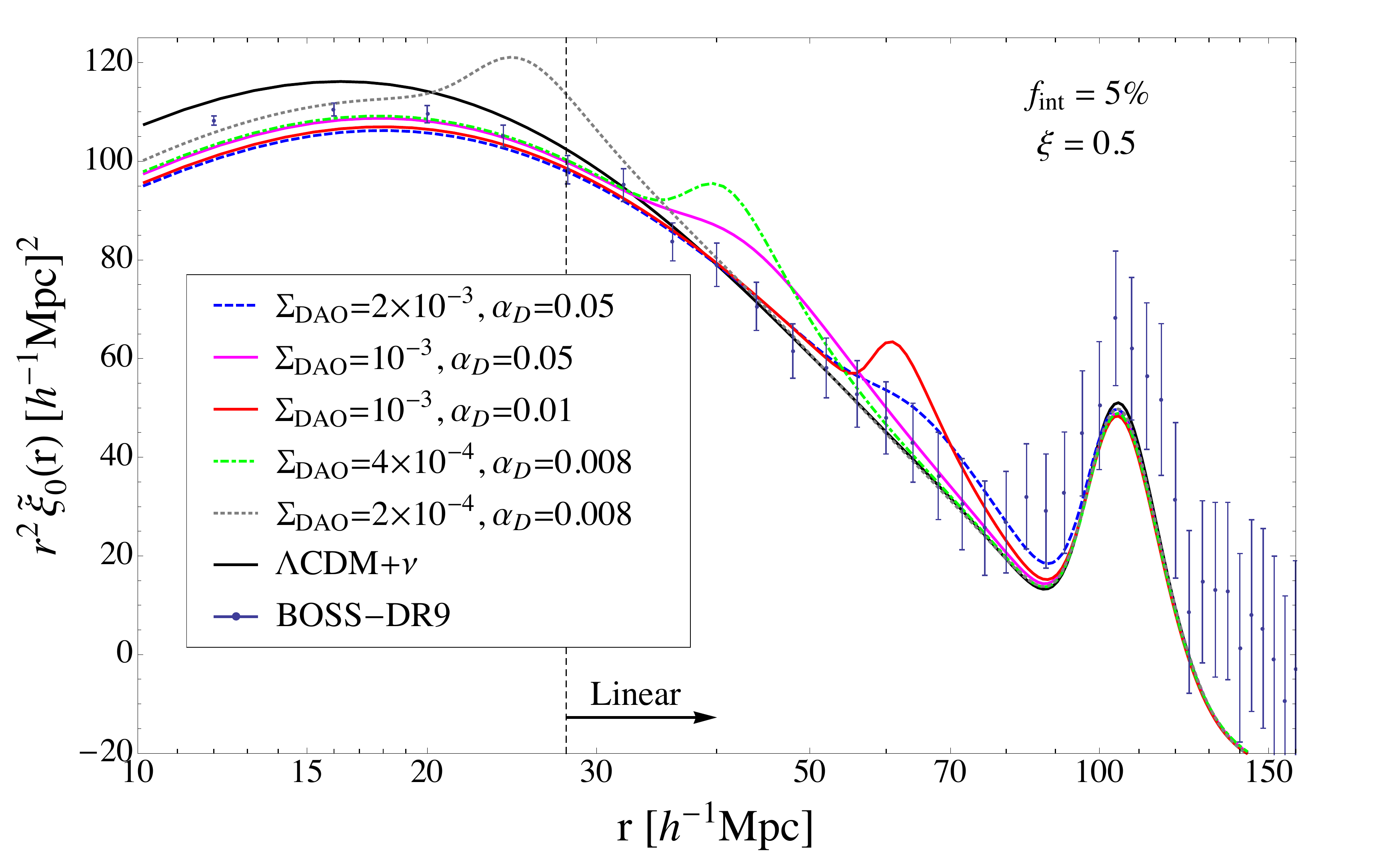}
\includegraphics[width=0.5\textwidth]{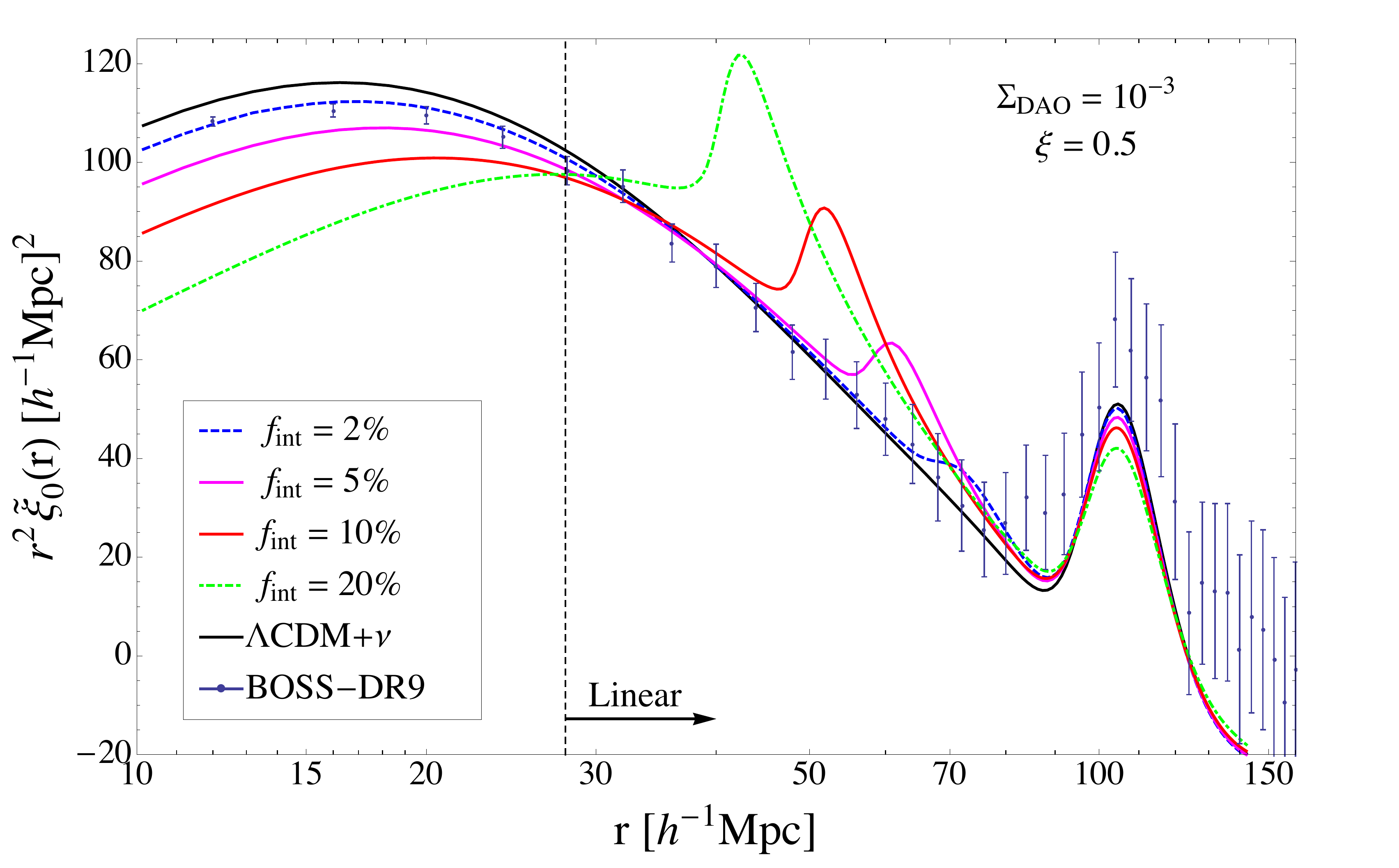}
\caption{Angle averaged galaxy correlation function $\tilde{\xi}_0(r)$ for different PIDM models. In the upper panel, we take $f_{\rm int}=5\%$, $\xi=0.5$ and vary $\Sigma_{\rm DAO}$ and $\alpha_D$. In the lower panel, we fix $\Sigma_{\rm DAO}=10^{-3}$, $\alpha_D=0.01$ and $\xi=0.5$, but let the fraction of interacting DM vary. We set the galaxy bias to $b=2.2$ and the dilation scale to $\alpha=1.016$. We compare theoretical predictions with BOSS-DR9 measurements from Ref.~\cite{Anderson:2012ve}, and we also show a standard $\Lambda$CDM model with an equivalent number of effective neutrinos. In this work, we focus uniquely on linear scales, which lie to the right of the dashed vertical line on the plot.}
\label{fig:correlation}
\end{centering}
\end{figure}

Similar to the baryon acoustic oscillation signature in the galaxy correlation function of the standard $\Lambda$CDM model, the DAO feature appears as a local enhancement of the correlation function at $r_{\rm DAO}$.  While the location of the DAO feature depends mostly on $\xi=(T_D/T_{\rm CMB})|_{z=0}$, $f_{\rm int}$, and $\Sigma_{\rm DAO}$ (it also depends somewhat on $\alpha_D$ through $\epsilon_D$, see Eq.~\ref{eq:S_D_w_epsilon}), the shape on the DAO feature does depend on the ``microphysical'' dark parameters such as $\alpha_D$, $m_D$, and $B_D$, with the feature being generally sharper for small values of $\alpha_D$. Moreover, since fluctuations on scales smaller than the DAO scale are suppressed by acoustic damping, the correlation function (and the matter power spectrum) will generally be damped on small scales as compared to a standard $\Lambda$CDM cosmology without a DAO feature. If it affects the linear cosmological scales, we expect this damping to play a major role in our constraints on PIDM.

We illustrate in Fig.~\ref{fig:correlation} the predicted galaxy correlation function for different PIDM models. We plot the galaxy linear correlation function, that is computed from the linear matter correlation as
\begin{equation}
\tilde{\xi}_g(r) =b^2 \tilde{\xi}_m(\alpha r) \, ,
\end{equation}
where $b$ is the galaxy bias and $\alpha$ is the scale dilation parameter compensating for the difference between the fiducial cosmology used to compute the correlation function from the data and the actual cosmology (see~\cite{Xu:2012uq} for more details). The matter correlation function is computed from the linear matter power spectrum, $P_m(k)$, via the relation
\begin{equation}
\tilde{\xi}_m (r) = \int \frac{k^2}{2 \pi^2} P_m(k) \, j_0 (kr)  \, dk.
\end{equation}

In this work, we focus exclusively on the linear cosmological scales (corresponding to comoving $k\leq0.12 h/{\rm Mpc}$, see Section~\ref{sec:pk_data} for more details). Nevertheless, we also plot in Fig.~\ref{fig:correlation} the predictions for smaller scales to highlight the considerable damping at small scales for PIDM models. This shows how important it would be, in order to further reduce the allowed parameter space, to be able to model the quasi-linear regime.

In the upper panel of Fig.~\ref{fig:correlation}, we vary $\Sigma_{\rm DAO}$ and $\alpha_D$ for a fixed fraction of interacting DM $f_{\rm int}=5\%$. In all cases considered the DAO feature is clearly visible, providing a characteristic signature for these models. We observe that for the majority of the models shown ($\Sigma_{\rm DAO}> 10^{-4}$), even such a small fraction of interacting DM is in tension with measurements of the galaxy correlation function from the BOSS survey. In the lower panel, we fix $\Sigma_{\rm DAO}= 10^{-3}$ and instead vary the fraction of interacting DM between $2\%$ and $20\%$. We observe the scaling of the DAO scale with $f_{\rm int}$, $r_{\rm DAO}\propto1/\sqrt{f_{\rm int}}$, and also that a $\sim2\%$ fraction of strongly interacting DM seems to be compatible with current data. As we discuss in section \ref{results}, these qualitative observations will turn out to be supported by quantitative analyses. 

For quantitative statistical analyses, it is usually computationally easier to consider the matter power spectrum directly. The signatures of interacting DM on the matter power spectrum has been extensively studied in Ref.~\cite{CyrRacine:2012fz} and we only review them briefly here. First, the presence of the DAO scale in PIDM models generally appears as extra oscillations in the matter power spectrum on scales with $k> k_{\rm DAO}\sim\pi/r_{\rm DAO}$ . Second, just as the correlation function is suppressed on small-scales due to acoustic damping in the dark plasma, the matter power spectrum displays less power at large wave numbers as compared with an equivalent $\Lambda$CDM model.

\begin{figure}[t!]
\begin{centering}
\includegraphics[width=0.5\textwidth]{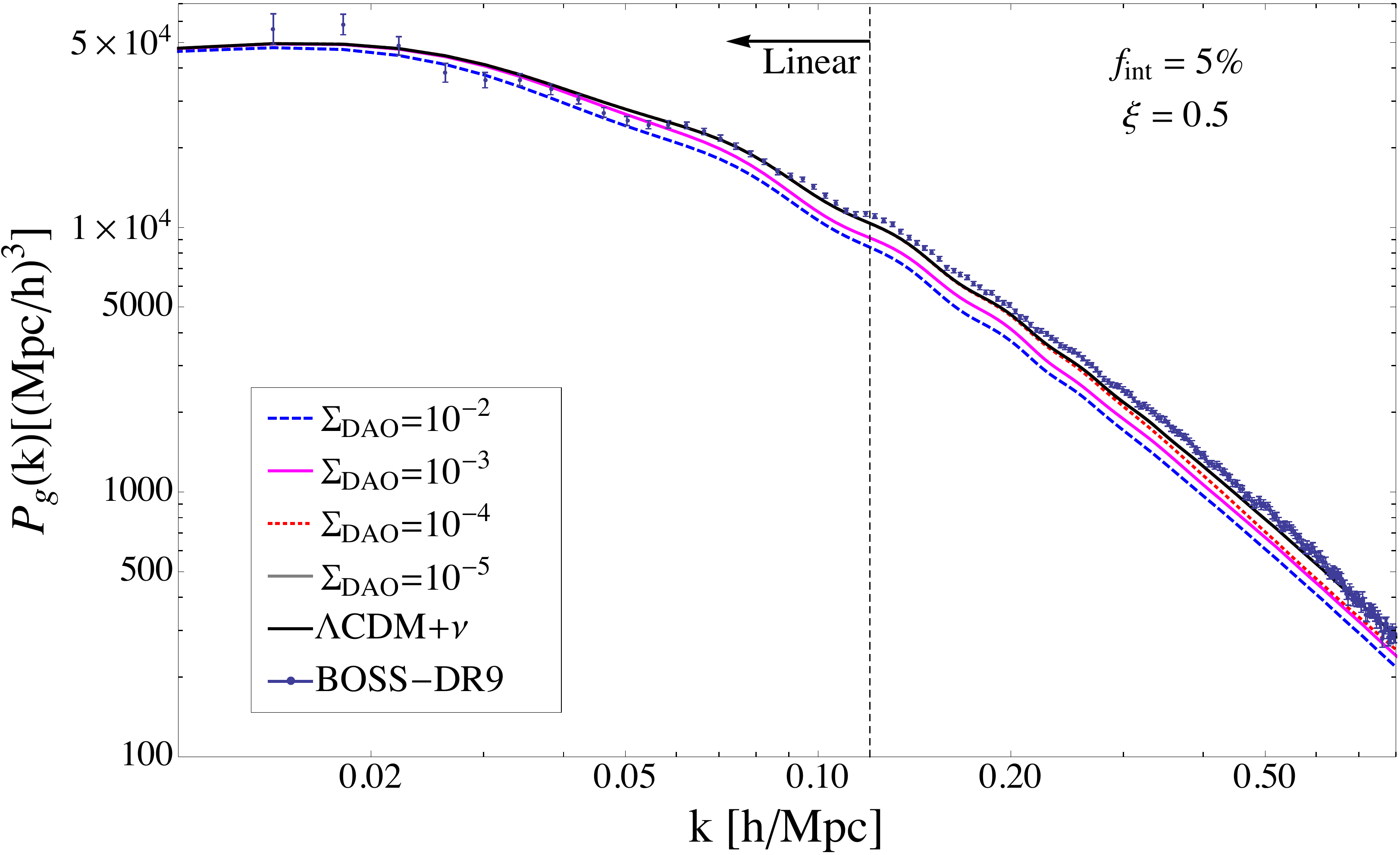}
\includegraphics[width=0.5\textwidth]{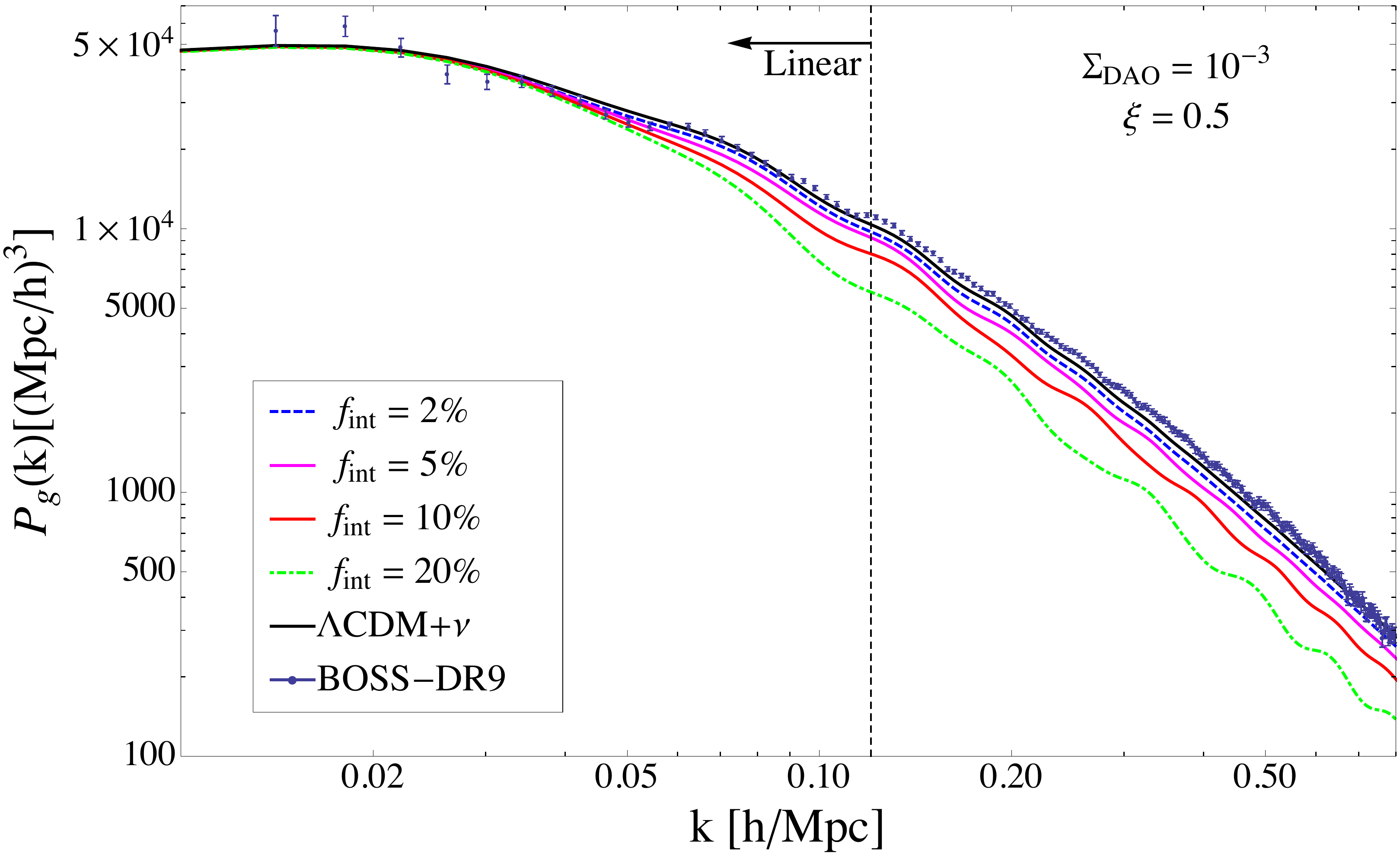}
\caption{Linear galaxy power spectra for different PIDM models. In the upper panel, we fix $f_{\rm int}=5\%$, $\xi=0.5$ and vary $\Sigma_{\rm DAO}$. The lower panel uses $\Sigma_{\rm DAO}=10^{-3}$ and $\xi=0.5$ but let the fraction of interacting DM vary. To compare with galaxy power spectrum from the CMASS catalogue, we have convolved our linear matter power spectra with the BOSS window function and multiplied the results by a scale-independent galaxy bias $b=2.01$ (see section \ref{sec:pk_data} for more details).  For comparison, we also show a standard $\Lambda$CDM model with an equivalent number of effective neutrinos. In this work, we focus uniquely on linear scales, which lie to the left of the dashed vertical line on the plot.}
\label{fig:matterpower}
\end{centering}
\end{figure}

In Fig.~\ref{fig:matterpower}, we show the linear galaxy power spectrum for different PIDM models, along with the measured power spectrum from the BOSS-CMASS data~\cite{Anderson:2012ve}. In Section~\ref{sec:pk_data}, we explain how we convert theoretical PIDM matter power spectra to the shown galaxy power spectra, and give more details on the measurement of the BOSS-CMASS power spectrum and the computation of its errors. The upper panel of Fig.~\ref{fig:matterpower} displays how the power spectrum varies as $\Sigma_{\rm DAO}$ changes for the case of only $5\%$ of interacting DM. The lower panel illustrates the variations in the power spectrum as $f_{\rm int}$ changes from $2\%$ to $20\%$ for a fixed $\Sigma_{\rm DAO}=10^{-3}$ (these are the same models as those plotted in the lower panel of Fig.~\ref{fig:correlation}). The most obvious signature of PIDM in these plots is the damping of small-scale power. The actual acoustic oscillations are only clearly visible for models with $f_{\rm int}\gtrsim10\%$, indicating that dark oscillations are probably better illustrated through the DAO scale in the correlation function for models with small interacting DM fraction (see Fig.~\ref{fig:correlation}).

For the purpose of this work we limit our analysis to linear scales and avoid modeling the small and very large scales, where the galaxy clustering needs to include corrections due to non-linearities~\cite{Taruya:2009bs, Taruya:2010ij, Kwan:2011fu, Jennings:2012fv, Torre:2012dz} and large-scale effects~\cite{Raccanelli:2010fk, Yoo:2010cr, Challinor:2011nx, Bonvin:2011oq, Bertacca:2012hc, Yoo:2012dq, Yoo:2012bh, Montanari:2012kl, Raccanelli:2013tg}, respectively. 
Our constraints on PIDM models using measurements of the galaxy power spectrum will be presented in section \ref{results}.
%
%%%%
\subsection{Cosmic Microwave Background}\label{section:CMB}
%%%%
The CMB probes cosmological fluctuations 380,000 years after the big bang. At that epoch, DM accounts for about $65\%$ of the energy budget of the Universe, hence making the CMB a particularly good probe of nonstandard DM physics. The PIDM scenario affects the CMB in three different ways. First, the presence of extra DR mimics the presence of extra neutrino species and affects the expansion history of the Universe, possibly modifying the epoch of matter-radiation equality, the CMB Silk damping tail, and the early Integrated Sachs-Wolfe effect. However, unlike standard free-streaming neutrinos, the DR forms a tightly-coupled fluid at early times, leading to distinct signatures on CMB fluctuations (see e.g. Ref.~\cite{Cyr-Racine:2013jua}). Second, the DR pressure prohibits the growth of interacting DM fluctuations on length scales entering the causal horizon before the epoch of DM kinematic decoupling. This weakens the depth of gravitational potential fluctuations on these scales, hence affecting the source term of CMB temperature fluctuations. Finally, as discussed in the previous subsection, the modified matter clustering in the Universe due to nonstandard DM properties will affect the lensing of the CMB as it travel from the last-scattering surface to us. We briefly review these signatures below but refer the reader to Ref.~\cite{CyrRacine:2012fz} for more detail. CMB lensing will be covered in the next subsection.

The impact of extra free-streaming relativistic species on the CMB has been well-studied in the literature \cite{Bashinsky:2003tk,Hou:2011ec}. In the following, we focus on the CMB signatures that arise due the modified evolution of cosmological fluctuations. These are sensitive to the actual physical properties of the relativistic species, hence allowing one to discriminate between, say, tightly-coupled DR or extra free-streaming neutrinos. In the standard $\Lambda$CDM universe, free-streaming neutrinos can establish gravitational potential perturbations beyond the sound horizon of the photon-baryon plasma, leading to a suppression of photon fluctuations as well as inducing a shift to their phase shorty after horizon entry. As shown in \cite{Bashinsky:2003tk}, these suppressions and phase shifts are directly proportional to the free-streaming fraction of the total amount of radiation in the Universe. In the PIDM scenario, this fraction is altered by the presence of a DR component that is tightly-coupled to the interacting DM at early times. This results in the CMB displaying a higher amplitude and having its peak structure shifted toward smaller scales (higher $l$) for multipoles that receive contributions from scales entering the horizon before the epoch of dark kinematic decoupling. On the other hand, CMB multipoles receiving most of their contributions from scales entering the horizon well after dark kinetic decoupling should be essentially indistinguishable from a $\Lambda$CDM model having an equivalent effective number of free-streaming species.

In addition to horizon-entry effects, the delayed onset of DR free-streaming as compared to neutrinos generally results in a modified evolution of sub-horizon shear stress perturbations. The modified anisotropic stress will in turn affect the amplitude of the photon quadrupole moment ($F_{\gamma2}$), leading to changes in the CMB temperature and polarization spectra. However, this effect can only be important if the impact on the photon quadrupole is significant near the peak of the CMB visibility function. We thus expect it to be largest for PIDM models with a kinematic decoupling epoch near or after the time of CMB last scattering. Indeed, in theses scenarios, the sudden growth of anisotropic stress due to the onset of DR free-streaming will modify the photon quadrupole just as the primary anisotropies are imprinted on the CMB. Similarly, the relative absence (as compared with an equivalent free-streaming neutrino model) of anisotropic stress near the peak of the visibility function in models with late kinetic decoupling will also result in a modified photon quadrupole, hence leading to different CMB anisotropies. The magnitude and sign of this effect depends on the relative phase between the photon quadrupole and the shear stress perturbation. Since the CMB polarization signal is very sensitive to the photon quadrupole moment near the epoch of last scattering, we expect this effect to be most important for the  polarization power spectrum.
\begin{figure}[t!]
\begin{centering}
\includegraphics[width=0.5\textwidth]{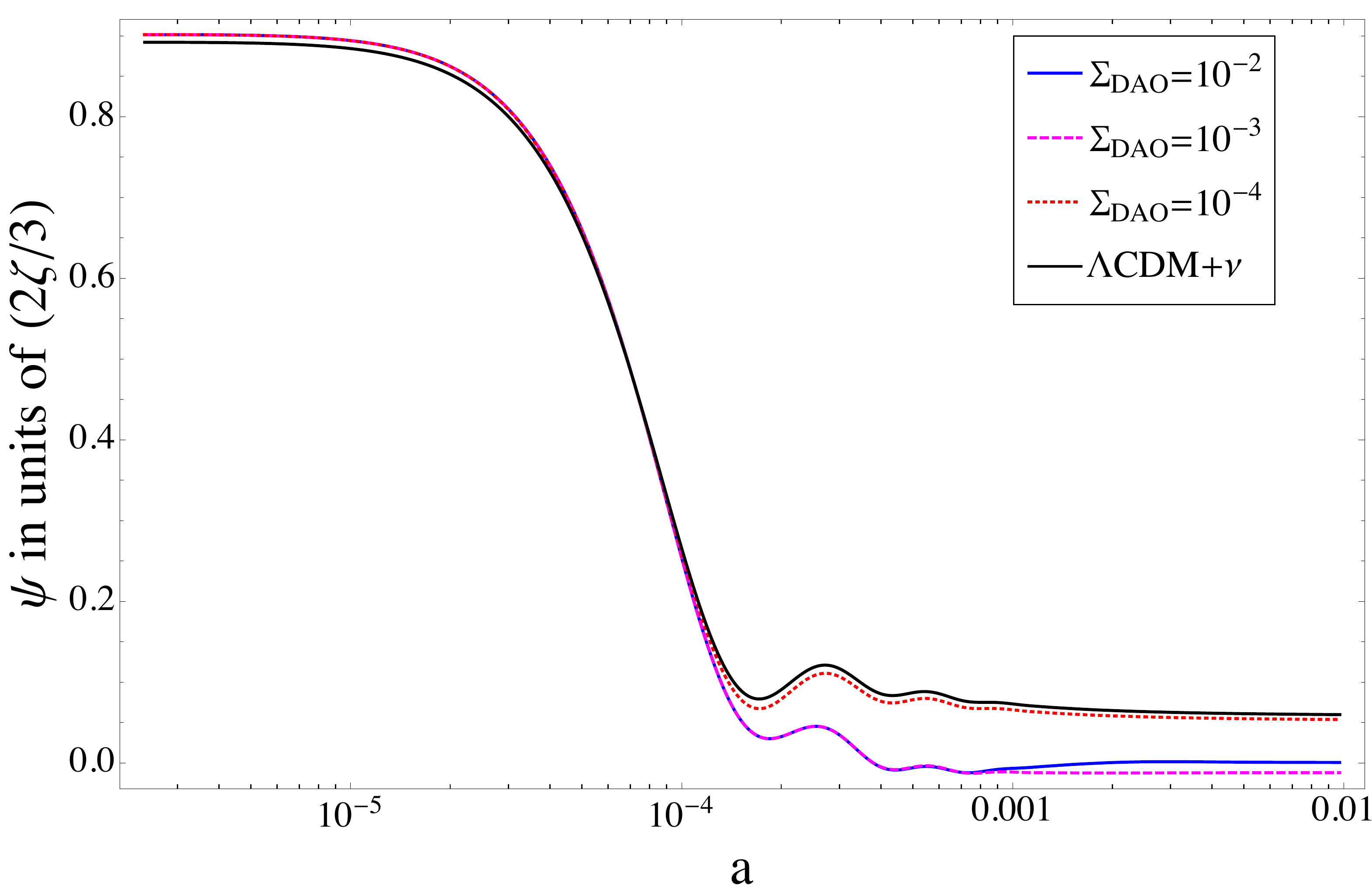}
\includegraphics[width=0.5\textwidth]{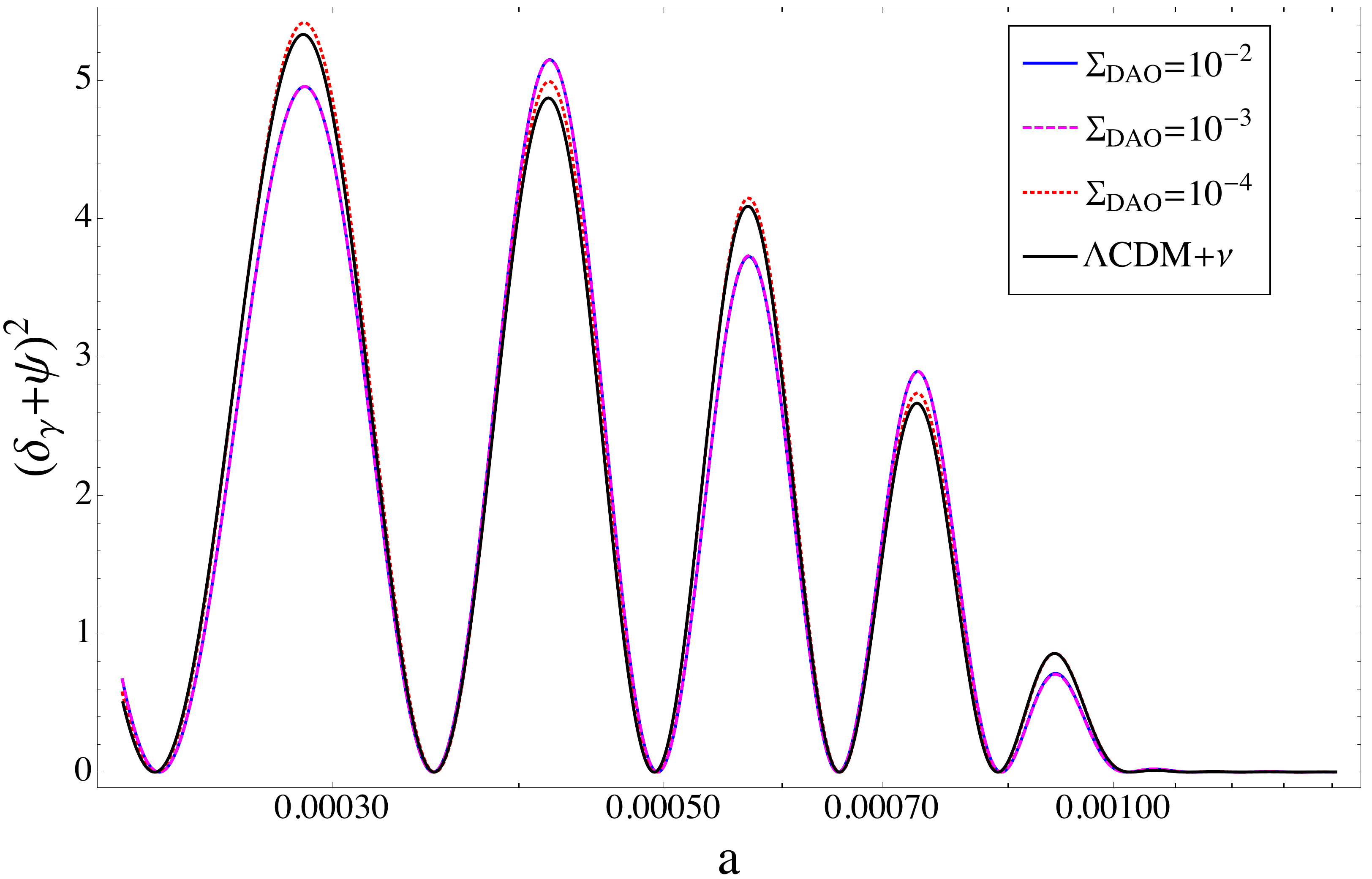}
\caption{{\it Upper panel:} Time-evolution of the gravitational potential $\psi$ for different values of $\Sigma_{\rm DAO}$. Here, $\psi$ is given in units of $2\zeta/3$, where $\zeta$ is the curvature perturbation on constant density hypersurfaces, which is conserved on super-horizon scale for pure adiabatic fluctuations. Before horizon entry, we have $\psi\rightarrow (2\zeta/3)(1+\frac{4}{15}R_{\rm f-s})^{-1}$, where $R_{\rm f-s}$ is the free-streaming fraction of the total radiation content of the Universe. {\it Lower panel:} Monopole source term for the CMB temperature anisotropies.  For both panels, we have taken $k = 0.15$ Mpc$^{-1}$, $\xi=0.5$, and $f_{\rm int}=100\%$. For comparison, we also show a standard $\Lambda$CDM model with an equivalent number of effective neutrinos. }
\label{fig:delta_gam+psi}
\end{centering}
\end{figure}

\begin{figure}[t!]
\begin{centering}
\includegraphics[width=0.49\textwidth]{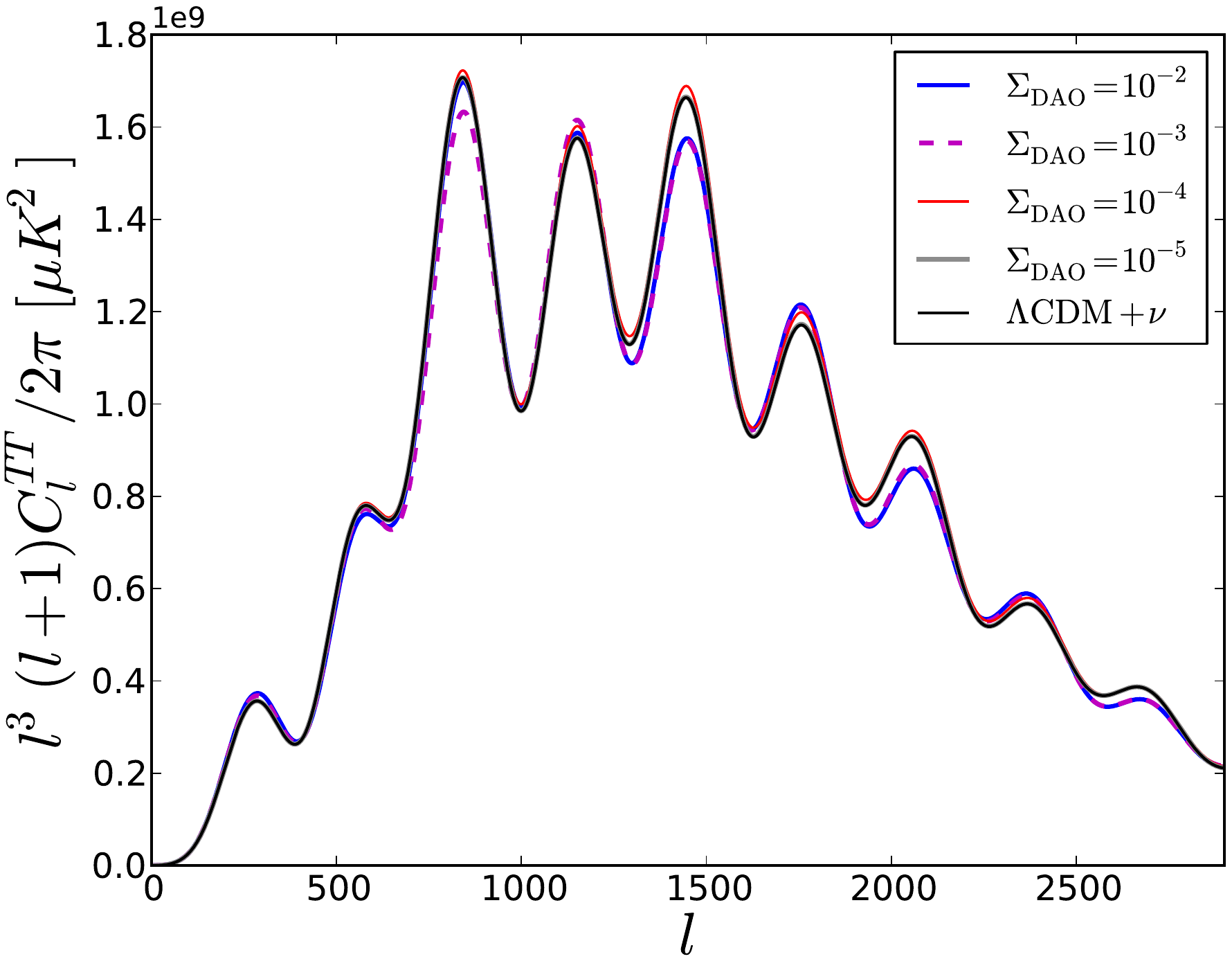}
\includegraphics[width=0.49\textwidth]{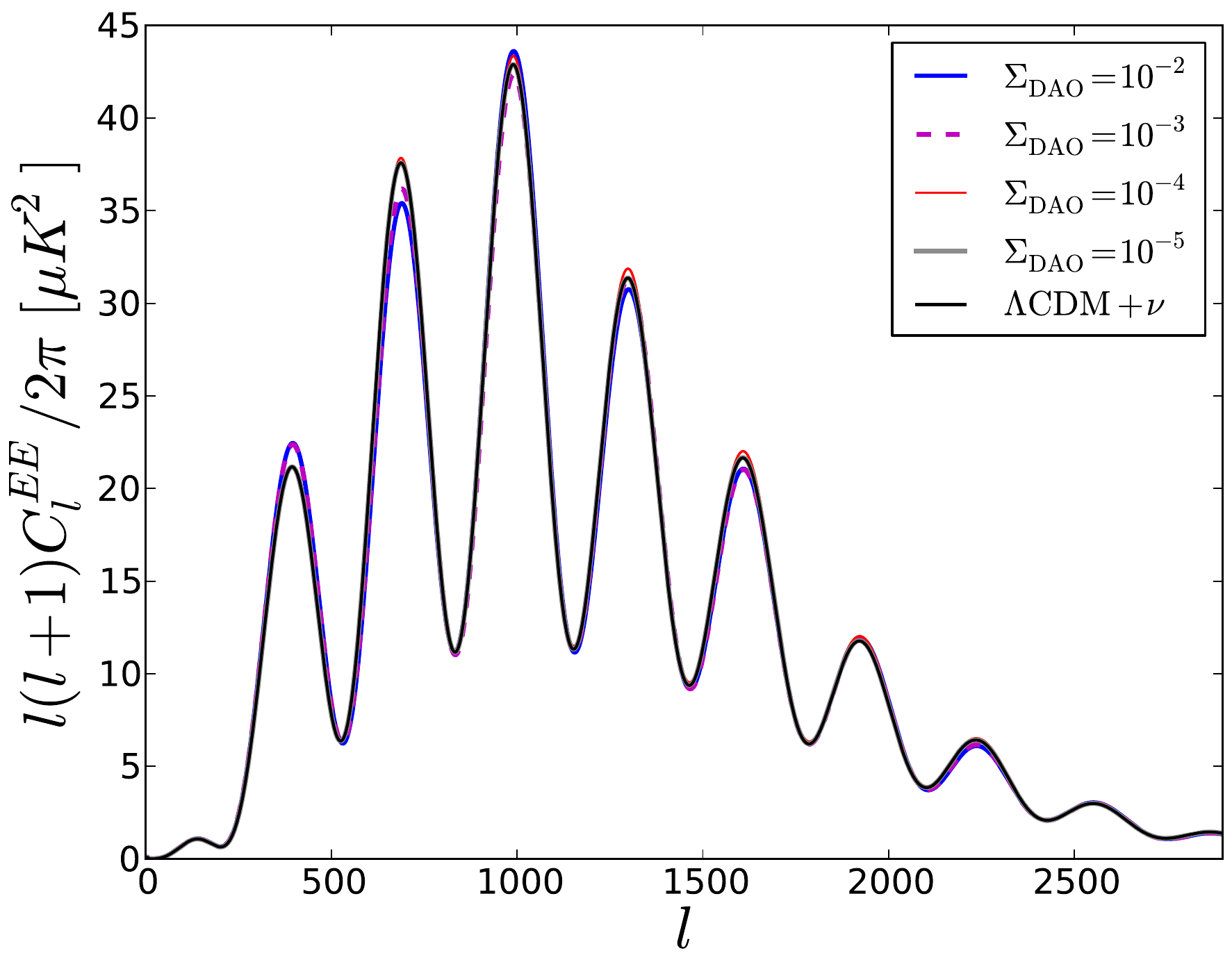}
\caption{CMB unlensed temperature (upper panel) and E polarization (lower panel) power spectra for four different PIDM models with $f_{\rm int}=100\%$. We have taken $\xi=0.5$. For comparison, we also show a standard $\Lambda$CDM model with an equivalent number of effective neutrinos. }
\label{fig:cmb_nolens}
\end{centering}
\end{figure}

Before interacting DM kinematically decouples from the DR, fluctuations in the former cannot grow, leading to substantially shallower gravitational potentials in the matter-dominated era on sub-horizon scales (see upper panel of Fig.~\ref{fig:delta_gam+psi}). Physically, since perturbations in the photon-baryon plasma essentially obey a harmonic oscillator equation driven by the force of gravity, this amounts to a severe damping of the driving term\footnote{Mathematically, this can be understood as a damping of the particular solution to the differential equation governing perturbations in the photon-baryon plasma, while leaving unchanged its homogeneous solution.}. This has for consequences of weakening the gravitationally-induced compression phase of the acoustic oscillations (corresponding to odd $C_l^{\rm TT}$ peaks) while strengthening the pressure-supported expansion phase (corresponding to even $C_l^{\rm TT}$ peaks) of the oscillations (see Ref.~\cite{Foot:2012ai} for a similar effect in a different context). In lower panel of Fig.~\ref{fig:delta_gam+psi}, we illustrate the time-evolution of the monopole source term of CMB temperature fluctuations $(\delta_{\gamma}+\psi)^2$ for different values of $\Sigma_{\rm DAO}$. The enhancement of the expansion peaks and the suppression of the compression peaks are clearly visible for models with $\Sigma_{\rm DAO}\gtrsim10^{-3}$. The effect is also clearly visible at $l\gtrsim600$ in the $C_l^{\rm TT}$ spectra shown in Fig.~\ref{fig:cmb_nolens} for $\Sigma_{\rm DAO}=10^{-3}$.
\begin{figure}[]
\begin{centering}
\includegraphics[width=0.5\textwidth]{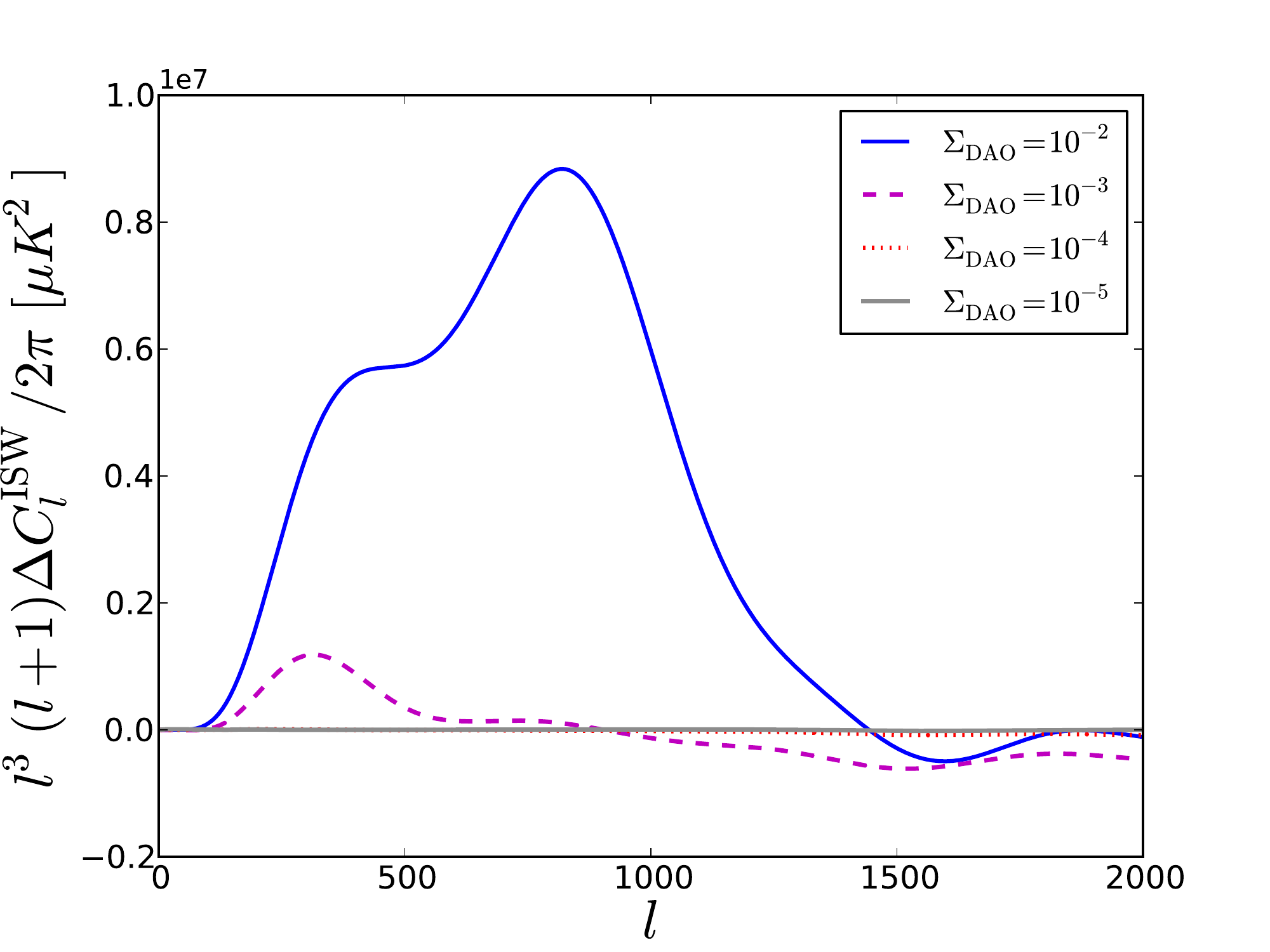}
\includegraphics[width=0.5\textwidth]{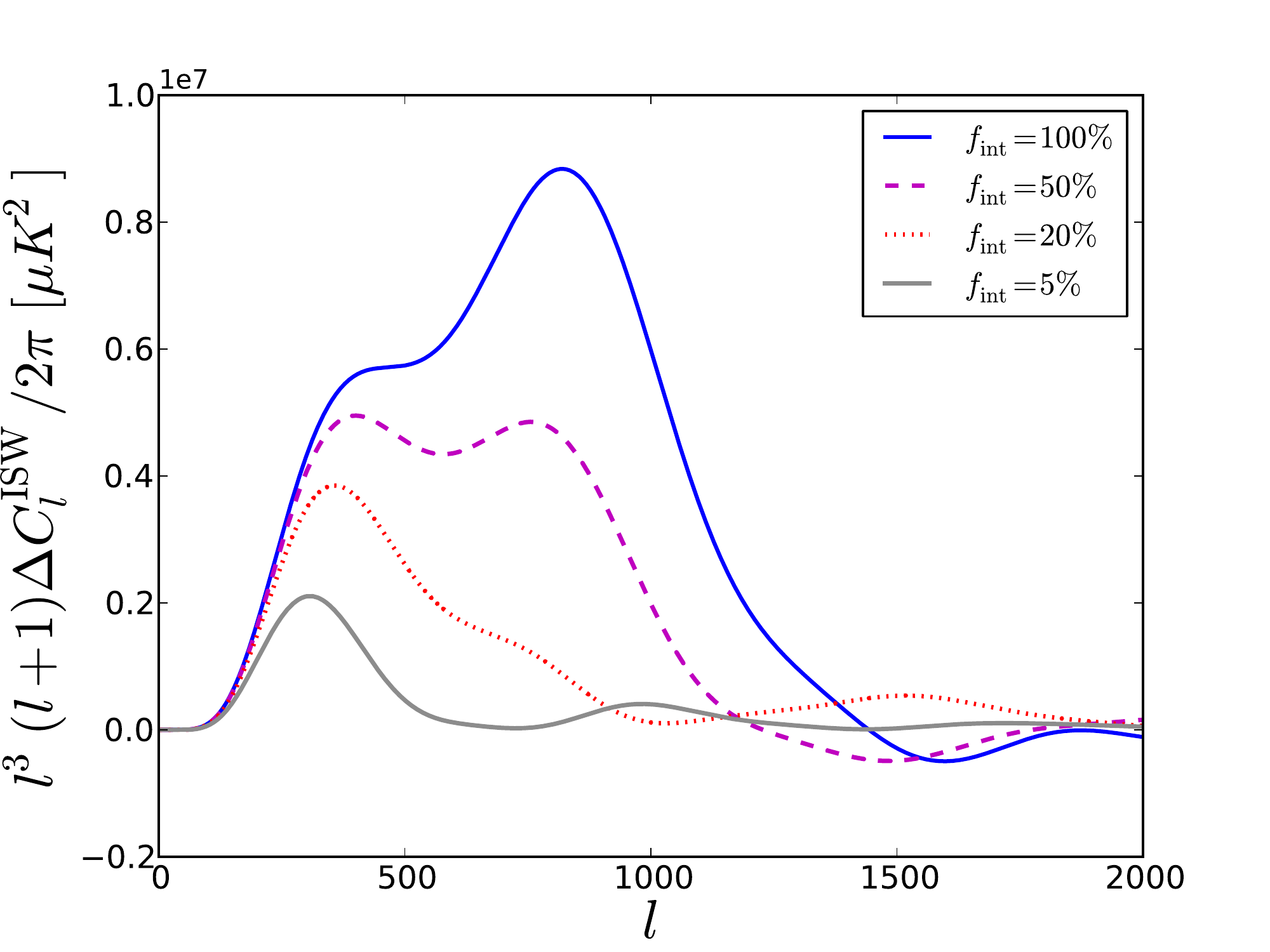}
\caption{Difference between PIDM models and a $\Lambda$CDM model with an equivalent number of neutrinos for the pure ISW contribution to $C_l^{\rm TT}$, $\Delta C_l^{\rm ISW} = C_l^{\rm ISW,PIDM}-C_l^{\rm ISW,CDM}$. In the upper panel, we take $\xi=0.5$, $f_{\rm int}=100\%$, and vary $\Sigma_{\rm DAO}$. In the lower panel, we take $\xi=0.5$, $\Sigma_{\rm DAO}=10^{-2}$, but let $f_{\rm int}$ vary.}
\label{fig:ISW}
\end{centering}
\end{figure}

For PIDM models that kinematically decouple after the last scattering of CMB photons, the continuous decay of the gravitational potential due to the DR pressure leads also to an enhanced early integrated Sachs-Wolfe effect (ISW). This is illustrated in Fig.~\ref{fig:ISW} where we show the difference in the pure early ISW contribution (that is, the ISW-ISW autocorrelation) to $C_l^{\rm TT}$ between a few PIDM models and a $\Lambda$CDM model with an equivalent number of neutrinos. We see that the temperature anisotropies of models with $\Sigma_{\rm DAO} > 10^{-3}$ can receive a large positive contribution from the early ISW effect for $200 \lesssim l\lesssim1200$. At larger multipoles however, the early ISW effect in PIDM models can actually be weaker than that of an equivalent $\Lambda$CDM model, although this effect is much less relevant since the ISW effect is subdominant on those scales.  We note that the actual impact on the temperature anisotropies can be larger than illustrated here since the ISW contribution adds in phase with the photon monopole source term. In the lower panel of Fig.~\ref{fig:ISW}, we observe that the impact of PIDM on the early ISW effect can be fairly large even if the fraction of interacting DM is subdominant.

Putting all these effects together, we obtain the unlensed temperature and polarization power spectra shown in Fig.~\ref{fig:cmb_nolens} where we have taken $f_{\rm int}=100\%$ to magnify the impact of PIDM on the CMB. First, we note that models with $\Sigma_{\rm DAO}\lesssim10^{-5}$ are essentially undistinguishable from a $\Lambda$CDM model with an equivalent number of neutrinos. As $\Sigma_{\rm DAO}$ is increased to $10^{-4}$, both temperature and polarization spectra begin to display the rise in amplitude and the phase shift associated with the DR being tightly-coupled at early times. If $\Sigma_{\rm DAO}$ is further increased to $10^{-3}$, the damping of the gravitational potential perturbations on small scales leads to the suppression of the odd peaks and the enhancement of the even peaks for the temperature anisotropies, which is clearly visible for $l\gtrsim600$. The early ISW also enhances the first acoustic peak of the temperature spectrum for this model. For the polarization spectrum, the impact of the DR shear stress on the photon quadrupole moment leads to a fairly complex variation, with some peak being enhanced, while some are suppressed as compared with an equivalent $\Lambda$CDM model.  Finally, as $\Sigma_{\rm DAO}$ is increased to $10^{-2}$, all the physical effects discussed above are present, with the early ISW leading to strong enhancement of the third temperature peak, which is almost sufficient to offset the suppression caused by the weak gravitational driving force. The early ISW also enhances the first temperature peak while the second peak is somewhat suppressed due the DR shear stress, which also affect the polarization spectrum in a non-trivial way. 

In summary, as illustrated in Fig.~\ref{fig:cmb_nolens}, PIDM models can lead to rich CMB signatures that cannot be easily mimicked by varying other standard cosmological parameters. This makes the CMB an excellent probe of interacting DM and DR physics. As we will see in Section \ref{results}, the latest CMB data does indeed provide strong constraints on PIDM models.    
%
%%%%
\subsection{CMB Lensing}
%%%%
As the CMB photons free-stream from the last-scattering surface to us, they encounter large DM structures which can deflect their path and rotate their polarization state. This CMB lensing (see \cite{LewChal06} for a review) by foreground matter structures has now been detected at high statistical significance ($\sim25\sigma$, \cite{planck17_2013}) and can be used to study the distribution of matter throughout the Universe. Since PIDM models generally predict a modified matter distribution as compared to a pure CDM model, CMB lensing can by itself provide useful constraints on interacting DM scenarios. 

The gravitational deflection potential $\phi$, of which the gradient gives the lensing displacement vector on the sky,
is related to the gravitational potential perturbation $\psi$ projected along the line of sight in the ${\bf \hat{n}}$ direction, via
\be
\phi({\bf \hat{n}}) = -2\int_0^{\chi_*}d\chi\,\psi(\chi {\bf \hat{n}};\eta_0-\chi)\frac{\chi_*-\chi}{\chi\chi_*},
\ee
where $\chi_*$ is the comoving distance to the last scattering surface and $\eta_0$ is the comoving size of the causal horizon today. The lensing potential power spectrum can be written as
\be
C_l^{\phi\phi} = 16\pi \int\frac{dk}{k}P_{\mathcal{R}}(k) |\Delta_{\psi}(k)|^2,
\ee
where
\be
\Delta_{\psi}(k)= \int_0^{\chi_*}d\chi T_{\psi}(k;\eta_0-\chi))j_l(k\chi)\frac{\chi_*-\chi}{\chi\chi_*},
\ee
and where $P_{\mathcal{R}}(k)$ is the primordial spectrum of comoving curvature fluctuations. The transfer function $T_{\psi}(k,\eta)$ is defined by $\psi(k,\eta) = T_{\psi}(k,\eta)\mathcal{R}(k)$, where $\mathcal{R}(k)$ stands for the comoving curvature fluctuation. 

\begin{figure}[t!]
\begin{centering}
\includegraphics[width=0.5\textwidth]{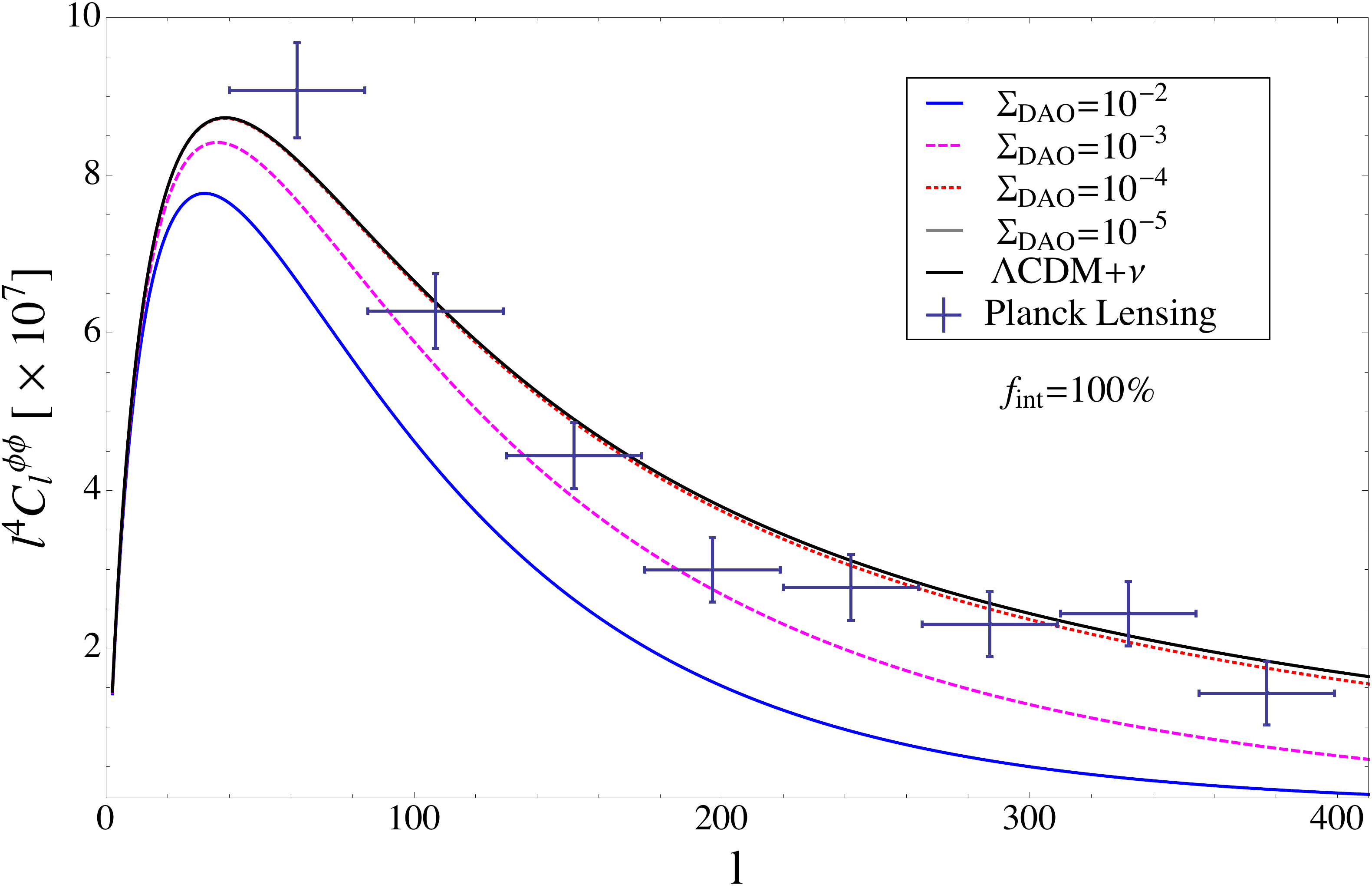}
\includegraphics[width=0.5\textwidth]{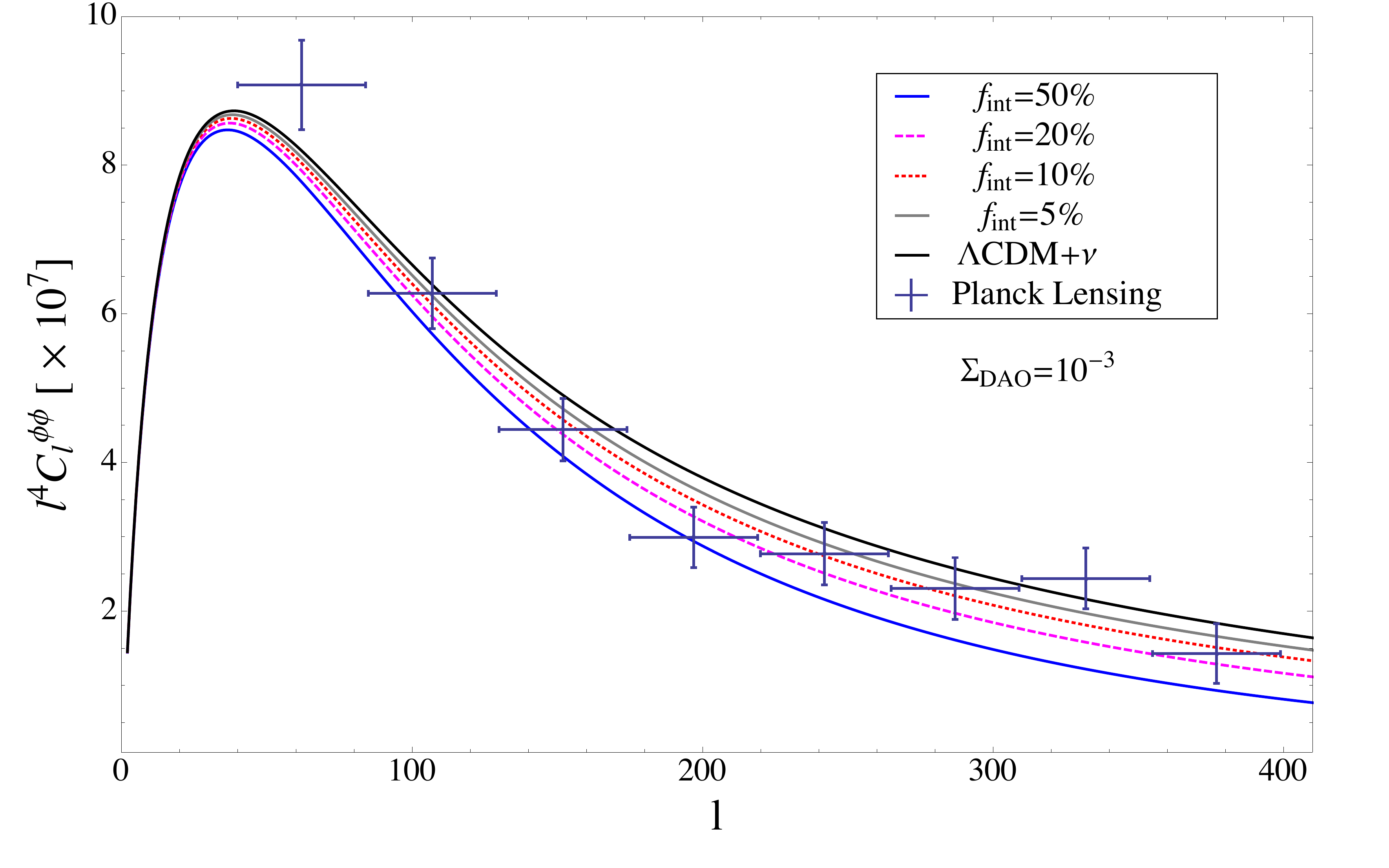}
\caption{CMB lensing power spectrum for different PIDM models. For both panels we use $\xi=0.5$. In the upper panel, we vary $\Sigma_{\rm DAO}$ while leaving $f_{\rm int}=100\%$ fixed. The model with $\Sigma_{\rm DAO}=10^{-5}$ is essentially undistinguishable from the $\Lambda$CDM+$\nu$ model. In the lower panel, we vary $f_{\rm int}$ but leave $\Sigma_{\rm DAO}=10^{-3}$ fixed. We show the eight band powers used in the Planck lensing likelihood.  For comparison, we also show a $\Lambda$CDM model with an equivalent number of neutrinos.}
\label{fig:lensing}
\end{centering}
\end{figure}

We show in Fig.~\ref{fig:lensing} the CMB lensing power spectrum for different PIDM models. In the upper panel, we display the spectra for increasing values of $\Sigma_{\rm DAO}$. It should be clear from this plot that the most extreme models with $\Sigma_{\rm DAO}\gtrsim10^{-3}$ are ruled out by current data if interacting DM forms the totality of the DM. In the lower panel of Fig.~\ref{fig:lensing}, we fix  $\Sigma_{\rm DAO}=10^{-3}$ but instead vary the fraction of interacting DM. We observe that even a fraction as small as $5\%$ can have a sizable effect on the lensing power spectrum. This indicates that current and future CMB lensing measurements could potentially be very sensitive probes of nonstandard DM physics. 

\begin{figure}[b!]
\begin{centering}
\includegraphics[width=0.5\textwidth]{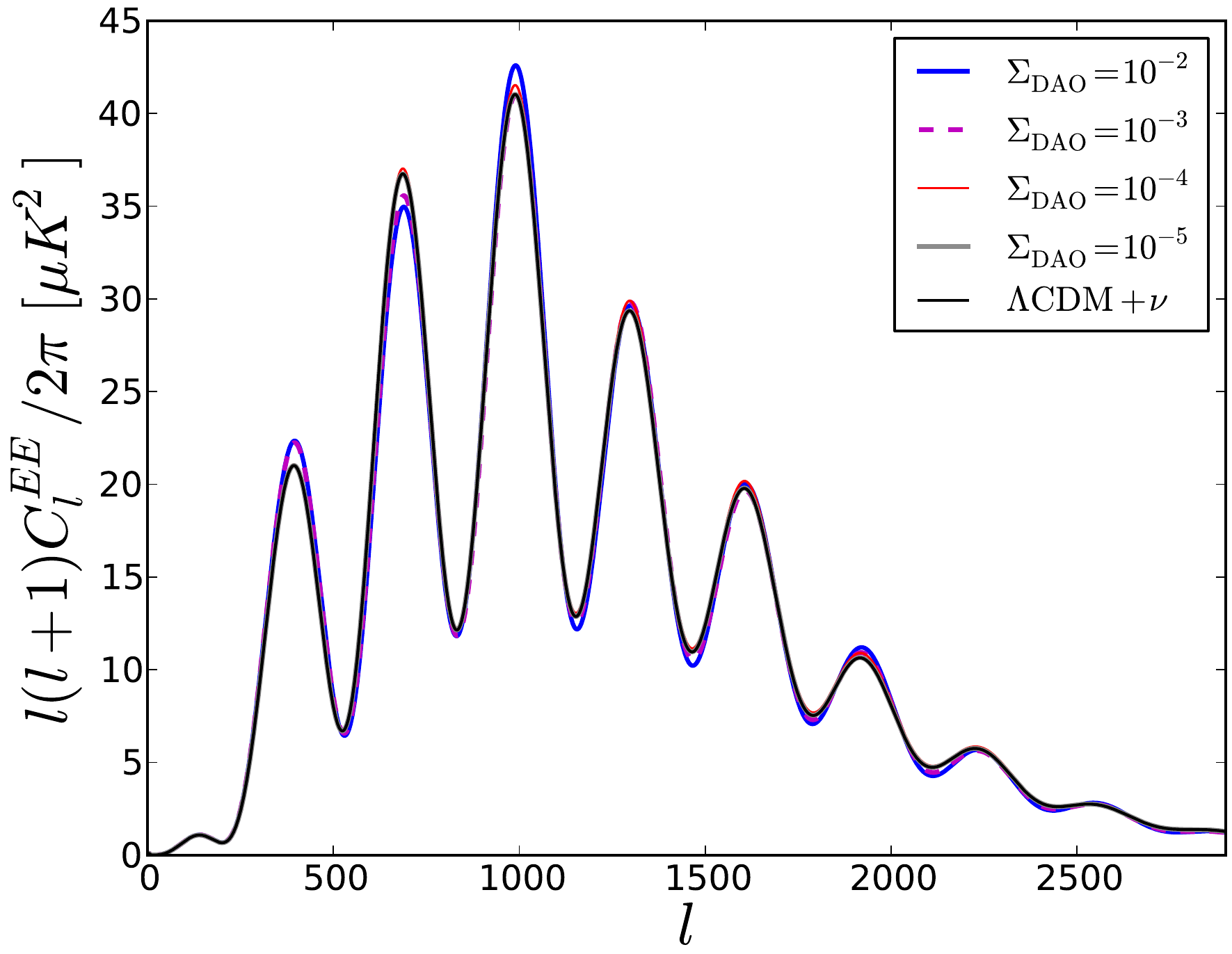}
\caption{CMB E polarization power spectrum for four different PIDM models with $f_{\rm int}=100\%$. and $\xi=0.5$.  For comparison, we also show a standard $\Lambda$CDM model with an equivalent number of effective neutrinos. }
\label{fig:cmbEE}
\end{centering}
\end{figure}

\begin{figure*}[]
\begin{centering}
\includegraphics[width=\textwidth]{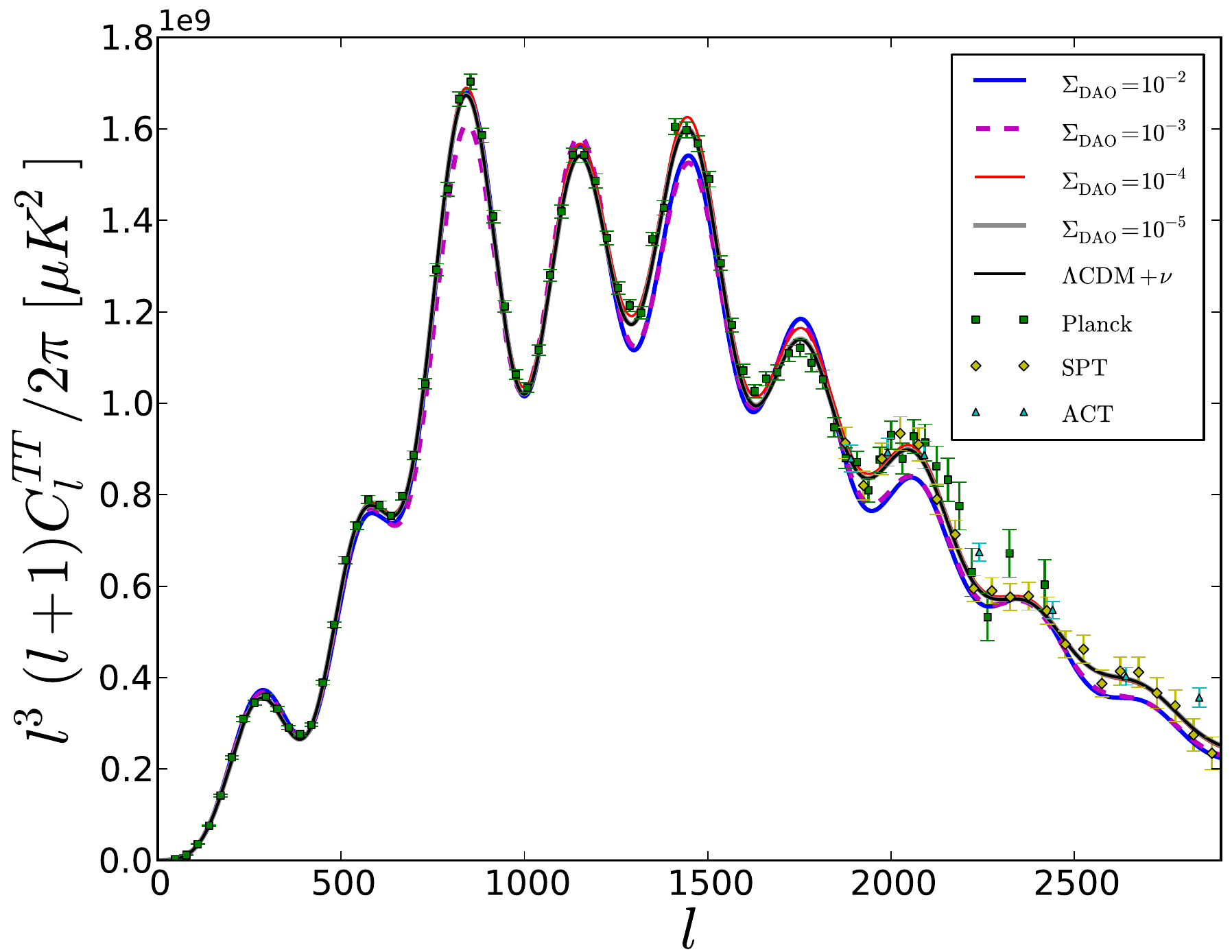}
\caption{CMB temperature power spectrum for four different PIDM models with $f_{\rm int}=100\%$. and $\xi=0.5$. We also show the Planck \cite{planckXVI}, SPT \cite{Story:2012wx}  and ACT  \cite{Dunkley:2013vu,Das:2013zf} band powers.  For comparison, we also show a standard $\Lambda$CDM model with an equivalent number of effective neutrinos.  }
\label{fig:cmbTT}
\end{centering}
\end{figure*}

Lensing by foreground matter structure also distorts the CMB temperature and polarization power spectra presented in Fig.~\ref{fig:cmb_nolens} above. Essentially, lensing acts to smooth out the oscillatory structure of the spectra, filling in the troughs and damping the peaks. As we discussed above, since PIDM models generally predict different amount of lensing, the associated smoothing of the CMB spectra provides yet another handle (albeit correlated with other CMB signatures) to constrain interacting DM. We illustrate lensed CMB spectra in Figs.~\ref{fig:cmbEE} and \ref{fig:cmbTT} for increasing values of $\Sigma_{\rm DAO}$ and for $f_{\rm int}=1$. Besides the PIDM signatures discussed in section \ref{section:CMB}, we observe that the TT and EE spectra display sharper peaks and troughs in the damping tail as $\Sigma_{\rm DAO}$ is increased, which is in line with our expectations that these models should be \emph{less} affected by gravitational lensing. We also note that the lensing signatures can obscure some of the effects discussed in section \ref{section:CMB}, especially the enhancement of the even acoustic peaks in the damping tail of the temperature spectrum.

Taken as a whole, it is clear that the CMB and its lensing by foreground matter structures provide an exquisite probe of DM physics and of its possible interaction with new relativistic species. Having described the CMB signatures predicted by PIDM models and the physics behind them, we now turn our attention to the data and what they can tell us about the physics of DM.
%

%%%%
\section{Data}\label{data}
%%%%

%%%
\subsection{Cosmic Microwave Background and Lensing}
%%%
To constrain interacting DM, we use the CMB data from the Planck satellite \cite{planckXVI}. We utilize both the low-multipole and high-multipole temperature data, incorporating the required ``nuisance'' parameters describing foregrounds and instrumental effects, and also include the WMAP low-$l$ polarization data.  We refer to this dataset as ``Planck+WP''. We also incorporate the high-resolution temperature data from the South Pole Telescope (SPT) and the Atacama Cosmology Telescope (ACT). As in the Planck analysis, we only include the ACT $148\!\times\!148$ spectra for $l\!\geq\!1000$, the ACT $148\!\times\!218$ and $218\!\times\!218$ spectra for $l\!\geq\!1500$ \cite{Dunkley:2013vu,Das:2013zf}, and the SPT data described in \cite{Story:2012wx} for $l\!\geq\!2000$. We fully incorporate the nuisance parameters describing foregrounds and calibration uncertainties for both SPT and ACT. We collectively refer to this dataset as ``High-$l$''.

Our likelihood also makes use of the measurement of the CMB lensing potential power spectrum
by the Planck collaboration \cite{planck17_2013}. This data set
consists of eight bandpower estimates of the lensing power spectrum,
covering the multipole range $l = 40 - 400$, as shown in Fig.~\ref{fig:lensing}.
For a fixed realization of the lensing potential,
CMB lensing causes correlations between different Fourier modes of the CMB temperature field.
The lensing potential can therefore be reconstructed
by averaging over products of pairs of temperature modes (see e.g.~Ref.~\cite{okahu03}).
The measurement of the power spectrum of this reconstructed lensing potential
is thus a temperature {\it trispectrum} measurement and is non-zero because when averaged over realizations
of the lensing potential, the lensed temperature field is non-Gaussian. We refer to the Planck lensing dataset as ``Lens''.
%%%%
\subsection{Baryon Acoustic Oscillation}
%%%%

We also include in our analysis baryon acoustic oscillation (BAO) data from a reanalysis
of the Sloan Digital Sky Survey DR7 \cite{Padmanabhan:2012hf}, from the 6-degree Field
survey \cite{Beutler:2011hx}, and from the Baryon Oscillation Spectroscopic Survey \cite{Anderson:2012ve}. 
To measure the BAO scale from the galaxy power spectrum (or correlation function),
typically a fit is performed to a template spectrum based on a $\Lambda$CDM fiducial
cosmology. The template is allowed to shift in the horizontal direction (corresponding
to a dilation of distances) and the best fit dilation parameter
gives a measurement of the ratio of the true BAO scale over the true distance to the galaxy sample, relative
to this ratio in the fiducial cosmology. To account for modeling systematics,
modifications of the broadband shape and amplitude of the power spectrum are allowed, and parametrized by nuisance parameters,
which are then marginalized over. This procedure is based on the assumption
that modifications to the template power spectrum are slowly varying, i.e.~that the deviation
of the true spectrum from the model does not have any sharp, or rapidly varying features.

While this assumption is very reasonable when the main systematics come from non-linear corrections, etc.,
to a $\Lambda$CDM power spectrum, one might worry that it breaks down in the presence of
DAO wiggles. For example, if the DAO scale is close to the BAO scale, the BAO measurement procedure may be biased or might even inadvertently pick out the DAO scale instead.
However, it turns out that most of the region of PIDM parameter space where the DAO scale is large and might cause such confusion is ruled out already by the CMB data alone, so that for the parameter space where the BAO measurement has any effect, the standard BAO measurement procedure is appropriate. Only for very small fractions of interacting DM can the DAO scale be close to the BAO scale and not be in tension with CMB data. For these cases, we confirm our results by using the full shape of the galaxy power spectrum which is free from the assumptions made in order to reconstruct the BAO feature. We can thus safely use the BAO measurements given in the literature as a simple set of low-redshift distance priors.

%%%%%%
\subsection{BOSS Galaxy Power Spectrum}
\label{sec:pk_data}
%%%%%%%%

We use the galaxy power spectrum from the Baryon Oscillation Spectroscopic
Survey (BOSS \cite{dawsonetal13}), which is a component of the Sloan Digital Sky Survey (SDSS \cite{yorketal00}),
specifically SDSS-III \cite{sdss3}.
The data set used here is the CMASS sample of luminous galaxies (see e.g.~\cite{whiteetal11,Anderson:2012ve}), released as part of
Data Release 9 \cite{DR9}.
This sample consists of $\sim 264,000$ galaxies in the redshift range $z = 0.43 - 0.7$ (effective redshift
$z_{\rm eff} = 0.57$), covers 3275 deg$^2$ of the sky, and has an effective volume
$V_{\rm eff} =  2.2$ Gpc$^3$.
We quantify galaxy clustering by the angle-averaged power spectrum of this sample,
which has also been used to obtain the strongest measurement to date of the baryon acoustic oscillation (BAO)
scale in \cite{Anderson:2012ve}, to constrain neutrino mass in \cite{Zhao:2012ly},
and to study primordial non-Gaussianity \cite{Ross:2012ys3}.
%Other works have used the correlation function and anisotropic power spectrum
%to derive cosmological constraints, e.g.~\cite{Reid:2012zr, Samushia:2013vn,sanchezetal12}

The power spectrum is measured using the Feldman-Kaiser-Peacock (FKP \cite{FKP}) estimator,
see \cite{reidetal10,Anderson:2012ve} for details.
To calculate constraints on the PIDM model, we use only the wave vector
range $k = 0.03 - 0.12 h/$Mpc. The upper limit $k_{\rm max} = 0.12 h/$Mpc serves to ensure
that the power spectrum can be modeled using linear perturbation theory (this will be discussed in more detail below).

While the lower limit $k_{\rm min} = 0.03 h/$Mpc minimizes its importance,
we include a template to subtract
any spurious clustering signal on large scales
due to systematics that have not been accounted for \cite{arossetal12,Ross:2012ys3},
\begin{equation}
\label{eq:sys}
P_{\rm meas}(k) = P_{\rm meas, w}(k) - S \, P_{\rm sys}(k).
\end{equation}
Here, $P_{\rm meas, w}(k)$ is the observed power spectrum described in \cite{Anderson:2012ve},
using the standard weights that take into account known systematics (including stellar density).
The template $P_{\rm sys}(k)$ is equal to the contribution to the galaxy power spectrum
due to the correlation of observed galaxy density with stellar density (the dominant known systematic on large scales),
if stellar density weights had not been included. It is argued in \cite{arossetal12,Ross:2012ys3} that
the $k$-dependence of the contribution from other systematics will be the same as that of $P_{\rm sys}(k)$,
so that, following these works, we model unknown systematics
by simply treating the amplitude $S$ in Eq.~(\ref{eq:sys}) as a free parameter and marginalizing over it. We restrict $S$ to be in the range from -1 to 1.

The likelihood for a given PIDM cosmology is obtained by comparing the measured power spectrum $P_{\rm meas}(k)$
to a model power spectrum, $P_{\rm model}(k)$, which we describe below.
To start, we model the true galaxy power spectrum by
\begin{equation}
\label{eq:Pg}
P_g(k) = b^2 P_m(k) + P_0.
\end{equation}
Here, $P_m(k)$ is the {\it linear} matter power spectrum, computed using a modified version
of CAMB (see Section \ref{CosmoPert}), and $b$ is a scale-independent, linear galaxy bias.
%$b$ is redshift space bias
The first term on the right hand side of Eq.~(\ref{eq:Pg})
describes the galaxy power spectrum in the linear regime.
While our choice $k_{\rm max} = 0.12 h/$Mpc limits the importance of non-linear corrections, 
we also include a nuisance parameter $P_0$ to model possible deviations from the linear description
due to scale-dependent galaxy bias, and/or due to imperfect shot-noise subtraction.
This simple galaxy bias model is motivated by the halo model \cite{seljak2000,seljak01,schulzwhite06,guziketal07}
and local bias \cite{scherrwein98,colesetal99,saitoetal09} approaches to galaxy clustering.
%Since these studies focus on cold dark matter cosmologies, the motivation for PIDM
%is less clear, and the model (\ref{eq:Pg}) can be considered a purely phenomenological parametrization.

The power spectrum model also needs to take into account how the galaxy power spectrum is estimated from data.
We follow closely the approach of, e.g., \cite{Anderson:2012ve,Zhao:2012ly} here.
First of all, to estimate the power spectrum, cosmic distances were calculated assuming
a fixed fiducial cosmology. To take this into account, we
dilate the wave vector $k$ appearing in the theory power spectrum by
a factor $\alpha = D_{V, {\rm fid}}/D_V$ (see e.g.~\cite{Tegetal2006}),
$k \to \alpha k$,
where $D_{V, {\rm fid}}$ is the volume weighted distance measure to $z_{\rm eff} = 0.57$
in the fiducial cosmology, and $D_V$ is the same quantity in the cosmology in which the
galaxy power spectum is modeled.
The amplitude of the power spectrum should also be rescaled by a factor $\alpha^3$, but we
absorb this shift into the galaxy bias parameters.
Secondly, we account for the effect of the survey geometry
by convolving the model power spectrum with the Fourier transform of the survey window
function \cite{percetal07,arossetal12}.
Schematically, we thus have the following model power spectrum,
\be
P_{\rm model}(k) = W * P_g(\alpha k),
\ee
where $P_g$ was defined in Eq.~(\ref{eq:Pg}).

To calculate the likelihood, we assume the power spectrum estimator follows a Gaussian distribution
and use a covariance matrix based on 600 CMASS mock catalogs \cite{maneraetal13}.
Since the galaxy bias parameter $b^2$ corresponds to an overall
scaling of the model power spectrum (after a trivial redefinition $P_0 \to P_0/b^2$),
we marginalize over it analytically, thus reducing by one the number of free parameters to sample. We shall refer to this dataset as ``DR9''.

%%%%
\section{Results}\label{results}
%%%%
We determine constraints on PIDM models using the publicly available Markov Chain Monte Carlo code \texttt{CosmoMC} \cite{Lewis:2002ah}. In all our chains, we let the six $\Lambda$CDM cosmological parameter vary ($\Omega_{\rm b}h^2$, $\Omega_{\rm DM}h^2$, $\theta_{\rm MC}$, $\tau$, $\ln(10^{10} A_{\rm s})$, and $n_{\rm s}$) in addition to some combinations of the dark parameters specified below. We use uniform priors on the standard cosmological parameters and nuisance parameters similar to those described in \cite{planckXVI}. We assume a flat Universe with massless neutrinos and fix $N_{\rm eff}=3.046$. We compute the helium abundance consistently \cite{Simha:2008zj}, taking into account the baryon density and the extra contributions to the radiation density from DR and dark electrons if the latter are relativistic at big bang nucleosynthesis. To explore the $\Sigma_{\rm DAO}$ constraints and ensure maximum chain mobility, we vary $\log_{10}(\Sigma_{\rm DAO})$ instead of $\Sigma_{\rm DAO}$ itself. The ranges of the uniform priors we use for the dark parameters are listed in Table \ref{table_prior}. To ensure chain convergence, we run 8 independent chains for each dataset combinations and make sure that the Gelman and Rubin criterion is $R-1 \leq 0.02$.
\begin{table}[t!]
\begin{center}
\begin{tabular}{|c|c|}
\hline
Parameters & Prior Range \\
\hline
\hline
$\log_{10}(\Sigma_{\rm DAO})$ & [-6,1] \\
$f_{\rm int}$ & [0,1] \\
$\xi$ & [$10^{-3}$,1] \\
\hline
\end{tabular}
\caption{\label{table_prior} Prior range of the dark parameters used in our analysis.\vspace{-0.7cm}}
\end{center}
\end{table}

Before looking at the quantitative results, it is important to note that PIDM models reduce to a standard CDM model in a few different limits. First, as $\xi$ is driven to small values, interacting DM looses the pressure support that prevents it from clustering and basically behaves like CDM. Second, as $\Sigma_{\rm DAO}$ is reduced, the epoch of DM kinematic decoupling is pushed toward earlier times, hence restricting the impact of interacting DM to small non-linear scales, unobservables with the CMB and (linear) galaxy clustering. Finally, the signatures of PIDM obviously become less and less important as $f_{\rm int}$ is reduced toward zero. The caveat here is that a smaller interacting DM fraction also implies a larger DAO scale, hence potentially bringing the small signatures of PIDM to potentially observable scales. Of course, as $f_{\rm int}\rightarrow0$, we recover the standard CDM observables. Since there are many ways to effectively ``hide'' the interacting DM component, it usually makes more sense to consider the cosmological constraints on specific slices of the DM parameter space. We discuss the most interesting slices in the next few subsections.

In the following subsections, we vary different subsets of the dark parameters to study their joint probability density function.

%%%%
\subsection{ $\Sigma_{\rm DAO}$ + $\xi$ + $f_{\rm int}$}
%%%%
%
\begin{figure}[b]
\begin{centering}
\includegraphics[width=0.5\textwidth]{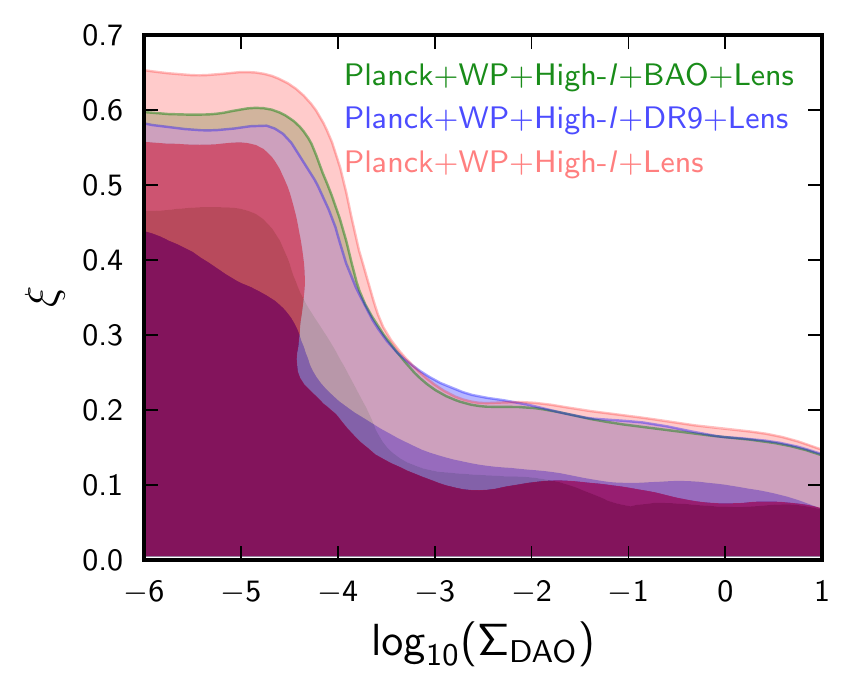}
\caption{Marginalized constraints on $\xi$ and $\Sigma_{\rm DAO}$ for three combination of datasets. Here, the fraction of interacting DM has been marginalized over. We display the $68\%$ and $95\%$ confidence regions.}
\label{GamDAOvsxi_1}
\end{centering}
\end{figure}

For these constraints, we let the trio of dark parameters $\{\Sigma_{\rm DAO},f_{\rm int},\xi\}$ freely vary within the prior ranges listed in Table \ref{table_prior}. Here, we focus on strongly-coupled models and fix $\alpha_D=0.05$ and $m_D=10$ GeV. Constraints for different values of $\alpha_D$ and $m_D$ can be obtained by appropriately rescaling $\Sigma_{\rm DAO}$, as long as $\alpha_D > 0.025$. Constraints for smaller values of the dark fine-structure constant will be discussed in Section \ref{sec:var_alphaD}. In Fig.~\ref{GamDAOvsxi_1}, we illustrate the 2D marginalized posterior in the $\xi-\log_{10}(\Sigma_{\rm DAO})$ plane for three different combinations of datasets.  We note that the two combinations of datasets that make use of galaxy clustering data (here identified as BAO and DR9) yield very similar limits, indicating that most of the constraining power from galaxy power spectrum data comes from the BAO feature. The red contours display the constraints if no galaxy clustering data is included in the analysis. We observe that for $\Sigma_{\rm DAO} > 10^{-3}$ most of the constraining power comes from the CMB data alone (including lensing), while the large-scale structure data act to strengthen the constraints for lower values of $\Sigma_{\rm DAO}$.

In all cases, we observe a sharp regime change around $\Sigma_{\rm DAO}=10^{-4.5}$. For $\Sigma_{\rm DAO}\lesssim10^{-4.5}$, the bound on $\xi$ reads $\xi \lesssim 0.6$ ($95\%$ C.L., Planck+WP+High-$l$+BAO+Lens) which is equivalent to current cosmological constraints on $N_{\rm eff}$ for the datasets shown. This indicates that in this region of parameter space, PIDM essentially behaves like CDM and DR is indistinguishable from extra species of free-streaming neutrinos, in agreement with our discussion from section \ref{observables}. On the other hand, the allowed range of $\xi$ values rapidly shrinks as $\Sigma_{\rm DAO}$ is increased beyond $10^{-4.5}$. The strong constraints on $\xi$ in this region of parameter space have very different origins than the standard constraints on extra relativistic species. Indeed, the suppressed DM fluctuations and the absence of early DR free-streaming for these values of $\Sigma_{\rm DAO}$ have a significant impact on the CMB and the matter power spectrum  which forces $\xi$ to take small values in order to hide these effects. The constraints shown in Fig.~\ref{GamDAOvsxi_1} implies that we must have $\xi<0.2$ for $\Sigma_{\rm DAO} > 10^{-3}$. To give a sense of scale, we note that $\xi=0.2$ corresponds, according to Eq.~\ref{delta_n_eff}, to $\Delta N_{\rm eff}\simeq0.007$, indicating that even such a small amount of DR can dramatically affect the evolution of cosmological perturbations if it couples strongly enough to DM. Since the PIDM models lying inside the $95\%$ confidence region are, for all practical purposes, undistinguishable from a $\Lambda$CDM universe, the fraction of interacting DM is largely unconstrained in the allowed region. Constraints on models lying on the edge of the $95\%$ confidence region will however depend on the value of $f_{\rm int}$ and we explore these limits in the next section.

%%%%
\subsection{$\Sigma_{\rm DAO}$ + $\xi$}
%%%%
%
\begin{figure}[b!]
\begin{centering}
\includegraphics[width=0.5\textwidth]{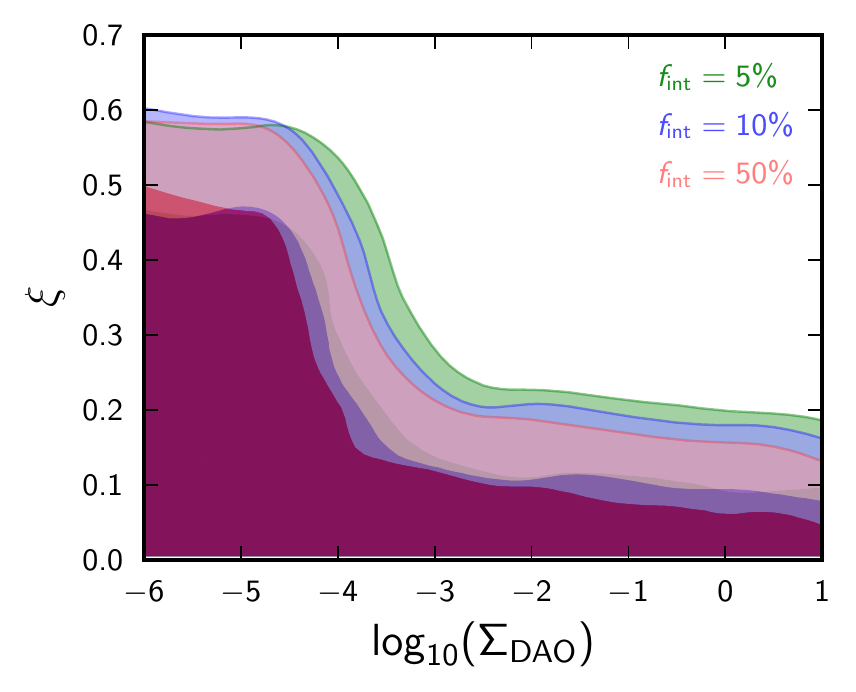}
\caption{Marginalized constraints on $\xi$ and $\Sigma_{\rm DAO}$ for three fixed values of $f_{\rm int}$. We display the $68\%$ and $95\%$ confidence regions for the dataset ``Planck+WP+High-$l$+BAO+Lens''.}
\label{GamDAOvsxi_2}
\end{centering}
\end{figure}
In this section, we keep the fraction of interacting DM fixed while letting $\xi$ and $\log_{10}(\Sigma_{\rm DAO})$ vary, allowing to determine how the constraint contours change as a function of the interacting DM fraction. As before, we fix $\alpha_D=0.05$ and $m_D=10$ GeV. We show in Fig.~\ref{GamDAOvsxi_2} the marginalized constraints for three different values of $f_{\rm int}$, using the dataset Planck+WP+High-$l$+BAO+Lens. While we observe the constraints becoming progressively weaker as $f_{\rm int}$ is reduced, the difference between the $f_{\rm int}=50\%$ and $f_{\rm int}=5\%$ limits is surprisingly modest. This indicates that our constraints are robust to changes in the interaction DM fraction (for $f_{\rm int}\gtrsim5\%$). It also show that it matters little if $f_{\rm int}=5\%$, $50\%$, or $100\%$ in the ruled out regions: there, PIDM affects the cosmological observables in a way that is incompatible with the current data and lowering the fraction of interacting DM only slowly improves the fit. This is in agreement with our discussion of sections \ref{CosmoPert} and \ref{observables} where we showed that shrinking the fraction of interacting DM does reduce the impact on the cosmological observables but at the price of \emph{increasing} the DAO scale and bringing it to linear observable scales.

%%%%
\subsection{Varying the Dark Fine-structure Constant}\label{sec:var_alphaD}
%%%%
%
\begin{figure}[t!]
\begin{centering}
\includegraphics[width=0.5\textwidth]{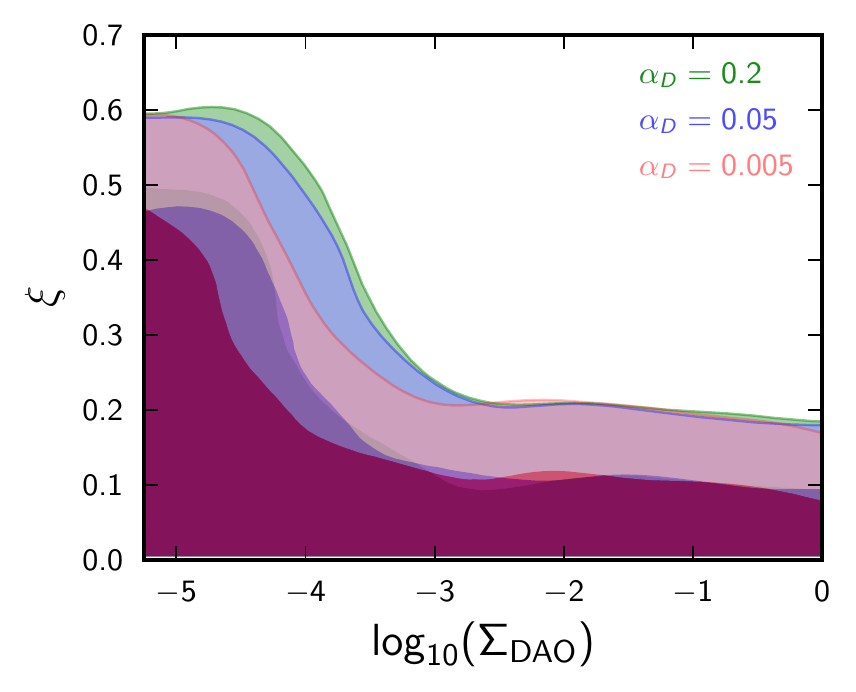}
\caption{Marginalized constraints on $\xi$ and $\Sigma_{\rm DAO}$ for three fixed values of $\alpha_D$. Here, we have fixed $f_{\rm int}=10\%$ and $m_D=10$ GeV. We display the $68\%$ and $95\%$ confidence regions for the dataset ``Planck+WP+High-$l$+BAO+Lens''.}
\label{GamDAOvsxi_3}
\end{centering}
\end{figure}
In this section, we study the effect of varying $\alpha_D$ on the cosmological limits on the $\xi$ and $\Sigma_{\rm DAO}$ parameters. We display the constraints on these two dark parameters in Fig.~\ref{GamDAOvsxi_3} for three values of the dark fine-structure constant. Here, we fix $f_{\rm int}=10\%$ which yields constraints representative of a broad range of interacting DM fraction (see previous section). For $\Sigma_{\rm DAO}> 10^{-2.5}$, we observe that the constraint on $\xi$ is largely independent of $\alpha_D$, indicating that our limits are robust to changes in the dark sector microphysics in that region of parameter space. At smaller values of $\Sigma_{\rm DAO}$, the constraints become \emph{stronger} as $\alpha_D$ is reduced. This somewhat counterintuitive result is a consequence of the definition of $\Sigma_{\rm DAO}$: at fixed $\Sigma_{\rm DAO}$, lowering $\alpha_D$ leads to a smaller values of the atomic binding energy, hence bringing dark recombination and kinematic decoupling closer to the last scattering surface of CMB photons and leading to a larger effects on the cosmological observables. From Fig.~\ref{GamDAOvsxi_3}, we observe that the main impact of varying $\alpha_D$ is to modify the shape of the $\xi$ constraint in the transition region delimiting the parameter space where the limit is similar to the $N_{\rm eff}$ bound (small $\Sigma_{\rm DAO}$) and where it is dominated by PIDM effects (large $\Sigma_{\rm DAO}$).

%%%%
\subsection{$\Sigma_{\rm DAO}$ + $f_{\rm int}$}\label{DAO+f}
%%%%
%
\begin{figure}[]
\begin{centering}
\includegraphics[width=0.5\textwidth]{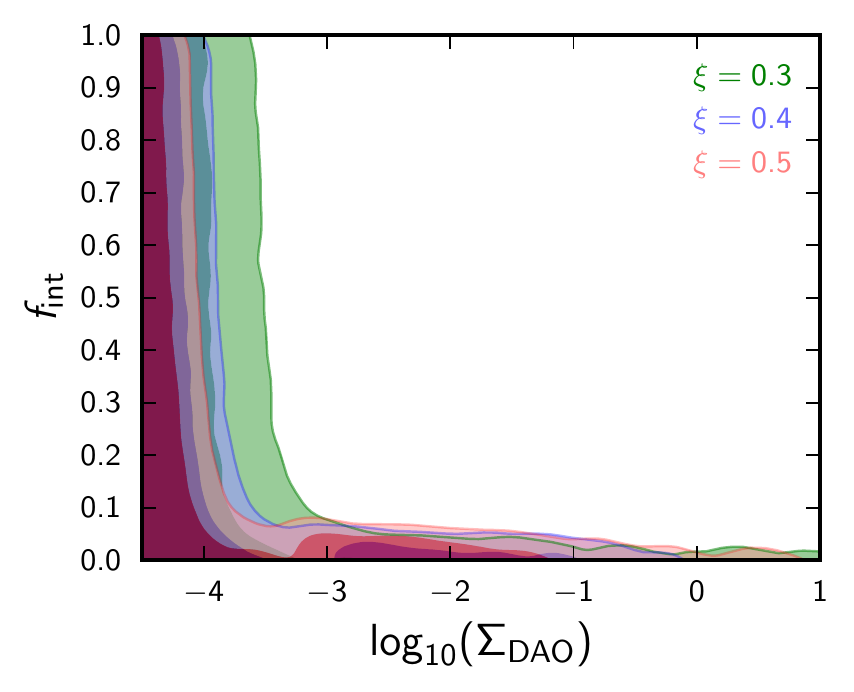}
\caption{Marginalized constraints on $f_{\rm int}$ and $\Sigma_{\rm DAO}$ for three values of $\xi$. Here, we have fixed $\alpha_D=0.05$ and $m_D=10$ GeV. We display the $68\%$ and $95\%$ confidence regions for the dataset ``Planck+WP+High-$l$+BAO+Lens''.}
\label{GamDAOvsxi_4}
\end{centering}
\end{figure}

\begin{figure}[t!]
\begin{centering}
\includegraphics[width=0.5\textwidth]{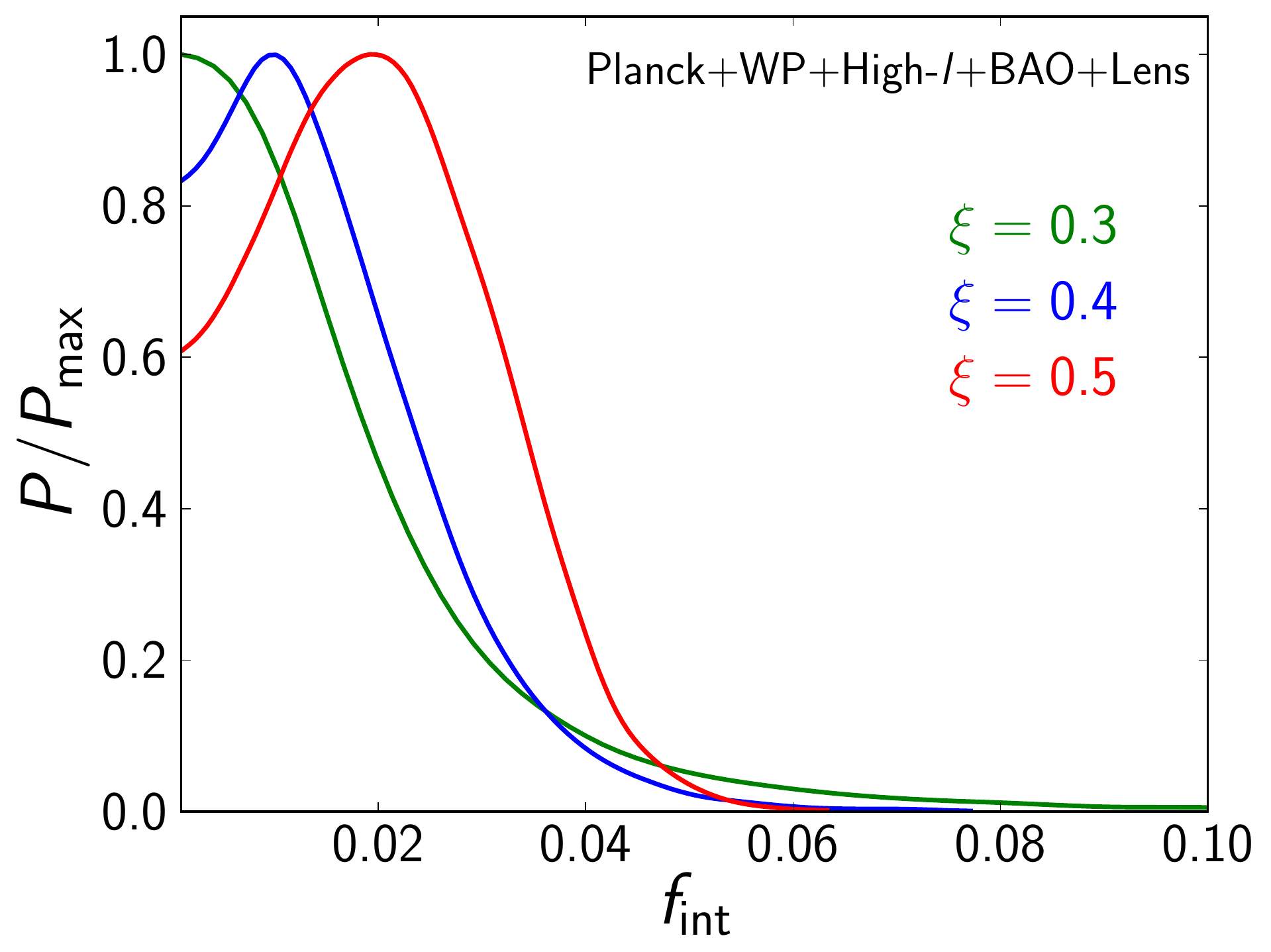}
\includegraphics[width=0.5\textwidth]{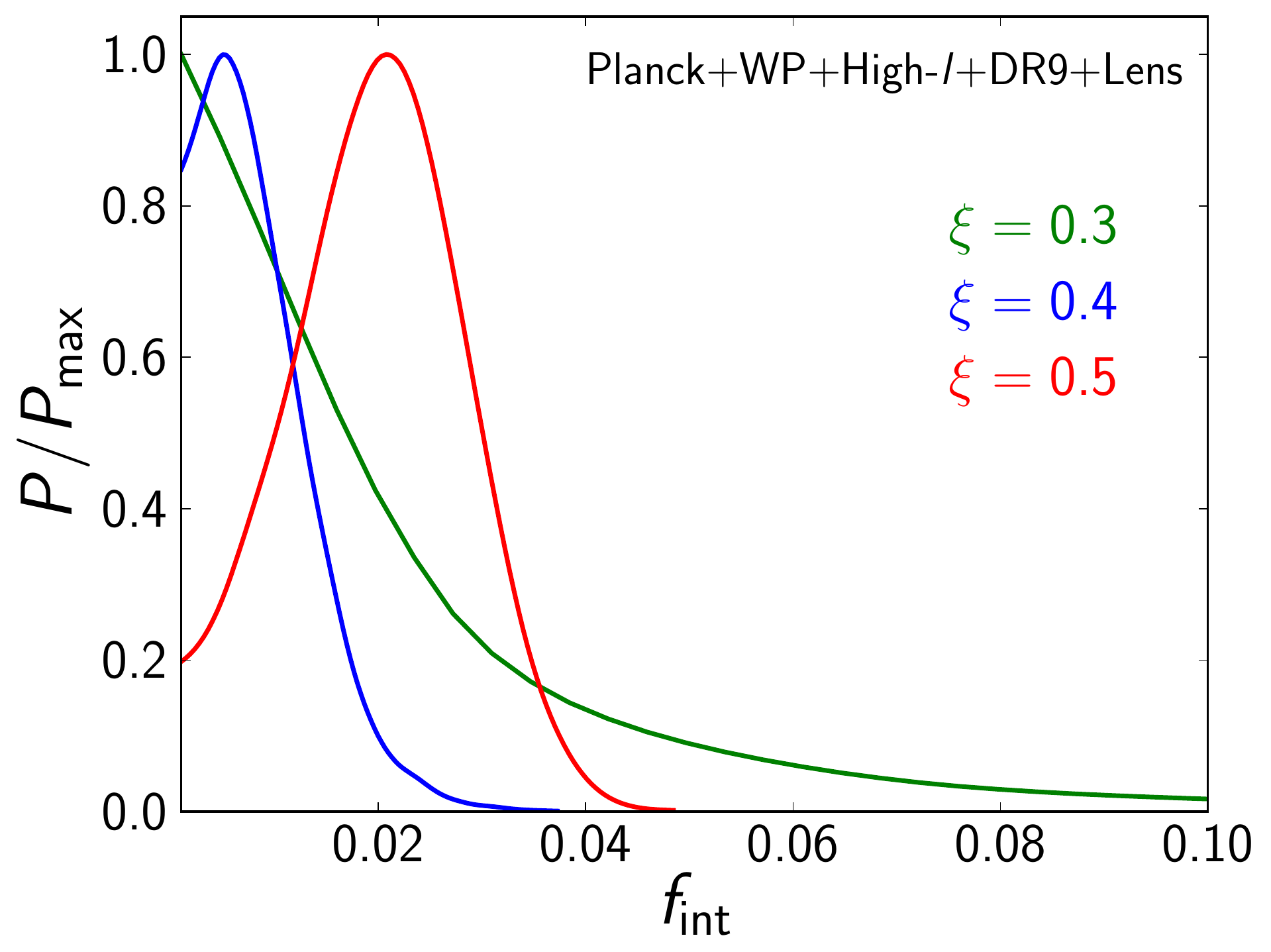}
\caption{Marginalized constraints on the fraction of very strongly interacting DM for three values of $\xi$. Here, we have fixed $\alpha_D=0.05$, $\Sigma_{\rm DAO}=10^{-2.5}$ and $m_D=10$ GeV. The top panel displays the constraints for the ``Planck+WP+High-$l$+BAO+Lens'' dataset, while the lower panel substitutes the BAO measurements for the BOSS DR9 CMASS galaxy power spectrum \cite{Anderson:2012ve}.}
\label{allowed_frac}
\end{centering}
\end{figure}

In this section, we explore the constraints in the $f_{\rm int}-\log_{10}(\Sigma_{\rm DAO})$ plane for fixed values of $\xi$. This analysis allows to determine \emph{how much} interacting DM is allowed by the current data given an interaction strength and a certain density of DR. We displays the constraints in Fig.~\ref{GamDAOvsxi_4} for three values of $\xi$. As in the bounds presented in the previous subsections, there is a sharp transition around $\Sigma_{\rm DAO}\sim10^{-3}-10^{-4}$ for which the fraction of interacting DM goes from being largely unconstrained to being tightly bounded with$f_{\rm int}\lesssim5\%$. The exact constraint depends somewhat on the value of $\Sigma_{\rm DAO}$, with larger values of the latter leading to smaller allowed interacting DM fraction, as we intuitively expect. For $\xi\sim0.4-0.5$, there is an intriguing high-probability region near $\Sigma_{\rm DAO}\sim10^{-2.5}$ for small values of $f_{\rm int}$. To obtain a better picture of what is going on there and to determine a rigorous constraint on the allowed fraction of very strongly interaction DM (that is, interacting with strength equal or stronger than regular baryons), we fix $\Sigma_{\rm DAO}=10^{-2.5}$ and let $f_{\rm int}$ vary freely for the three values of $\xi$ shown in Fig.~\ref{GamDAOvsxi_4}. The resulting marginalized posteriors are shown in Fig.~\ref{allowed_frac} for both the ``Planck+WP+High-$l$+BAO+Lens'' and ``Planck+WP+High-$l$+DR9+Lens'' datasets. 

We indeed observe that both datasets display a mild preference for a nonzero fraction of interacting DM for $\xi=0.5$, while this preference largely goes away as $\xi$ is decreased. It is interesting that this preference becomes stronger when the full shape of the BOSS DR9 galaxy power spectrum is included in the analysis. The penchant for $f_{\rm int}=(2.0\pm1.6)\%$ ($95\%$ C.L.) when $\xi=0.5$ and $\Sigma_{\rm DAO}=10^{-2.5}$ can be understood by looking at the DAO scale for these models. Indeed, in this corner of parameter space the DAO scale lies very close to the standard BAO scale (see Fig.~\ref{fig:rDAO}) leading to substantial overlap and interaction between the DAO and BAO features. The results is effectively a modified amplitude and shape of the BAO feature which, incidentally, improve the fit to the data. However, given the uncertainties in modeling and reconstructing the BAO bump in the galaxy correlation function, it is possible that this preference for a non-vanishing interacting DM fraction is purely coincidental. It is nonetheless intriguing that a $\sim 2\%$ DM fraction interacting very strongly with a DR bath at a temperature $T_D\sim0.5 T_{\rm CMB,0}$ provides an excellent fit to current cosmological data. We emphasize that this allowed PIDM model is particularly interesting since it departs significantly from a pure CDM scenario, in contrast with other allowed regions of the PIDM parameter space which have low values of $\Sigma_{\rm DAO}$ and/or $\xi$, and thus for which PIDM is for all practical purposes indistinguishable from CDM on linear cosmological scales.

We list in Table~\ref{table_f_ADM} the $95\%$ confidence limits for the three values of $\xi$ shown in Figs.~\ref{GamDAOvsxi_4} and \ref{allowed_frac}. While the exact numbers somewhat vary depending on the dataset considered, we observe that current data bound the deviation from a pure CDM scenario to be at most $\sim5\%$ of the overall cosmological DM density for a DM candidate that strongly interacts  with a cosmologically significant amount\footnote{We note that $\xi=0.3$ corresponds to $\Delta N_{\rm eff}\simeq0.036$ which is only marginally cosmologically significant.} of DR. This is a key result of our paper. Since it is based purely on the gravitational effects of interacting DM with the rest of the cosmological constituents, this result is very general and model-independent. It only relies on having a fraction of the DM interacting with a cosmologically significant amount of DR (which can be any type of relativistic particles) and assumes no particle coupling between the visible and dark sector.

\begin{table}[]
\begin{center}
\begin{tabular}{|c|c|c|}
\hline
$\xi$ &CMB+DR9&CMB+BAO  \\
\hline
\hline
$0.3$ & $< 0.085$ & $< 0.049$ \\
\hline
$0.4$ & $< 0.018$ & $<0.035$\\
\hline
$0.5$ & $0.02\pm0.016$ & $< 0.038$\\
\hline
\end{tabular}
\caption{\label{table_f_ADM} Limits on $f_{\rm int}$ for the ``Planck+WP+High-$l$+DR9+Lens'' (middle column) and  ``Planck+WP+High-$l$+BAO+Lens'' (right column) dataset. Here, we have fixed $\alpha_D=0.05$, $\Sigma_{\rm DAO}=10^{-2.5}$ and $m_D=10$ GeV. We display the $95\%$ confidence limits. The severe constraint for $\xi=0.4$ arising when the shape of the BOSS DR9 galaxy power spectrum is taken into account is caused by the DAO scale being just below the BAO scale, leading to a gravitational damping of the BAO bump which is incompatible with data. 
\vspace{-0.7cm}}
\end{center}
\end{table}
%
%%%%
\subsection{Limits on the DAO Scale}
%%%%%

We can translate our constraints on the dark parameters to an upper limit on the size of the sound horizon of interacting DM when it kinematically decouples from the DR bath.  Fixing the value of the interacting DM fraction, we obtain the following limits on the comoving value $r_{\rm DAO}$ 
\be
r_{\rm DAO} < 3.7 h^{-1} {\rm Mpc}\quad(f_{\rm int}=100\%),
\ee
\be
r_{\rm DAO} < 5.3 h^{-1} {\rm Mpc}\quad(f_{\rm int}=50\%),
\ee
\be
r_{\rm DAO} < 15.2 h^{-1} {\rm Mpc}\quad(f_{\rm int}=10\%),\\
\ee
\be
r_{\rm DAO} < 27.9 h^{-1} {\rm Mpc}\quad(f_{\rm int}=5\%),
\ee
where we are giving the $95\%$ confidence limits for the ``Planck+WP+High-$l$+BAO+Lens'' dataset and where we used Eq.~\ref{r_DAO_theory} to compute the DAO scale. These constraints are valid for $\alpha_D>0.025$, but only become slightly stronger for lower values of the dark fine-structure constant. For $f_{\rm int}\geq5\%$, our constraints imply that the DAO scale must lie on relatively small scales where nonlinear effects can be important. Improving upon these constraints will therefore necessitate a prescription to model small-scale nonlinearities in PIDM models. One might worry that some of the bounds listed above are on comoving scales smaller than those probed by the data used in our analysis. The resolution to this apparent paradox lies in the shape and width of the DAO feature, which can be quite broad and affect a large range of scales (see Fig.~\ref{fig:correlation}). Therefore, even if the peak of the DAO feature (that is, $r_{\rm DAO}$) is outside the reach of the data considered, the tail of the DAO bump can have an effect on observable scales, hence the above limits.  For  $f_{\rm int}\lesssim2\%$, the DAO feature becomes very small and the DAO scale is thus largely unconstrained. If instead of fixing $f_{\rm int}$ we marginalize over it, the constraint reads
\be
r_{\rm DAO} < 8.5h^{-1} {\rm Mpc}\quad(f_{\rm int} \text{ marginalized}),
\ee
where again we are displaying the $95\%$ confidence limits for the ``Planck+WP+High-$l$+BAO+Lens'' dataset.

%%%%%
\section{Impact on Galaxy Formation: Double-Disk Dark Matter}\label{galform}
%%%%%
In this section, we turn our attention to the late-time consequences of having a fraction of the dark matter interacting directly with a massless gauge boson.  Much like the baryonic case, the coupling between DM and DR allows the interacting DM to cool via the emission of DR. This cooling can have a large impact on the structure of DM halos around galaxies and can possibly lead to the formation of a DM disk, as was shown in Refs.~\cite{Fan:2013tia,Fan:2013yva}. Here, we point out that the parameters required to obtain the large amount of cooling necessary to form a dark disk can also lead to large-scale cosmological signatures that might be incompatible with current data. 
\begin{figure*}[t]
\begin{centering}
\subfigure{\includegraphics[width=0.49\textwidth]{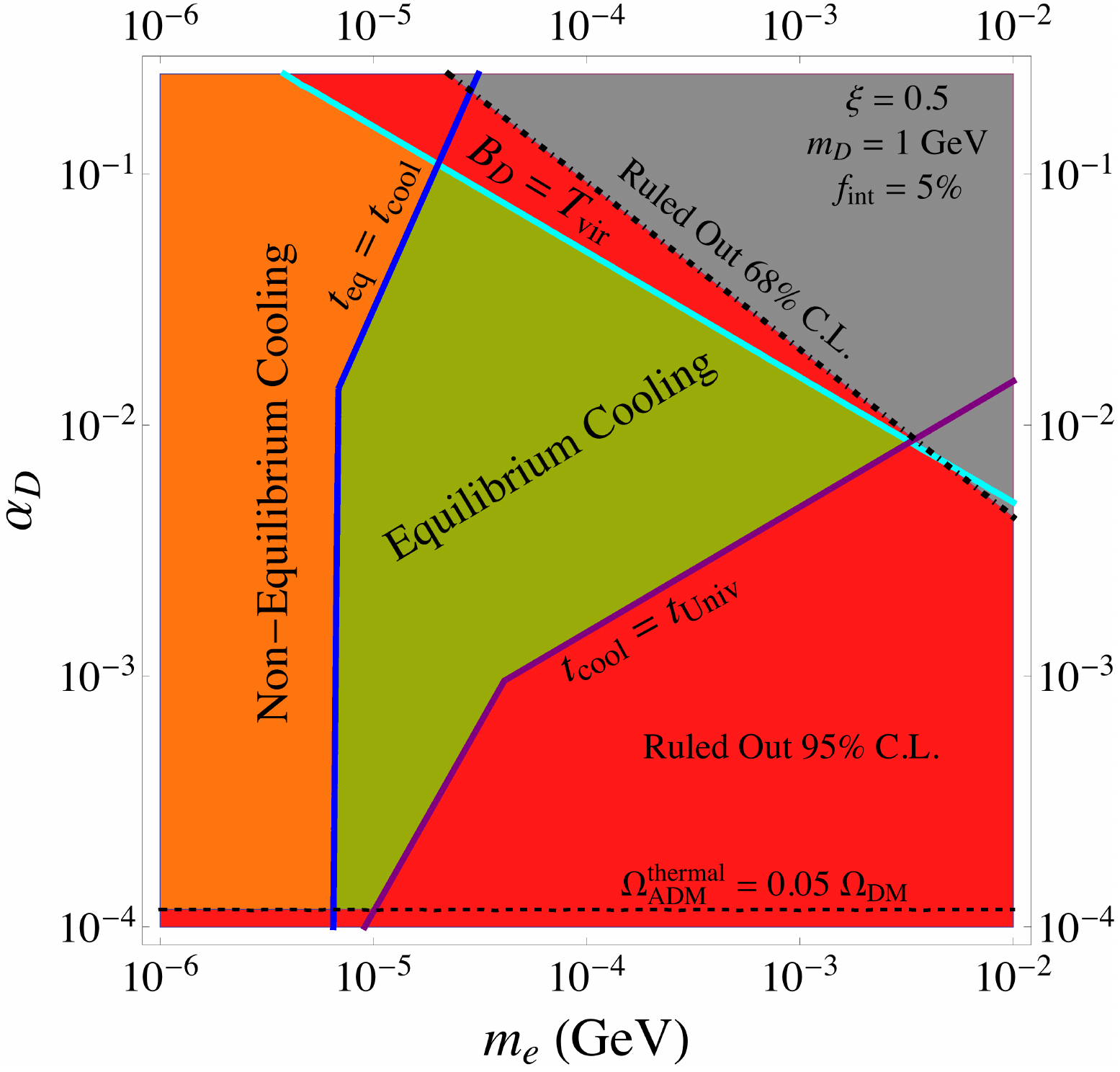}}
\subfigure{\includegraphics[width=0.49\textwidth]{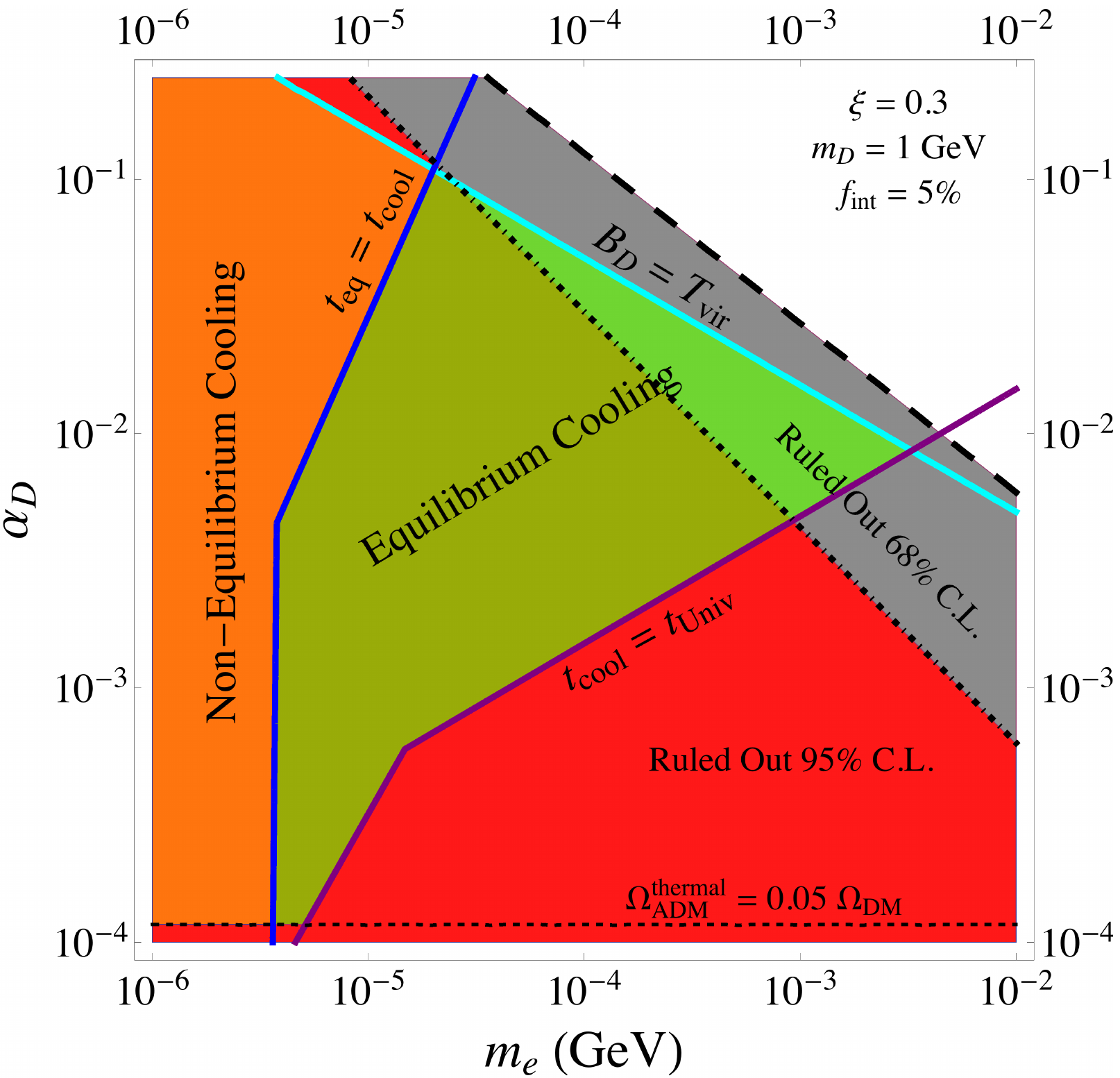}}\\
\subfigure{\includegraphics[width=0.49\textwidth]{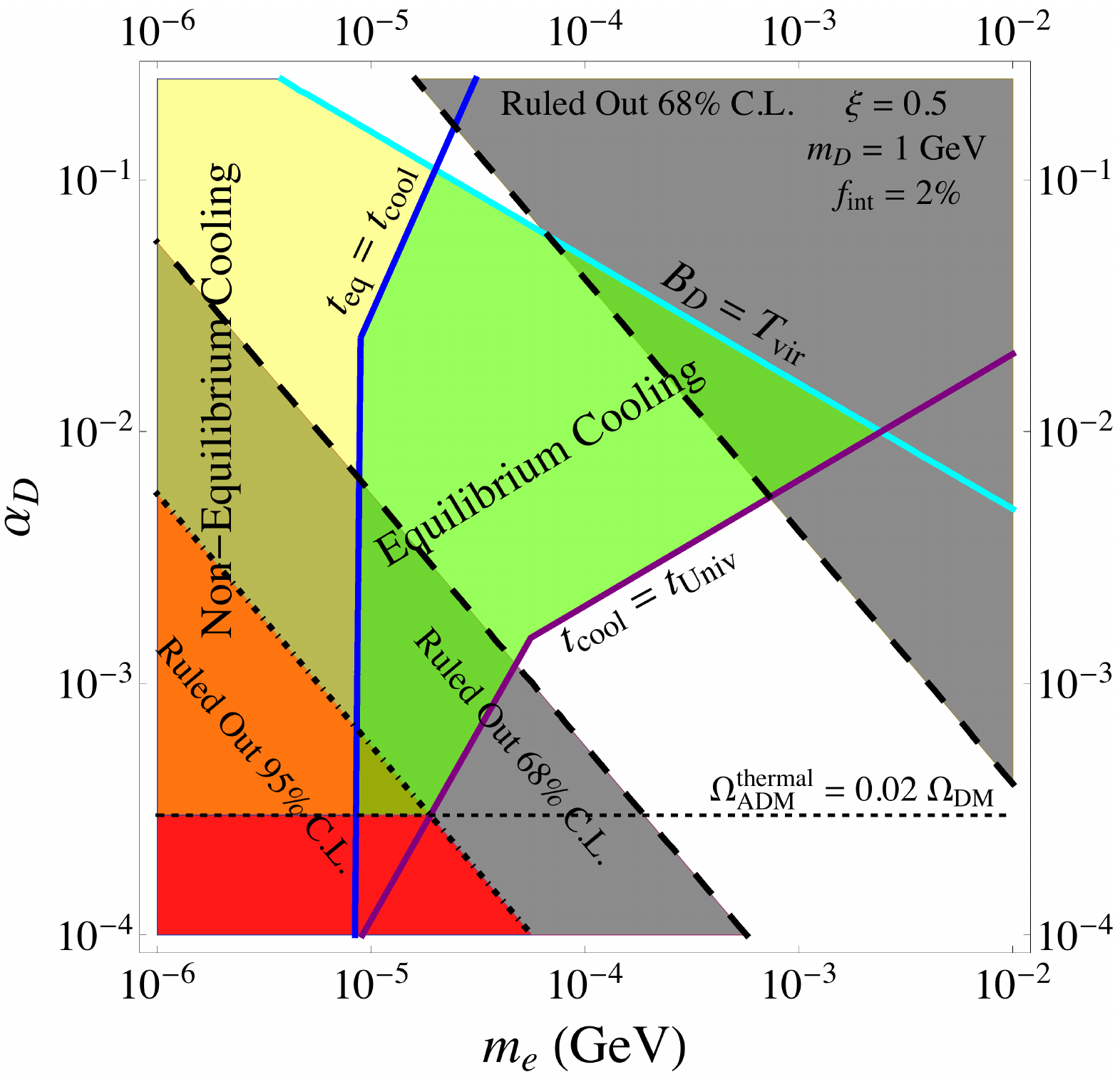}}
\subfigure{\includegraphics[width=0.49\textwidth]{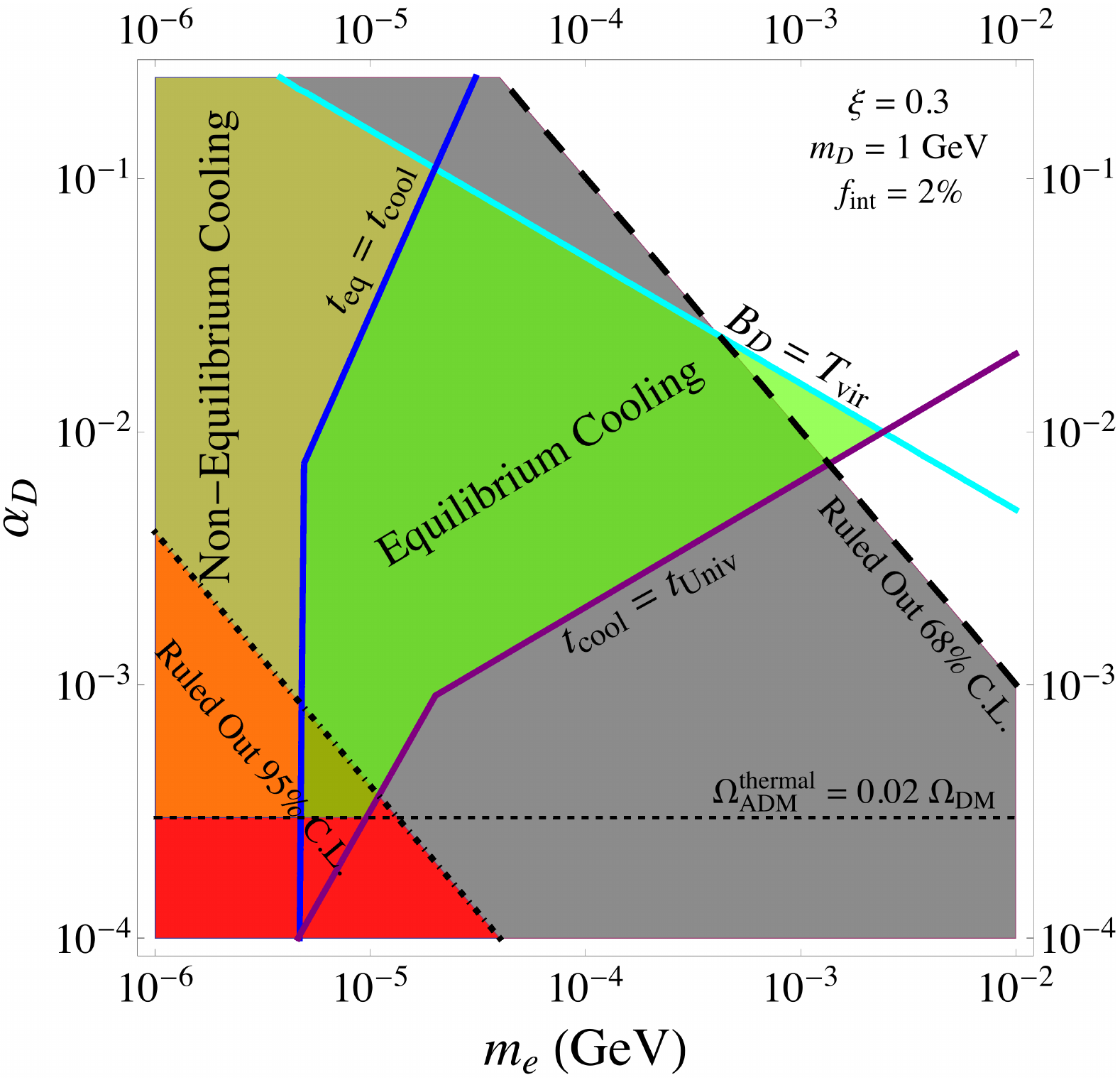}}\\
\caption{PIDM parameter space where a galactic dark disk is likely to form superimposed on the cosmological constraints from the ``Planck+WP+High-$l$+BAO+Lens'' dataset. The red regions below the black dot-dashed line are ruled out at $95\%$ confidence level, while the gray regions delimitated by the black long-dashed line are ruled out at $68\%$ confidence level. As indicated, interacting DM can cool in equilibrium in the green regions while only out-of-equilibrium cooling is possible in the yellow regions. Along the solid purple line, the cooling timescale ($t_{\rm cool}$, the minimum of either bremsstrahlung or Compton scattering) is equal to the age of the Universe ($t_{\rm Univ}$), while the solid blue line denotes the parameters for which the Coulomb equilibration time ($t_{\rm eq}$) is equal to the cooling time. The cyan solid line denotes the parameters for which the virial temperature ($T_{\rm vir}$) of the halo is equal to the dark atomic binding energy. The short-dashed black line shows the value of $\alpha_D$ for which the thermal relic abundance of atomic DM is equal to $f_{\rm int}\Omega_{\rm DM}$. Throughout, we take $n_{\bf e}=n_{\bf p}=7.3\times10^{-3}(f_{\rm int}/0.05)(m_D/{\rm GeV})^{-1}$ cm$^{-3}$ and assume a DM halo mass $M_{\rm DM}=10^{12}M_{\odot}$.}
\label{DDDM1}
\end{centering}
\end{figure*}

The formation of a dark disk in the PIDM scenario requires DM to cool on a fast enough time scale through either the emission of soft dark photons (bremsstrahlung) or through Compton scattering off colder dark photons\footnote{If dark atoms behave at all similarly to regular baryonic hydrogen, we note that molecular cooling could also be very important for the formation of the dark disk.}. Both mechanisms are only effective if the dark atoms are ionized which generally requires the virial temperature of the halo to be larger than the dark atomic binding energy. Since bremsstrahlung and Compton scattering are mostly effective at cooling the lighter dark electron, the timescale for dark proton and dark electron to equilibrate through Coulomb scattering must also be folded into the analysis. As long as the Coulomb scattering rate is faster than the overall cooling rate (through either bremsstrahlung or Compton scattering), the interacting DM cools in equilibrium and a dark disk generally arises. In the opposite scenario, the dark electrons cool faster than they can equilibrate with the heavy dark protons. Ref.~\cite{Fan:2013yva} argues that the overall interacting DM sector could also cool in this regime, but actual simulations will be required to determine the exact outcome. The expressions for the bremsstrahlung cooling, Compton cooling, and Coulomb equilibration times are respectively given in Eqs.~[23], [24], and [27] of Ref.~\cite{Fan:2013yva}.

We illustrate in Fig.~\ref{DDDM1} both the PIDM parameter space where the formation of a dark disk is plausible and our cosmological constraints on the model derived in the previous section. To ease the comparison with the work of Ref.~\cite{Fan:2013yva}, we display the constraints in the $m_{\bf e}-\alpha_D$ plane, where $m_{\bf e}$ is the mass of the dark electron\footnote{In terms of these variables, $\Sigma_{\rm DAO}\simeq\frac{2}{\alpha_D}\left(\frac{m_{\bf e}}{\rm eV}\right)^{-1}\left(\frac{m_D}{\rm GeV}\right)^{-1/6}$.}. We illustrate the double-disk DM parameter space for the optimistic case that DM is distributed according to a Navarro-Frenk-White profile with a characteristic scale $R_s=20$ kpc, leading to central number density $n_{\bf e}=n_{\bf p}=7.3\times10^{-3}(f_{\rm int}/0.05)(m_D/{\rm GeV})^{-1}$ cm$^{-3}$. Note that we have taken the interacting DM fraction inside the halo to be equal to the cosmological mean value, but in general the former might differ from the latter. The green and yellow (appearing as orange when superimposed on the ruled out red regions) regions show the parameter space where equilibrium and non-equilibrium cooling happen, respectively. In each panel, the red region below the black dot-dashed line shows the parameter space that is ruled out at $95\%$ confidence level by the ``Planck+WP+High-$l$+BAO+Lens'' dataset, while the gray regions denote PIDM models ruled out at the $68\%$ confidence level.

We first note that all of the double-disk parameter space with $\xi=0.5$ and $f_{\rm int}=5\%$ (top-left panel of Fig.~\ref{DDDM1}) is ruled out by cosmological data with high confidence. Models with larger values of the interacting DM fraction or of $\xi$ would result is even \emph{stronger} constraints and are therefore also ruled out. As $\xi$ is decreased, the cosmological constraints slowly weaken, opening up some interesting parameter space where a dark disk could form. The top-right panel of Fig.~\ref{DDDM1} illustrates the case with $\xi=0.3$ where we see that a model with $\alpha_D\sim0.01$ and $B_D\sim50$ eV has the right parameters to form a dark disk while lying within the two-sigma contour of current cosmological data. For $\xi\lesssim0.2$ and $f_{\rm int}=5\%$, the double-disk DM parameter space is largely unconstrained by cosmological data, as indicated by the green contours in Fig.~\ref{GamDAOvsxi_2}.

Another avenue to weaken the bounds on double-disk DM is to further reduce the interacting DM fraction below $5\%$. The lower panels of Fig.~\ref{DDDM1} show the constraints on the double-disk parameter space for $f_{\rm int}=2\%$. For $\xi=0.5$ (left bottom panel), we observe that a large swath of the parameter space where a dark disk could form is in good agreement with current cosmological data. As we discussed in Section~\ref{DAO+f}, the CMB and galaxy clustering data are well fitted by a model with $\xi=0.5$, $f_{\rm int}=2\%$, and $\Sigma_{\rm DAO}\sim10^{-2.5}$, which explain the large allowed region overlapping with the double-disk DM parameter space. As $\xi$ is decreased, this preferred region closes up, but most parameter values where DM can cool and form a disk remain within the allowed $95\%$ confidence region.    

In summary, most PIDM models that could lead to a dark matter disk within galaxies are ruled out if $f_{\rm int}\gtrsim5\%$ or $\xi\gtrsim0.2$. Nevertheless, some interesting models remain viable if $f_{\rm int}\sim 2\%$. However, given the small fraction of interacting DM in these scenarios, it remains to be seen if such models could lead to a significant impact on galactic dynamics and on direct and indirect DM searches. Simulations will be necessary to assess the relevance of these allowed models on galactic scales. In any case, our results compellingly highlight the complementarity between the largest cosmological scales and the much smaller galactic scales in pinpointing the nature of dark matter.

%%%%
\section{Discussion}\label{sec:conc}
%%%%
In this paper we have shown that if all or a fraction of the DM were coupled to a bath of DR in the early Universe we expect the combined DM-DR system to give rise to acoustic oscillations of the dark matter until it decouples from the DR.  Much like the standard baryon acoustic oscillations, these DAO imprint a characteristic scale, the sound horizon of dark matter, on the matter distribution in the Universe. We have seen that a having such a fraction of interacting DM can lead to potentially unique signatures on the CMB and large-scale structure data. Although we have modeled the interacting DM and DR system as dark atoms coupled to a bath of dark photons, our results can be straightforwardly applied to a broad class of models that couple DM particles to various light relativistic species. These include, for instance, models where dark matter is coupled to light scalar states or models where the dark sector couples to light states via heavy mediators (analogous to the neutrinos coupling via the weak force).  

We have determined that PIDM models with $\Sigma_{\rm DAO}\gtrsim 10^{-3}$, $\xi\gtrsim 0.2$, and $f_{\rm int}\gtrsim 5\%$ are generally in severe tension with the most recent cosmological data. For much lower values of $\Sigma_{\rm DAO}$, the fraction of interacting DM becomes largely unconstrained while the bounds on $\xi$ reflects the current limits on the effective number of relativistic species, $N_{\rm eff}$. It is particularly interesting that the transition between this last regime and the regime where $\xi$ is severely constrained happens for values of $\Sigma_{\rm DAO}$ similar to that of standard baryons (remember that for baryons, $\Sigma_{\rm BAO}\sim10^{-3.3}$). This is not a coincidence. For $\Sigma_{\rm DAO}\ll\Sigma_{\rm BAO}$, the kinematic decoupling of interacting DM happens much before the epoch of CMB last scattering and any change to the matter power spectrum is limited to scales smaller than the BAO scale. On the other hand, for $\Sigma_{\rm DAO}\gg\Sigma_{\rm BAO}$, interacting DM stays coupled to the DR bath after the epoch of CMB last scattering and the clustering of matter is affected on large cosmological scales, leading to severe constraints on these PIDM models.

For PIDM models with interaction strength equal or greater than that of baryons, we have determined that at most $\sim5\%$  of the DM could be interacting with a cosmologically-significant ($\xi\gtrsim0.3$) DR bath. For $\xi\gtrsim0.4$, the constraint is even more restrictive with $f_{\rm int}\lesssim4\%$. To our knowledge, this is the first time that the allowed deviation from a pure collisionless CDM scenario on large scales is rigorously quantified. The surprise here is that there exists a class of models with $\xi\sim0.5$, $\Sigma_{\rm DAO}\sim10^{-2.5}$, and $f_{\rm int}\sim2\%$ which provides a very good fit to the data, although the improvement to the fit is marginal compared to a simple $\Lambda$CDM model.  This class of models is nonetheless interesting since $\xi\sim0.5$ is the ``natural'' value that we expect if the visible and dark sectors were coupled above the weak scale. Moreover, these are the only PIDM models where the data actually prefers a non-vanishing value of the interacting DM, albeit only at the $\sim2\sigma$ level. Finally, these models are generally expected to impact galactic and possibly cluster dynamics due to the expected cooling of interacting DM via the emission of DR, implying that these models could be probed on a large range of scales. 

We have also determined that current cosmological data allow a large fraction of DM interacting with strength less than that of standard baryons. For $\Sigma_{\rm DAO}\lesssim10^{-4.5}$, the fraction of interacting DM is largely unconstrained (see Fig.~\ref{GamDAOvsxi_4}) and the latter could therefore form all of the DM. Improving upon these constraints will require a prescription to model the small-scale non-linearities in PIDM scenarios. Since these models generally predict a different overall shape for the matter power spectrum, it will likely be necessary to run $N$-body simulations to determine how non-linear structures form and evolve. Depending on the exact PIDM model considered, these simulations could be a lot more involved than standard CDM simulations because of the interacting nature of a fraction (or all) of the DM. Even if we restrict the parameter space to regions where radiative processes such as cooling are inefficient, PIDM will generally be self-interacting at some level inside halos, which can affect their central density profiles. Encouragingly, recent work on $N$-body simulations \cite{Vogelsberger:2012ku,Rocha:2012jg,Peter:2012jh,Zavala:2012us} have started exploring self-interacting DM for some simple cases, and hence are building up the knowledge necessary to eventually conduct and interpret realistic PIDM simulations.  
\begin{figure}[t!]
\begin{centering}
\includegraphics[width=0.5\textwidth]{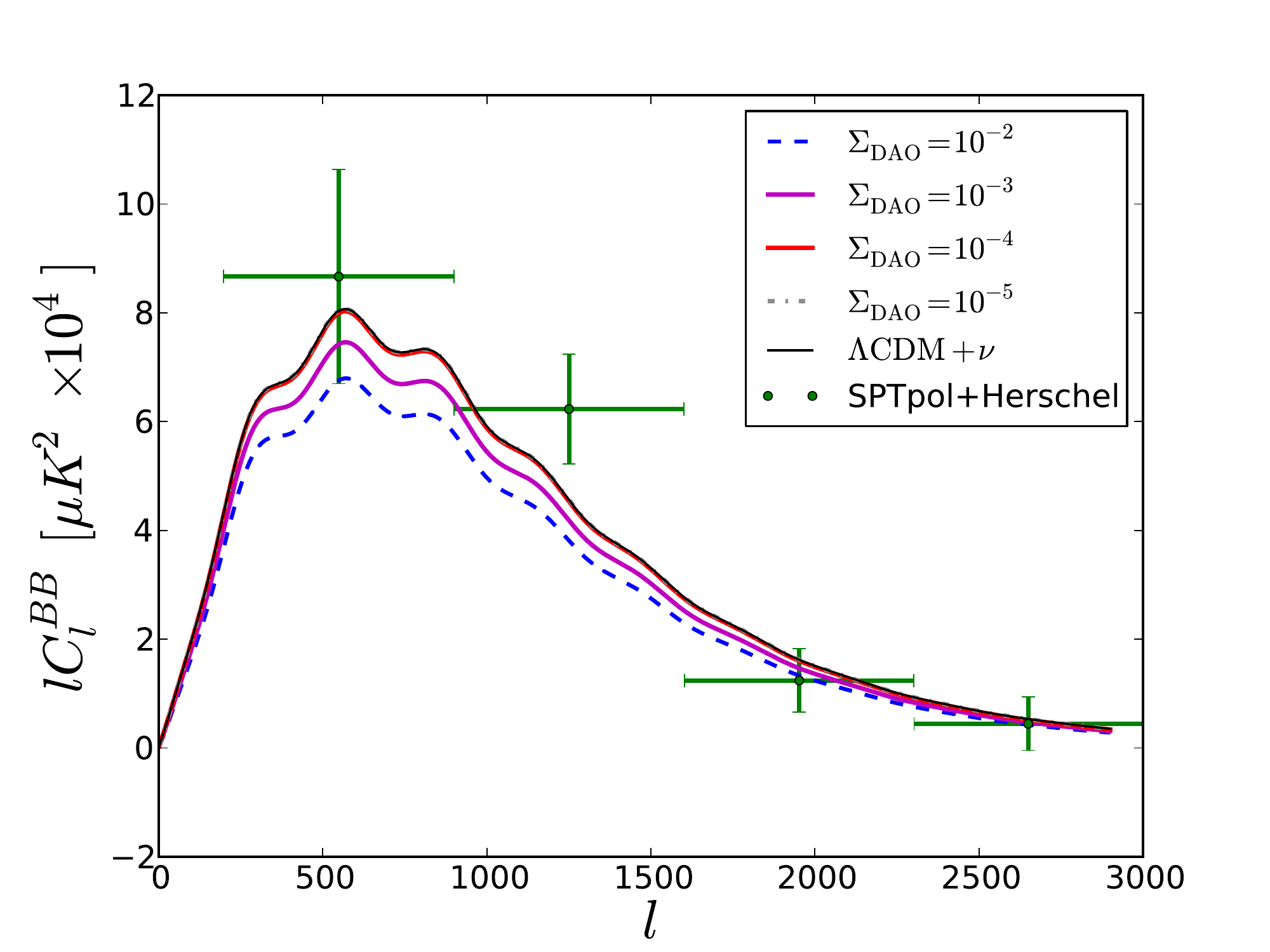}
\includegraphics[width=0.5\textwidth]{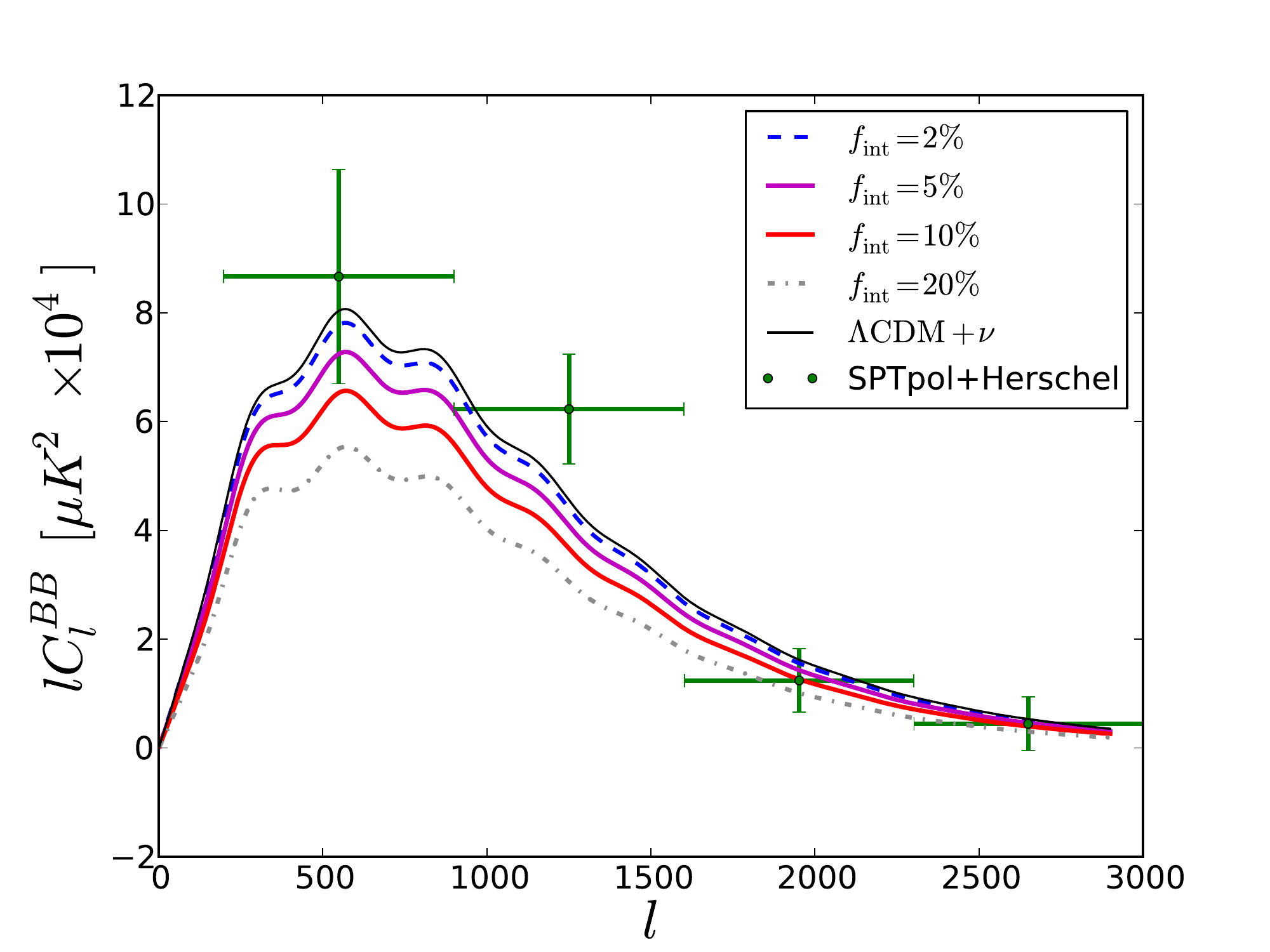}
\caption{Lensed CMB B-mode spectra for various PIDM models. The top panel fixes $\xi=0.5$ and $f_{\rm int}=5\%$ while letting $\Sigma_{\rm DAO}$ varies. The bottom panel fixes $\xi=0.5$ and $\Sigma_{\rm DAO}=10^{-3}$ while varying the fraction of interacting DM. We also show the data from Ref.~\cite{Hanson:2013hsb} obtained by combining data from SPTpol \cite{Austermann:2012ga} and Herschel \cite{Pilbratt:2010mv} (more specifically, we illustrate the $(\hat{E}^{150}\hat{\phi}^{\rm CIB})\times\hat{B}^{150}$ cross correlation). }
\label{fig:b-modes}
\end{centering}
\end{figure}

A promising avenue to improve the constraints on PIDM is CMB lensing. Upcoming CMB polarization data from Planck and from ground-based telescopes will dramatically improve the reconstruction of the lensing potential, which itself depends strongly on the matter power spectrum. Since it can probe the lensing potential down to smaller angular scales \cite{Smith:2006nk}, the conversion of polarization E-modes into B-modes via gravitational lensing may well provide the strongest bounds on PIDM scenarios.  Moreover, since lensed B-modes are more sensitive to the matter distribution at redshifts higher than those probed by current and near-future galaxy surveys \cite{Smith:2006nk}, it is less affected by non-linearities and therefore has the potential to provide more reliable limits on PIDM. We illustrate a few lensed B-mode spectra in Fig.~\ref{fig:b-modes} as well as the first-detection data from Ref.~\cite{Hanson:2013hsb}. We observe that even a $2\%$ fraction of interacting DM can have a sizable effect on the B-mode spectrum.

One topic that we have not touched upon in this work is whether PIDM can address the apparent discrepancies between state-of-the-art CMB data and the Hubble parameter inferred from local measurements and supernovae on the one hand, and between the CMB data and the Sunyaev-Zel'dovich (SZ) cluster count on the other hand (see Refs.~\cite{planckXVI,Ade:2013lmv}). Since PIDM models naturally contain a DR component, they tend to prefer a higher value of the Hubble parameter which could ease the tension between the CMB and other probes of the local expansion rate. Moreover, since PIDM generically predicts a damping of small-scale power, it could potentially reconcile the SZ cluster count with the CMB data. We leave such an analysis for future work.

On that note, we point out that it would be interesting to explore the possible degeneracy between PIDM and models with massive neutrinos (active or sterile). While PIDM can indeed mimic some of the signatures of massive neutrinos like the damping of the matter power spectrum on small scales, the CMB could provide enough discriminatory power to distinguish the two scenarios. For instance, in the PIDM case, the DR radiation transitions from being tightly-coupled to the DM to a free-streaming state at dark kinematic decoupling while in the massive neutrino case, the neutrinos transition from a free-streaming state to a cold non-relativistic state when their temperature falls below their mass. Since these two types of transition impact the CMB differently, we expect the PIDM and the massive neutrino scenarios to be be distinguishable to some degree. We leave the exploration of these degeneracies to future work.

\section{Conclusion}\label{sec:conc}

In this work we have found that if at least $5\%$ of the dark matter was coupled to a bath of dark radiation in the early Universe, its sound horizon must lie on small non-linear scales. For a smaller interacting dark matter fraction, the DAO scale becomes progressively unconstrained as the DAO feature shrinks in importance. 
Our results imply that more than $95\%$ of the dark matter must behave like collisionless CDM, long decoupled from any radiation component, on large cosmological scales.  
Like most knowledge we have gleaned about dark matter since inferring its existence \cite{1933AcHPh...6..110Z,1937ApJ....86..217Z,1980ApJ...238..471R}, this constraint rests on the  gravitational influence of dark matter (or dark radiation) on the observable Universe.
Provided gravity is universal the pull of dark matter betrays its distribution even though it remains otherwise elusive and invisible.
Our conclusions further rely on the extraordinarily detailed description of the physics of cosmological perturbations in the early Universe that are now required to enable cosmologists to make precise predictions for the CMB and LSS observables --- and reveal ever more about the nature of our Universe.\\

%%%%
{\noindent \bf Acknowledgements---} We thank  JiJi Fan, Andrey Katz, Lisa Randall, Matthew Reece, and Manoj Kaplinghat for useful discussions. We further thank Abhilash Mishra, Marius Millea, and Matthew Reece for insightful comments on an earlier version of this draft. We are also grateful to Olga Mena and Shun Saito for generously providing an initial version of the code for the BOSS galaxy power spectrum likelihood. This work was performed in part at the California Institute of Technology for the Keck Institute for Space Studies, which is funded by the W. M. Keck Foundation. The research of KS is supported in part by a National Science and Engineering Research Council (NSERC) of Canada Discovery Grant.  FYCR and KS thank the Aspen Center for Physics, where part of this work was completed, for their hospitality.  This research was supported in part by the National Science Foundation under Grant No. NSF PHY11-25915.  Part of the research described in this paper was carried out at the Jet Propulsion Laboratory, California Institute of Technology, under a contract with the National Aeronautics and Space Administration. Part of this work is supported by NASA ATP grant 11-ATP-090.

%%%%%
\appendix
%%%%%

%%%%%
\section{Scale Factor at DM Kinematic Decoupling}\label{app:a_D}
%%%%%
The value of the scale factor at the epoch of dark kinematic decoupling can be approximately obtained by solving the criterion
\be
n_{\rm ADM}x_D\sigma_{{\rm T},D}= H.
\ee
In a matter-radiation Universe, this equation can be rewritten as an algebraic equation for $a_D$
\be\label{eq:a_D_cubic}
a_D^3+\omega_{\rm r}a_D^2
=S_D,
\ee
where
\be
S_D=\frac{1}{\Omega_{\rm m}H_0^2}\left(\frac{2\pi\Omega_{\rm DM}\rho_{\rm crit}}{3}\frac{\alpha_D^6f_{\rm int}x_D(a_D)}{m_DB_D^2}\right)^2,
\ee
and where $\rho_{\rm crit}$ is the critical density of the Universe, $\omega_{\rm r}\equiv\Omega_{\rm r}/\Omega_{\rm m}$, and other symbols are described in the 
section \ref{CosmoPert}. We note that $S_D$ implicitly depends on $a_D$ itself through its dependence on the ionized fraction $x_D$. The exact time-evolution of $x_D$ needs to be solved numerically, but Ref.~\cite{CyrRacine:2012fz} derives an approximate scaling for $x_D$ as a function of the dark parameters
\be
x_D(a_D)\propto \frac{m_D^{5/6} B_D\xi}{\Omega_{\rm DM}f_{\rm int}\alpha_D^5},
\ee
where we used $x_D(a_D)\propto\alpha_D\bar{x}_D/(B_Dm_D^{1/6})$, $\bar{x}_D$ being the asymptotic value of the dark atom ionized fraction at late times. We can thus rewrite $S_D$ as
\be\label{eq:S_D_w_epsilon}
S_D = \frac{1}{\Omega_{\rm m}h^2}\left(\epsilon_D \frac{\alpha_D\xi}{(B_D/{\rm eV})(m_D/{\rm GeV})^{1/6}}\right)^2,
\ee
where $\epsilon_D$ is a fitting constant that can be determined by solving numerically the ionization and thermal history of dark atoms. For strongly-coupled models with $\alpha_D\gtrsim 0.025$, we find  $\epsilon_D\sim8\times10^{-3}$, while for $\alpha_D < 0.025$, $\epsilon\sim1.7\times10^{-2}$ provides a better fit. We note that $S_D$ can be written in terms of the quantity $\Sigma_{\rm DAO}$ defined in Eq.~\ref{eq:pidm_mod} above
\be
S_D=\frac{1}{\Omega_{\rm m}h^2}\left(\epsilon_D\xi\Sigma_{\rm DAO}\right)^2.
\ee
Equation \ref{eq:a_D_cubic} can then be solved exactly. Keeping only the real positive definite root, we obtain
\be\label{eq:sol_a_D}
a_D = \frac{1}{12}\left(2^{5/3}\Xi_D+4\omega_{\rm r}\left(\frac{2^{1/3}\omega_{\rm r}}{\Xi_D}-1\right)\right),
\ee
where
\be
\Xi_D = \left(27 S_D-2\omega_{\rm r}^3-3\sqrt{3}\sqrt{S_D(27S_D-4\omega_{\rm r}^3)}\right)^{1/3}
\ee
Deep into the matter-dominated era, Eq.~\ref{eq:sol_a_D} reduces to
\be
a_D\simeq \left(\frac{1}{\Omega_{\rm m}h^2}\right)^{1/3}\left(\epsilon_D\xi\Sigma_{\rm DAO}\right)^{2/3}\quad(a_D\gg a_{\rm eq}),
\ee
while in the radiation-dominated era, it takes the form
\be
a_D\simeq  \left(\frac{1}{\Omega_{\rm r}h^2}\right)^{1/2}\left(\epsilon_D\xi\Sigma_{\rm DAO}\right)\quad(a_D\ll a_{\rm eq}),
\ee
where $a_{\rm eq}$ is the scale factor at radiation-matter equality.

%%%%%%%%%
\bibliography{PIDM_DAO}

\begin{thebibliography}{139}
\expandafter\ifx\csname natexlab\endcsname\relax\def\natexlab#1{#1}\fi
\expandafter\ifx\csname bibnamefont\endcsname\relax
  \def\bibnamefont#1{#1}\fi
\expandafter\ifx\csname bibfnamefont\endcsname\relax
  \def\bibfnamefont#1{#1}\fi
\expandafter\ifx\csname citenamefont\endcsname\relax
  \def\citenamefont#1{#1}\fi
\expandafter\ifx\csname url\endcsname\relax
  \def\url#1{\texttt{#1}}\fi
\expandafter\ifx\csname urlprefix\endcsname\relax\def\urlprefix{URL }\fi
\providecommand{\bibinfo}[2]{#2}
\providecommand{\eprint}[2][]{\url{#2}}

\bibitem[{\citenamefont{{Zwicky}}(1933)}]{1933AcHPh...6..110Z}
\bibinfo{author}{\bibfnamefont{F.}~\bibnamefont{{Zwicky}}},
  \bibinfo{journal}{Helvetica Physica Acta} \textbf{\bibinfo{volume}{6}},
  \bibinfo{pages}{110} (\bibinfo{year}{1933}).

\bibitem[{\citenamefont{{Zwicky}}(1937)}]{1937ApJ....86..217Z}
\bibinfo{author}{\bibfnamefont{F.}~\bibnamefont{{Zwicky}}},
  \bibinfo{journal}{\apj} \textbf{\bibinfo{volume}{86}}, \bibinfo{pages}{217}
  (\bibinfo{year}{1937}).

\bibitem[{\citenamefont{{Rubin} et~al.}(1980)\citenamefont{{Rubin}, {Ford}, and
  {.~Thonnard}}}]{1980ApJ...238..471R}
\bibinfo{author}{\bibfnamefont{V.~C.} \bibnamefont{{Rubin}}},
  \bibinfo{author}{\bibfnamefont{W.~K.~J.} \bibnamefont{{Ford}}},
  \bibnamefont{and}
  \bibinfo{author}{\bibfnamefont{N.}~\bibnamefont{{.~Thonnard}}},
  \bibinfo{journal}{\apj} \textbf{\bibinfo{volume}{238}}, \bibinfo{pages}{471}
  (\bibinfo{year}{1980}).

\bibitem[{\citenamefont{Aalseth et~al.}(2011{\natexlab{a}})}]{Aalseth:2010vx}
\bibinfo{author}{\bibfnamefont{C.}~\bibnamefont{Aalseth}} \bibnamefont{et~al.}
  (\bibinfo{collaboration}{CoGeNT collaboration}),
  \bibinfo{journal}{Phys.Rev.Lett.} \textbf{\bibinfo{volume}{106}},
  \bibinfo{pages}{131301} (\bibinfo{year}{2011}{\natexlab{a}}),
  \eprint{1002.4703}.

\bibitem[{\citenamefont{Aalseth
  et~al.}(2011{\natexlab{b}})\citenamefont{Aalseth, Barbeau, Colaresi, Collar,
  Diaz~Leon et~al.}}]{Aalseth:2011wp}
\bibinfo{author}{\bibfnamefont{C.}~\bibnamefont{Aalseth}},
  \bibinfo{author}{\bibfnamefont{P.}~\bibnamefont{Barbeau}},
  \bibinfo{author}{\bibfnamefont{J.}~\bibnamefont{Colaresi}},
  \bibinfo{author}{\bibfnamefont{J.}~\bibnamefont{Collar}},
  \bibinfo{author}{\bibfnamefont{J.}~\bibnamefont{Diaz~Leon}},
  \bibnamefont{et~al.}, \bibinfo{journal}{Phys.Rev.Lett.}
  \textbf{\bibinfo{volume}{107}}, \bibinfo{pages}{141301}
  (\bibinfo{year}{2011}{\natexlab{b}}), \eprint{1106.0650}.

\bibitem[{\citenamefont{Angloher et~al.}(2012)\citenamefont{Angloher, Bauer,
  Bavykina, Bento, Bucci et~al.}}]{Angloher:2011uu}
\bibinfo{author}{\bibfnamefont{G.}~\bibnamefont{Angloher}},
  \bibinfo{author}{\bibfnamefont{M.}~\bibnamefont{Bauer}},
  \bibinfo{author}{\bibfnamefont{I.}~\bibnamefont{Bavykina}},
  \bibinfo{author}{\bibfnamefont{A.}~\bibnamefont{Bento}},
  \bibinfo{author}{\bibfnamefont{C.}~\bibnamefont{Bucci}},
  \bibnamefont{et~al.}, \bibinfo{journal}{Eur.Phys.J.}
  \textbf{\bibinfo{volume}{C72}}, \bibinfo{pages}{1971} (\bibinfo{year}{2012}),
  \eprint{1109.0702}.

\bibitem[{\citenamefont{Agnese et~al.}(2013)}]{Agnese:2013rvf}
\bibinfo{author}{\bibfnamefont{R.}~\bibnamefont{Agnese}} \bibnamefont{et~al.}
  (\bibinfo{collaboration}{CDMS Collaboration}),
  \bibinfo{journal}{Phys.Rev.Lett.}  (\bibinfo{year}{2013}),
  \eprint{1304.4279}.

\bibitem[{\citenamefont{Bernabei et~al.}(2013)\citenamefont{Bernabei, Belli,
  Cappella, Caracciolo, Castellano et~al.}}]{Bernabei:2013xsa}
\bibinfo{author}{\bibfnamefont{R.}~\bibnamefont{Bernabei}},
  \bibinfo{author}{\bibfnamefont{P.}~\bibnamefont{Belli}},
  \bibinfo{author}{\bibfnamefont{F.}~\bibnamefont{Cappella}},
  \bibinfo{author}{\bibfnamefont{V.}~\bibnamefont{Caracciolo}},
  \bibinfo{author}{\bibfnamefont{S.}~\bibnamefont{Castellano}},
  \bibnamefont{et~al.} (\bibinfo{year}{2013}), \eprint{1308.5109}.

\bibitem[{\citenamefont{Walker and Penarrubia}(2011)}]{Walker:2011zu}
\bibinfo{author}{\bibfnamefont{M.~G.} \bibnamefont{Walker}} \bibnamefont{and}
  \bibinfo{author}{\bibfnamefont{J.}~\bibnamefont{Penarrubia}},
  \bibinfo{journal}{Astrophys.J.} \textbf{\bibinfo{volume}{742}},
  \bibinfo{pages}{20} (\bibinfo{year}{2011}), \eprint{1108.2404}.

\bibitem[{\citenamefont{{Salucci} et~al.}(2012)\citenamefont{{Salucci},
  {Wilkinson}, {Walker}, {Gilmore}, {Grebel}, {Koch}, {Frigerio Martins}, and
  {Wyse}}}]{2012MNRAS.420.2034S}
\bibinfo{author}{\bibfnamefont{P.}~\bibnamefont{{Salucci}}},
  \bibinfo{author}{\bibfnamefont{M.~I.} \bibnamefont{{Wilkinson}}},
  \bibinfo{author}{\bibfnamefont{M.~G.} \bibnamefont{{Walker}}},
  \bibinfo{author}{\bibfnamefont{G.~F.} \bibnamefont{{Gilmore}}},
  \bibinfo{author}{\bibfnamefont{E.~K.} \bibnamefont{{Grebel}}},
  \bibinfo{author}{\bibfnamefont{A.}~\bibnamefont{{Koch}}},
  \bibinfo{author}{\bibfnamefont{C.}~\bibnamefont{{Frigerio Martins}}},
  \bibnamefont{and} \bibinfo{author}{\bibfnamefont{R.~F.~G.}
  \bibnamefont{{Wyse}}}, \bibinfo{journal}{\mnras}
  \textbf{\bibinfo{volume}{420}}, \bibinfo{pages}{2034} (\bibinfo{year}{2012}),
  \eprint{1111.1165}.

\bibitem[{\citenamefont{Boylan-Kolchin
  et~al.}(2011)\citenamefont{Boylan-Kolchin, Bullock, and
  Kaplinghat}}]{BoylanKolchin:2011de}
\bibinfo{author}{\bibfnamefont{M.}~\bibnamefont{Boylan-Kolchin}},
  \bibinfo{author}{\bibfnamefont{J.~S.} \bibnamefont{Bullock}},
  \bibnamefont{and}
  \bibinfo{author}{\bibfnamefont{M.}~\bibnamefont{Kaplinghat}},
  \bibinfo{journal}{Mon.Not.Roy.Astron.Soc.} \textbf{\bibinfo{volume}{415}},
  \bibinfo{pages}{L40} (\bibinfo{year}{2011}), \eprint{1103.0007}.

\bibitem[{\citenamefont{Boylan-Kolchin
  et~al.}(2012)\citenamefont{Boylan-Kolchin, Bullock, and
  Kaplinghat}}]{BoylanKolchin:2011dk}
\bibinfo{author}{\bibfnamefont{M.}~\bibnamefont{Boylan-Kolchin}},
  \bibinfo{author}{\bibfnamefont{J.~S.} \bibnamefont{Bullock}},
  \bibnamefont{and}
  \bibinfo{author}{\bibfnamefont{M.}~\bibnamefont{Kaplinghat}},
  \bibinfo{journal}{Mon.Not.Roy.Astron.Soc.} \textbf{\bibinfo{volume}{422}},
  \bibinfo{pages}{1203} (\bibinfo{year}{2012}), \eprint{1111.2048}.

\bibitem[{\citenamefont{Walker}(2012)}]{Walker:2012td}
\bibinfo{author}{\bibfnamefont{M.~G.} \bibnamefont{Walker}}
  (\bibinfo{year}{2012}), \eprint{1205.0311}.

\bibitem[{\citenamefont{Zavala et~al.}(2012)\citenamefont{Zavala, Vogelsberger,
  and Walker}}]{Zavala:2012us}
\bibinfo{author}{\bibfnamefont{J.}~\bibnamefont{Zavala}},
  \bibinfo{author}{\bibfnamefont{M.}~\bibnamefont{Vogelsberger}},
  \bibnamefont{and} \bibinfo{author}{\bibfnamefont{M.~G.} \bibnamefont{Walker}}
  (\bibinfo{year}{2012}), \eprint{1211.6426}.

\bibitem[{\citenamefont{Laporte et~al.}(2013)\citenamefont{Laporte, Walker, and
  Pe–arrubia}}]{Laporte:2013fwa}
\bibinfo{author}{\bibfnamefont{C.~F.~P.} \bibnamefont{Laporte}},
  \bibinfo{author}{\bibfnamefont{M.~G.} \bibnamefont{Walker}},
  \bibnamefont{and}
  \bibinfo{author}{\bibfnamefont{J.}~\bibnamefont{Pe–arrubia}}
  (\bibinfo{year}{2013}), \eprint{1303.1534}.

\bibitem[{\citenamefont{Amorisco et~al.}(2013)\citenamefont{Amorisco, Zavala,
  and de~Boer}}]{Amorisco:2013uwa}
\bibinfo{author}{\bibfnamefont{N.~C.} \bibnamefont{Amorisco}},
  \bibinfo{author}{\bibfnamefont{J.}~\bibnamefont{Zavala}}, \bibnamefont{and}
  \bibinfo{author}{\bibfnamefont{T.~J.~L.} \bibnamefont{de~Boer}}
  (\bibinfo{year}{2013}), \eprint{1309.5958}.

\bibitem[{\citenamefont{{Gentile} et~al.}(2004)\citenamefont{{Gentile},
  {Salucci}, {Klein}, {Vergani}, and {Kalberla}}}]{2004MNRAS.351..903G}
\bibinfo{author}{\bibfnamefont{G.}~\bibnamefont{{Gentile}}},
  \bibinfo{author}{\bibfnamefont{P.}~\bibnamefont{{Salucci}}},
  \bibinfo{author}{\bibfnamefont{U.}~\bibnamefont{{Klein}}},
  \bibinfo{author}{\bibfnamefont{D.}~\bibnamefont{{Vergani}}},
  \bibnamefont{and}
  \bibinfo{author}{\bibfnamefont{P.}~\bibnamefont{{Kalberla}}},
  \bibinfo{journal}{\mnras} \textbf{\bibinfo{volume}{351}},
  \bibinfo{pages}{903} (\bibinfo{year}{2004}), \eprint{arXiv:astro-ph/0403154}.

\bibitem[{\citenamefont{de~Naray et~al.}(2009)\citenamefont{de~Naray, Martinez,
  Bullock, and Kaplinghat}}]{deNaray:2009xj}
\bibinfo{author}{\bibfnamefont{R.~K.} \bibnamefont{de~Naray}},
  \bibinfo{author}{\bibfnamefont{G.~D.} \bibnamefont{Martinez}},
  \bibinfo{author}{\bibfnamefont{J.~S.} \bibnamefont{Bullock}},
  \bibnamefont{and}
  \bibinfo{author}{\bibfnamefont{M.}~\bibnamefont{Kaplinghat}}
  (\bibinfo{year}{2009}), \eprint{0912.3518}.

\bibitem[{\citenamefont{de~Naray and Spekkens}(2011)}]{deNaray:2011hy}
\bibinfo{author}{\bibfnamefont{R.~K.} \bibnamefont{de~Naray}} \bibnamefont{and}
  \bibinfo{author}{\bibfnamefont{K.}~\bibnamefont{Spekkens}},
  \bibinfo{journal}{Astrophys.J.} \textbf{\bibinfo{volume}{741}},
  \bibinfo{pages}{L29} (\bibinfo{year}{2011}), \eprint{1109.1288}.

\bibitem[{\citenamefont{Vogelsberger et~al.}(2012)\citenamefont{Vogelsberger,
  Zavala, and Loeb}}]{Vogelsberger:2012ku}
\bibinfo{author}{\bibfnamefont{M.}~\bibnamefont{Vogelsberger}},
  \bibinfo{author}{\bibfnamefont{J.}~\bibnamefont{Zavala}}, \bibnamefont{and}
  \bibinfo{author}{\bibfnamefont{A.}~\bibnamefont{Loeb}},
  \bibinfo{journal}{Mon.Not.Roy.Astron.Soc.} \textbf{\bibinfo{volume}{423}},
  \bibinfo{pages}{3740} (\bibinfo{year}{2012}), \eprint{1201.5892}.

\bibitem[{\citenamefont{Markevitch et~al.}(2004)\citenamefont{Markevitch,
  Gonzalez, Clowe, Vikhlinin, David et~al.}}]{Markevitch:2003at}
\bibinfo{author}{\bibfnamefont{M.}~\bibnamefont{Markevitch}},
  \bibinfo{author}{\bibfnamefont{A.}~\bibnamefont{Gonzalez}},
  \bibinfo{author}{\bibfnamefont{D.}~\bibnamefont{Clowe}},
  \bibinfo{author}{\bibfnamefont{A.}~\bibnamefont{Vikhlinin}},
  \bibinfo{author}{\bibfnamefont{L.}~\bibnamefont{David}},
  \bibnamefont{et~al.}, \bibinfo{journal}{Astrophys.J.}
  \textbf{\bibinfo{volume}{606}}, \bibinfo{pages}{819} (\bibinfo{year}{2004}),
  \eprint{astro-ph/0309303}.

\bibitem[{\citenamefont{Randall et~al.}(2008)\citenamefont{Randall, Markevitch,
  Clowe, Gonzalez, and Bradac}}]{Randall:2007ph}
\bibinfo{author}{\bibfnamefont{S.~W.} \bibnamefont{Randall}},
  \bibinfo{author}{\bibfnamefont{M.}~\bibnamefont{Markevitch}},
  \bibinfo{author}{\bibfnamefont{D.}~\bibnamefont{Clowe}},
  \bibinfo{author}{\bibfnamefont{A.~H.} \bibnamefont{Gonzalez}},
  \bibnamefont{and} \bibinfo{author}{\bibfnamefont{M.}~\bibnamefont{Bradac}},
  \bibinfo{journal}{Astrophys.J.} \textbf{\bibinfo{volume}{679}},
  \bibinfo{pages}{1173} (\bibinfo{year}{2008}), \eprint{0704.0261}.

\bibitem[{\citenamefont{Merten et~al.}(2011)\citenamefont{Merten, Coe, Dupke,
  Massey, Zitrin et~al.}}]{Merten:2011wj}
\bibinfo{author}{\bibfnamefont{J.}~\bibnamefont{Merten}},
  \bibinfo{author}{\bibfnamefont{D.}~\bibnamefont{Coe}},
  \bibinfo{author}{\bibfnamefont{R.}~\bibnamefont{Dupke}},
  \bibinfo{author}{\bibfnamefont{R.}~\bibnamefont{Massey}},
  \bibinfo{author}{\bibfnamefont{A.}~\bibnamefont{Zitrin}},
  \bibnamefont{et~al.}, \bibinfo{journal}{Mon.Not.Roy.Astron.Soc.}
  \textbf{\bibinfo{volume}{417}}, \bibinfo{pages}{333} (\bibinfo{year}{2011}),
  \eprint{1103.2772}.

\bibitem[{\citenamefont{Dawson et~al.}(2012)\citenamefont{Dawson, Wittman, Jee,
  Gee, Hughes et~al.}}]{Dawson:2011kf}
\bibinfo{author}{\bibfnamefont{W.~A.} \bibnamefont{Dawson}},
  \bibinfo{author}{\bibfnamefont{D.}~\bibnamefont{Wittman}},
  \bibinfo{author}{\bibfnamefont{M.}~\bibnamefont{Jee}},
  \bibinfo{author}{\bibfnamefont{P.}~\bibnamefont{Gee}},
  \bibinfo{author}{\bibfnamefont{J.~P.} \bibnamefont{Hughes}},
  \bibnamefont{et~al.}, \bibinfo{journal}{Astrophys.J.}
  \textbf{\bibinfo{volume}{747}}, \bibinfo{pages}{L42} (\bibinfo{year}{2012}),
  \eprint{1110.4391}.

\bibitem[{\citenamefont{Dawson}(2013)}]{Dawson:2012fx}
\bibinfo{author}{\bibfnamefont{W.~A.} \bibnamefont{Dawson}},
  \bibinfo{journal}{ApJ} \textbf{\bibinfo{volume}{772}}, \bibinfo{pages}{131}
  (\bibinfo{year}{2013}), \eprint{1210.0014}.

\bibitem[{\citenamefont{Oh et~al.}(2010)\citenamefont{Oh, Brook, Governato,
  Brinks, Mayer et~al.}}]{Oh:2010mc}
\bibinfo{author}{\bibfnamefont{S.-H.} \bibnamefont{Oh}},
  \bibinfo{author}{\bibfnamefont{C.}~\bibnamefont{Brook}},
  \bibinfo{author}{\bibfnamefont{F.}~\bibnamefont{Governato}},
  \bibinfo{author}{\bibfnamefont{E.}~\bibnamefont{Brinks}},
  \bibinfo{author}{\bibfnamefont{L.}~\bibnamefont{Mayer}}, \bibnamefont{et~al.}
  (\bibinfo{year}{2010}), \eprint{1011.2777}.

\bibitem[{\citenamefont{Pontzen and Governato}(2011)}]{Pontzen:2011ty}
\bibinfo{author}{\bibfnamefont{A.}~\bibnamefont{Pontzen}} \bibnamefont{and}
  \bibinfo{author}{\bibfnamefont{F.}~\bibnamefont{Governato}}
  (\bibinfo{year}{2011}), \eprint{1106.0499}.

\bibitem[{\citenamefont{Governato et~al.}(2012)\citenamefont{Governato,
  Zolotov, Pontzen, Christensen, Oh et~al.}}]{Governato:2012fa}
\bibinfo{author}{\bibfnamefont{F.}~\bibnamefont{Governato}},
  \bibinfo{author}{\bibfnamefont{A.}~\bibnamefont{Zolotov}},
  \bibinfo{author}{\bibfnamefont{A.}~\bibnamefont{Pontzen}},
  \bibinfo{author}{\bibfnamefont{C.}~\bibnamefont{Christensen}},
  \bibinfo{author}{\bibfnamefont{S.}~\bibnamefont{Oh}}, \bibnamefont{et~al.},
  \bibinfo{journal}{Mon.Not.Roy.Astron.Soc.} \textbf{\bibinfo{volume}{422}},
  \bibinfo{pages}{1231} (\bibinfo{year}{2012}), \eprint{1202.0554}.

\bibitem[{\citenamefont{Zolotov et~al.}(2012)\citenamefont{Zolotov, Brooks,
  Willman, Governato, Pontzen et~al.}}]{Zolotov:2012xd}
\bibinfo{author}{\bibfnamefont{A.}~\bibnamefont{Zolotov}},
  \bibinfo{author}{\bibfnamefont{A.~M.} \bibnamefont{Brooks}},
  \bibinfo{author}{\bibfnamefont{B.}~\bibnamefont{Willman}},
  \bibinfo{author}{\bibfnamefont{F.}~\bibnamefont{Governato}},
  \bibinfo{author}{\bibfnamefont{A.}~\bibnamefont{Pontzen}},
  \bibnamefont{et~al.}, \bibinfo{journal}{Astrophys.J.}
  \textbf{\bibinfo{volume}{761}}, \bibinfo{pages}{71} (\bibinfo{year}{2012}),
  \eprint{1207.0007}.

\bibitem[{\citenamefont{Brooks and Zolotov}(2012)}]{Brooks:2012vi}
\bibinfo{author}{\bibfnamefont{A.~M.} \bibnamefont{Brooks}} \bibnamefont{and}
  \bibinfo{author}{\bibfnamefont{A.}~\bibnamefont{Zolotov}}
  (\bibinfo{year}{2012}), \eprint{1207.2468}.

\bibitem[{\citenamefont{Shen et~al.}(2013)\citenamefont{Shen, Madau, Conroy,
  Governato, and Mayer}}]{Shen:2013wva}
\bibinfo{author}{\bibfnamefont{S.}~\bibnamefont{Shen}},
  \bibinfo{author}{\bibfnamefont{P.}~\bibnamefont{Madau}},
  \bibinfo{author}{\bibfnamefont{C.}~\bibnamefont{Conroy}},
  \bibinfo{author}{\bibfnamefont{F.}~\bibnamefont{Governato}},
  \bibnamefont{and} \bibinfo{author}{\bibfnamefont{L.}~\bibnamefont{Mayer}}
  (\bibinfo{year}{2013}), \eprint{1308.4131}.

\bibitem[{\citenamefont{Blumenthal et~al.}(1984)\citenamefont{Blumenthal,
  Faber, Primack, and Rees}}]{Blumenthal:1984bp}
\bibinfo{author}{\bibfnamefont{G.~R.} \bibnamefont{Blumenthal}},
  \bibinfo{author}{\bibfnamefont{S.}~\bibnamefont{Faber}},
  \bibinfo{author}{\bibfnamefont{J.~R.} \bibnamefont{Primack}},
  \bibnamefont{and} \bibinfo{author}{\bibfnamefont{M.~J.} \bibnamefont{Rees}},
  \bibinfo{journal}{Nature} \textbf{\bibinfo{volume}{311}},
  \bibinfo{pages}{517} (\bibinfo{year}{1984}).

\bibitem[{\citenamefont{Davis et~al.}(1985)\citenamefont{Davis, Efstathiou,
  Frenk, and White}}]{Davis:1985rj}
\bibinfo{author}{\bibfnamefont{M.}~\bibnamefont{Davis}},
  \bibinfo{author}{\bibfnamefont{G.}~\bibnamefont{Efstathiou}},
  \bibinfo{author}{\bibfnamefont{C.~S.} \bibnamefont{Frenk}}, \bibnamefont{and}
  \bibinfo{author}{\bibfnamefont{S.~D.} \bibnamefont{White}},
  \bibinfo{journal}{Astrophys.J.} \textbf{\bibinfo{volume}{292}},
  \bibinfo{pages}{371} (\bibinfo{year}{1985}).

\bibitem[{\citenamefont{Foot}(2004)}]{Foot:2004pa}
\bibinfo{author}{\bibfnamefont{R.}~\bibnamefont{Foot}},
  \bibinfo{journal}{Int.J.Mod.Phys.} \textbf{\bibinfo{volume}{D13}},
  \bibinfo{pages}{2161} (\bibinfo{year}{2004}), \eprint{astro-ph/0407623}.

\bibitem[{\citenamefont{Arkani-Hamed et~al.}(2009)\citenamefont{Arkani-Hamed,
  Finkbeiner, Slatyer, and Weiner}}]{ArkaniHamed:2008qn}
\bibinfo{author}{\bibfnamefont{N.}~\bibnamefont{Arkani-Hamed}},
  \bibinfo{author}{\bibfnamefont{D.~P.} \bibnamefont{Finkbeiner}},
  \bibinfo{author}{\bibfnamefont{T.~R.} \bibnamefont{Slatyer}},
  \bibnamefont{and} \bibinfo{author}{\bibfnamefont{N.}~\bibnamefont{Weiner}},
  \bibinfo{journal}{Phys.Rev.} \textbf{\bibinfo{volume}{D79}},
  \bibinfo{pages}{015014} (\bibinfo{year}{2009}), \eprint{0810.0713}.

\bibitem[{\citenamefont{Ackerman et~al.}(2009)\citenamefont{Ackerman, Buckley,
  Carroll, and Kamionkowski}}]{Ackerman:2008gi}
\bibinfo{author}{\bibfnamefont{L.}~\bibnamefont{Ackerman}},
  \bibinfo{author}{\bibfnamefont{M.~R.} \bibnamefont{Buckley}},
  \bibinfo{author}{\bibfnamefont{S.~M.} \bibnamefont{Carroll}},
  \bibnamefont{and}
  \bibinfo{author}{\bibfnamefont{M.}~\bibnamefont{Kamionkowski}},
  \bibinfo{journal}{Phys.Rev.} \textbf{\bibinfo{volume}{D79}},
  \bibinfo{pages}{023519} (\bibinfo{year}{2009}), \eprint{0810.5126}.

\bibitem[{\citenamefont{Feng et~al.}(2009)\citenamefont{Feng, Kaplinghat, Tu,
  and Yu}}]{Feng:2009mn}
\bibinfo{author}{\bibfnamefont{J.~L.} \bibnamefont{Feng}},
  \bibinfo{author}{\bibfnamefont{M.}~\bibnamefont{Kaplinghat}},
  \bibinfo{author}{\bibfnamefont{H.}~\bibnamefont{Tu}}, \bibnamefont{and}
  \bibinfo{author}{\bibfnamefont{H.-B.} \bibnamefont{Yu}},
  \bibinfo{journal}{JCAP} \textbf{\bibinfo{volume}{0907}}, \bibinfo{pages}{004}
  (\bibinfo{year}{2009}), \eprint{0905.3039}.

\bibitem[{\citenamefont{Baldi}(2012)}]{Baldi:2012ua}
\bibinfo{author}{\bibfnamefont{M.}~\bibnamefont{Baldi}} (\bibinfo{year}{2012}),
  \eprint{1206.2348}.

\bibitem[{\citenamefont{Tulin et~al.}(2013{\natexlab{a}})\citenamefont{Tulin,
  Yu, and Zurek}}]{Tulin:2012wi}
\bibinfo{author}{\bibfnamefont{S.}~\bibnamefont{Tulin}},
  \bibinfo{author}{\bibfnamefont{H.-B.} \bibnamefont{Yu}}, \bibnamefont{and}
  \bibinfo{author}{\bibfnamefont{K.~M.} \bibnamefont{Zurek}},
  \bibinfo{journal}{Phys.Rev.Lett.} \textbf{\bibinfo{volume}{110}},
  \bibinfo{pages}{111301} (\bibinfo{year}{2013}{\natexlab{a}}),
  \eprint{1210.0900}.

\bibitem[{\citenamefont{Hooper et~al.}(2012)\citenamefont{Hooper, Weiner, and
  Xue}}]{Hooper:2012cw}
\bibinfo{author}{\bibfnamefont{D.}~\bibnamefont{Hooper}},
  \bibinfo{author}{\bibfnamefont{N.}~\bibnamefont{Weiner}}, \bibnamefont{and}
  \bibinfo{author}{\bibfnamefont{W.}~\bibnamefont{Xue}},
  \bibinfo{journal}{Phys.Rev.} \textbf{\bibinfo{volume}{D86}},
  \bibinfo{pages}{056009} (\bibinfo{year}{2012}), \eprint{1206.2929}.

\bibitem[{\citenamefont{van~den Aarssen et~al.}(2012)\citenamefont{van~den
  Aarssen, Bringmann, and Pfrommer}}]{Aarssen:2012fx}
\bibinfo{author}{\bibfnamefont{L.~G.} \bibnamefont{van~den Aarssen}},
  \bibinfo{author}{\bibfnamefont{T.}~\bibnamefont{Bringmann}},
  \bibnamefont{and} \bibinfo{author}{\bibfnamefont{C.}~\bibnamefont{Pfrommer}},
  \bibinfo{journal}{Phys.Rev.Lett.} \textbf{\bibinfo{volume}{109}},
  \bibinfo{pages}{231301} (\bibinfo{year}{2012}), \eprint{1205.5809}.

\bibitem[{\citenamefont{Tulin et~al.}(2013{\natexlab{b}})\citenamefont{Tulin,
  Yu, and Zurek}}]{Tulin:2013teo}
\bibinfo{author}{\bibfnamefont{S.}~\bibnamefont{Tulin}},
  \bibinfo{author}{\bibfnamefont{H.-B.} \bibnamefont{Yu}}, \bibnamefont{and}
  \bibinfo{author}{\bibfnamefont{K.~M.} \bibnamefont{Zurek}},
  \bibinfo{journal}{Phys.Rev.} \textbf{\bibinfo{volume}{D87}},
  \bibinfo{pages}{115007} (\bibinfo{year}{2013}{\natexlab{b}}),
  \eprint{1302.3898}.

\bibitem[{\citenamefont{Bode et~al.}(2001)\citenamefont{Bode, Ostriker, and
  Turok}}]{Bode:2000gq}
\bibinfo{author}{\bibfnamefont{P.}~\bibnamefont{Bode}},
  \bibinfo{author}{\bibfnamefont{J.~P.} \bibnamefont{Ostriker}},
  \bibnamefont{and} \bibinfo{author}{\bibfnamefont{N.}~\bibnamefont{Turok}},
  \bibinfo{journal}{Astrophys.J.} \textbf{\bibinfo{volume}{556}},
  \bibinfo{pages}{93} (\bibinfo{year}{2001}), \eprint{astro-ph/0010389}.

\bibitem[{\citenamefont{Dalcanton and Hogan}(2001)}]{Dalcanton:2000hn}
\bibinfo{author}{\bibfnamefont{J.~J.} \bibnamefont{Dalcanton}}
  \bibnamefont{and} \bibinfo{author}{\bibfnamefont{C.~J.} \bibnamefont{Hogan}},
  \bibinfo{journal}{Astrophys.J.} \textbf{\bibinfo{volume}{561}},
  \bibinfo{pages}{35} (\bibinfo{year}{2001}), \eprint{astro-ph/0004381}.

\bibitem[{\citenamefont{Zentner and Bullock}(2003)}]{Zentner:2003yd}
\bibinfo{author}{\bibfnamefont{A.~R.} \bibnamefont{Zentner}} \bibnamefont{and}
  \bibinfo{author}{\bibfnamefont{J.~S.} \bibnamefont{Bullock}},
  \bibinfo{journal}{Astrophys.J.} \textbf{\bibinfo{volume}{598}},
  \bibinfo{pages}{49} (\bibinfo{year}{2003}), \eprint{astro-ph/0304292}.

\bibitem[{\citenamefont{Smith and Markovic}(2011)}]{Smith:2011ev}
\bibinfo{author}{\bibfnamefont{R.~E.} \bibnamefont{Smith}} \bibnamefont{and}
  \bibinfo{author}{\bibfnamefont{K.}~\bibnamefont{Markovic}},
  \bibinfo{journal}{Phys.Rev.} \textbf{\bibinfo{volume}{D84}},
  \bibinfo{pages}{063507} (\bibinfo{year}{2011}), \eprint{1103.2134}.

\bibitem[{\citenamefont{Cyr-Racine and
  Sigurdson}(2013{\natexlab{a}})}]{CyrRacine:2012fz}
\bibinfo{author}{\bibfnamefont{F.-Y.} \bibnamefont{Cyr-Racine}}
  \bibnamefont{and}
  \bibinfo{author}{\bibfnamefont{K.}~\bibnamefont{Sigurdson}},
  \bibinfo{journal}{Phys.Rev.} \textbf{\bibinfo{volume}{D87}},
  \bibinfo{pages}{103515} (\bibinfo{year}{2013}{\natexlab{a}}),
  \eprint{1209.5752}.

\bibitem[{\citenamefont{Mangano et~al.}(2006)\citenamefont{Mangano, Melchiorri,
  Serra, Cooray, and Kamionkowski}}]{Mangano:2006mp}
\bibinfo{author}{\bibfnamefont{G.}~\bibnamefont{Mangano}},
  \bibinfo{author}{\bibfnamefont{A.}~\bibnamefont{Melchiorri}},
  \bibinfo{author}{\bibfnamefont{P.}~\bibnamefont{Serra}},
  \bibinfo{author}{\bibfnamefont{A.}~\bibnamefont{Cooray}}, \bibnamefont{and}
  \bibinfo{author}{\bibfnamefont{M.}~\bibnamefont{Kamionkowski}},
  \bibinfo{journal}{Phys.Rev.} \textbf{\bibinfo{volume}{D74}},
  \bibinfo{pages}{043517} (\bibinfo{year}{2006}), \eprint{astro-ph/0606190}.

\bibitem[{\citenamefont{Serra et~al.}(2010)\citenamefont{Serra, Zalamea,
  Cooray, Mangano, and Melchiorri}}]{Serra:2009uu}
\bibinfo{author}{\bibfnamefont{P.}~\bibnamefont{Serra}},
  \bibinfo{author}{\bibfnamefont{F.}~\bibnamefont{Zalamea}},
  \bibinfo{author}{\bibfnamefont{A.}~\bibnamefont{Cooray}},
  \bibinfo{author}{\bibfnamefont{G.}~\bibnamefont{Mangano}}, \bibnamefont{and}
  \bibinfo{author}{\bibfnamefont{A.}~\bibnamefont{Melchiorri}},
  \bibinfo{journal}{Phys.Rev.} \textbf{\bibinfo{volume}{D81}},
  \bibinfo{pages}{043507} (\bibinfo{year}{2010}), \eprint{0911.4411}.

\bibitem[{\citenamefont{Diamanti et~al.}(2013)\citenamefont{Diamanti, Giusarma,
  Mena, Archidiacono, and Melchiorri}}]{Diamanti:2012tg}
\bibinfo{author}{\bibfnamefont{R.}~\bibnamefont{Diamanti}},
  \bibinfo{author}{\bibfnamefont{E.}~\bibnamefont{Giusarma}},
  \bibinfo{author}{\bibfnamefont{O.}~\bibnamefont{Mena}},
  \bibinfo{author}{\bibfnamefont{M.}~\bibnamefont{Archidiacono}},
  \bibnamefont{and}
  \bibinfo{author}{\bibfnamefont{A.}~\bibnamefont{Melchiorri}},
  \bibinfo{journal}{Phys.Rev.} \textbf{\bibinfo{volume}{D87}},
  \bibinfo{pages}{063509} (\bibinfo{year}{2013}), \eprint{1212.6007}.

\bibitem[{\citenamefont{Boehm et~al.}(2002)\citenamefont{Boehm, Riazuelo,
  Hansen, and Schaeffer}}]{Boehm:2001hm}
\bibinfo{author}{\bibfnamefont{C.}~\bibnamefont{Boehm}},
  \bibinfo{author}{\bibfnamefont{A.}~\bibnamefont{Riazuelo}},
  \bibinfo{author}{\bibfnamefont{S.~H.} \bibnamefont{Hansen}},
  \bibnamefont{and}
  \bibinfo{author}{\bibfnamefont{R.}~\bibnamefont{Schaeffer}},
  \bibinfo{journal}{Phys.Rev.} \textbf{\bibinfo{volume}{D66}},
  \bibinfo{pages}{083505} (\bibinfo{year}{2002}), \eprint{astro-ph/0112522}.

\bibitem[{\citenamefont{McDermott et~al.}(2011)\citenamefont{McDermott, Yu, and
  Zurek}}]{McDermott:2010pa}
\bibinfo{author}{\bibfnamefont{S.~D.} \bibnamefont{McDermott}},
  \bibinfo{author}{\bibfnamefont{H.-B.} \bibnamefont{Yu}}, \bibnamefont{and}
  \bibinfo{author}{\bibfnamefont{K.~M.} \bibnamefont{Zurek}},
  \bibinfo{journal}{Phys.Rev.} \textbf{\bibinfo{volume}{D83}},
  \bibinfo{pages}{063509} (\bibinfo{year}{2011}), \eprint{1011.2907}.

\bibitem[{\citenamefont{Wilkinson et~al.}(2013)\citenamefont{Wilkinson,
  Lesgourgues, and Boehm}}]{Wilkinson:2013kia}
\bibinfo{author}{\bibfnamefont{R.~J.} \bibnamefont{Wilkinson}},
  \bibinfo{author}{\bibfnamefont{J.}~\bibnamefont{Lesgourgues}},
  \bibnamefont{and} \bibinfo{author}{\bibfnamefont{C.}~\bibnamefont{Boehm}}
  (\bibinfo{year}{2013}), \eprint{1309.7588}.

\bibitem[{\citenamefont{Kaplan et~al.}(2010)\citenamefont{Kaplan, Krnjaic,
  Rehermann, and Wells}}]{Kaplan:2009de}
\bibinfo{author}{\bibfnamefont{D.~E.} \bibnamefont{Kaplan}},
  \bibinfo{author}{\bibfnamefont{G.~Z.} \bibnamefont{Krnjaic}},
  \bibinfo{author}{\bibfnamefont{K.~R.} \bibnamefont{Rehermann}},
  \bibnamefont{and} \bibinfo{author}{\bibfnamefont{C.~M.} \bibnamefont{Wells}},
  \bibinfo{journal}{JCAP} \textbf{\bibinfo{volume}{1005}}, \bibinfo{pages}{021}
  (\bibinfo{year}{2010}), \eprint{0909.0753}.

\bibitem[{\citenamefont{Kaplan et~al.}(2011)\citenamefont{Kaplan, Krnjaic,
  Rehermann, and Wells}}]{Kaplan:2011yj}
\bibinfo{author}{\bibfnamefont{D.~E.} \bibnamefont{Kaplan}},
  \bibinfo{author}{\bibfnamefont{G.~Z.} \bibnamefont{Krnjaic}},
  \bibinfo{author}{\bibfnamefont{K.~R.} \bibnamefont{Rehermann}},
  \bibnamefont{and} \bibinfo{author}{\bibfnamefont{C.~M.} \bibnamefont{Wells}},
  \bibinfo{journal}{JCAP} \textbf{\bibinfo{volume}{1110}}, \bibinfo{pages}{011}
  (\bibinfo{year}{2011}), \eprint{1105.2073}.

\bibitem[{\citenamefont{{Gradwohl} and {Frieman}}(1992)}]{1992ApJ...398..407G}
\bibinfo{author}{\bibfnamefont{B.-A.} \bibnamefont{{Gradwohl}}}
  \bibnamefont{and} \bibinfo{author}{\bibfnamefont{J.~A.}
  \bibnamefont{{Frieman}}}, \bibinfo{journal}{\apj}
  \textbf{\bibinfo{volume}{398}}, \bibinfo{pages}{407} (\bibinfo{year}{1992}).

\bibitem[{\citenamefont{{Carlson} et~al.}(1992)\citenamefont{{Carlson},
  {Machacek}, and {Hall}}}]{1992ApJ...398...43C}
\bibinfo{author}{\bibfnamefont{E.~D.} \bibnamefont{{Carlson}}},
  \bibinfo{author}{\bibfnamefont{M.~E.} \bibnamefont{{Machacek}}},
  \bibnamefont{and} \bibinfo{author}{\bibfnamefont{L.~J.}
  \bibnamefont{{Hall}}}, \bibinfo{journal}{\apj}
  \textbf{\bibinfo{volume}{398}}, \bibinfo{pages}{43} (\bibinfo{year}{1992}).

\bibitem[{\citenamefont{{Machacek}}(1994)}]{1994ApJ...431...41M}
\bibinfo{author}{\bibfnamefont{M.~E.} \bibnamefont{{Machacek}}},
  \bibinfo{journal}{\apj} \textbf{\bibinfo{volume}{431}}, \bibinfo{pages}{41}
  (\bibinfo{year}{1994}).

\bibitem[{\citenamefont{{de Laix} et~al.}(1995)\citenamefont{{de Laix},
  {Scherrer}, and {Schaefer}}}]{1995ApJ...452..495D}
\bibinfo{author}{\bibfnamefont{A.~A.} \bibnamefont{{de Laix}}},
  \bibinfo{author}{\bibfnamefont{R.~J.} \bibnamefont{{Scherrer}}},
  \bibnamefont{and} \bibinfo{author}{\bibfnamefont{R.~K.}
  \bibnamefont{{Schaefer}}}, \bibinfo{journal}{\apj}
  \textbf{\bibinfo{volume}{452}}, \bibinfo{pages}{495} (\bibinfo{year}{1995}),
  \eprint{arXiv:astro-ph/9502087}.

\bibitem[{\citenamefont{Fan et~al.}(2013{\natexlab{a}})\citenamefont{Fan, Katz,
  Randall, and Reece}}]{Fan:2013tia}
\bibinfo{author}{\bibfnamefont{J.}~\bibnamefont{Fan}},
  \bibinfo{author}{\bibfnamefont{A.}~\bibnamefont{Katz}},
  \bibinfo{author}{\bibfnamefont{L.}~\bibnamefont{Randall}}, \bibnamefont{and}
  \bibinfo{author}{\bibfnamefont{M.}~\bibnamefont{Reece}},
  \bibinfo{journal}{Phys.Rev.Lett.} \textbf{\bibinfo{volume}{110}},
  \bibinfo{pages}{211302} (\bibinfo{year}{2013}{\natexlab{a}}),
  \eprint{1303.3271}.

\bibitem[{\citenamefont{Fan et~al.}(2013{\natexlab{b}})\citenamefont{Fan, Katz,
  Randall, and Reece}}]{Fan:2013yva}
\bibinfo{author}{\bibfnamefont{J.}~\bibnamefont{Fan}},
  \bibinfo{author}{\bibfnamefont{A.}~\bibnamefont{Katz}},
  \bibinfo{author}{\bibfnamefont{L.}~\bibnamefont{Randall}}, \bibnamefont{and}
  \bibinfo{author}{\bibfnamefont{M.}~\bibnamefont{Reece}}
  (\bibinfo{year}{2013}{\natexlab{b}}), \eprint{1303.1521}.

\bibitem[{\citenamefont{McCullough and Randall}(2013)}]{McCullough:2013jma}
\bibinfo{author}{\bibfnamefont{M.}~\bibnamefont{McCullough}} \bibnamefont{and}
  \bibinfo{author}{\bibfnamefont{L.}~\bibnamefont{Randall}}
  (\bibinfo{year}{2013}), \eprint{1307.4095}.

\bibitem[{\citenamefont{Rocha et~al.}(2013)\citenamefont{Rocha, Peter, Bullock,
  Kaplinghat, Garrison-Kimmel et~al.}}]{Rocha:2012jg}
\bibinfo{author}{\bibfnamefont{M.}~\bibnamefont{Rocha}},
  \bibinfo{author}{\bibfnamefont{A.~H.} \bibnamefont{Peter}},
  \bibinfo{author}{\bibfnamefont{J.~S.} \bibnamefont{Bullock}},
  \bibinfo{author}{\bibfnamefont{M.}~\bibnamefont{Kaplinghat}},
  \bibinfo{author}{\bibfnamefont{S.}~\bibnamefont{Garrison-Kimmel}},
  \bibnamefont{et~al.}, \bibinfo{journal}{Mon.Not.Roy.Astron.Soc.}
  \textbf{\bibinfo{volume}{430}}, \bibinfo{pages}{81} (\bibinfo{year}{2013}),
  \eprint{1208.3025}.

\bibitem[{\citenamefont{Peter et~al.}(2012)\citenamefont{Peter, Rocha, Bullock,
  and Kaplinghat}}]{Peter:2012jh}
\bibinfo{author}{\bibfnamefont{A.~H.} \bibnamefont{Peter}},
  \bibinfo{author}{\bibfnamefont{M.}~\bibnamefont{Rocha}},
  \bibinfo{author}{\bibfnamefont{J.~S.} \bibnamefont{Bullock}},
  \bibnamefont{and}
  \bibinfo{author}{\bibfnamefont{M.}~\bibnamefont{Kaplinghat}}
  (\bibinfo{year}{2012}), \eprint{1208.3026}.

\bibitem[{\citenamefont{Hinshaw et~al.}(2012)}]{Hinshaw:2012aka}
\bibinfo{author}{\bibfnamefont{G.}~\bibnamefont{Hinshaw}} \bibnamefont{et~al.}
  (\bibinfo{collaboration}{WMAP Collaboration}) (\bibinfo{year}{2012}),
  \eprint{1212.5226}.

\bibitem[{\citenamefont{Ade et~al.}(2013{\natexlab{a}})}]{planckXVI}
\bibinfo{author}{\bibfnamefont{P.}~\bibnamefont{Ade}} \bibnamefont{et~al.}
  (\bibinfo{collaboration}{Planck Collaboration})
  (\bibinfo{year}{2013}{\natexlab{a}}), \eprint{1303.5076}.

\bibitem[{\citenamefont{Riess et~al.}(2011)\citenamefont{Riess, Macri,
  Casertano, Lampeitl, Ferguson et~al.}}]{Riess:2011yx}
\bibinfo{author}{\bibfnamefont{A.~G.} \bibnamefont{Riess}},
  \bibinfo{author}{\bibfnamefont{L.}~\bibnamefont{Macri}},
  \bibinfo{author}{\bibfnamefont{S.}~\bibnamefont{Casertano}},
  \bibinfo{author}{\bibfnamefont{H.}~\bibnamefont{Lampeitl}},
  \bibinfo{author}{\bibfnamefont{H.~C.} \bibnamefont{Ferguson}},
  \bibnamefont{et~al.}, \bibinfo{journal}{Astrophys.J.}
  \textbf{\bibinfo{volume}{730}}, \bibinfo{pages}{119} (\bibinfo{year}{2011}),
  \eprint{1103.2976}.

\bibitem[{\citenamefont{Freedman et~al.}(2012)\citenamefont{Freedman, Madore,
  Scowcroft, Burns, Monson et~al.}}]{Freedman:2012ny}
\bibinfo{author}{\bibfnamefont{W.~L.} \bibnamefont{Freedman}},
  \bibinfo{author}{\bibfnamefont{B.~F.} \bibnamefont{Madore}},
  \bibinfo{author}{\bibfnamefont{V.}~\bibnamefont{Scowcroft}},
  \bibinfo{author}{\bibfnamefont{C.}~\bibnamefont{Burns}},
  \bibinfo{author}{\bibfnamefont{A.}~\bibnamefont{Monson}},
  \bibnamefont{et~al.}, \bibinfo{journal}{Astrophys.J.}
  \textbf{\bibinfo{volume}{758}}, \bibinfo{pages}{24} (\bibinfo{year}{2012}),
  \eprint{1208.3281}.

\bibitem[{\citenamefont{Steigman}(2013)}]{Steigman:2013yua}
\bibinfo{author}{\bibfnamefont{G.}~\bibnamefont{Steigman}},
  \bibinfo{journal}{Phys.Rev.} \textbf{\bibinfo{volume}{D87}},
  \bibinfo{pages}{103517} (\bibinfo{year}{2013}), \eprint{1303.0049}.

\bibitem[{\citenamefont{Goldberg and Hall}(1986)}]{Goldberg:1986nk}
\bibinfo{author}{\bibfnamefont{H.}~\bibnamefont{Goldberg}} \bibnamefont{and}
  \bibinfo{author}{\bibfnamefont{L.~J.} \bibnamefont{Hall}},
  \bibinfo{journal}{Phys.Lett.} \textbf{\bibinfo{volume}{B174}},
  \bibinfo{pages}{151} (\bibinfo{year}{1986}).

\bibitem[{\citenamefont{Behbahani et~al.}(2011)\citenamefont{Behbahani,
  Jankowiak, Rube, and Wacker}}]{Behbahani:2010xa}
\bibinfo{author}{\bibfnamefont{S.~R.} \bibnamefont{Behbahani}},
  \bibinfo{author}{\bibfnamefont{M.}~\bibnamefont{Jankowiak}},
  \bibinfo{author}{\bibfnamefont{T.}~\bibnamefont{Rube}}, \bibnamefont{and}
  \bibinfo{author}{\bibfnamefont{J.~G.} \bibnamefont{Wacker}},
  \bibinfo{journal}{Adv.High Energy Phys.} \textbf{\bibinfo{volume}{2011}},
  \bibinfo{pages}{709492} (\bibinfo{year}{2011}), \eprint{1009.3523}.

\bibitem[{\citenamefont{Petraki et~al.}(2012)\citenamefont{Petraki, Trodden,
  and Volkas}}]{Petraki:2011mv}
\bibinfo{author}{\bibfnamefont{K.}~\bibnamefont{Petraki}},
  \bibinfo{author}{\bibfnamefont{M.}~\bibnamefont{Trodden}}, \bibnamefont{and}
  \bibinfo{author}{\bibfnamefont{R.~R.} \bibnamefont{Volkas}},
  \bibinfo{journal}{JCAP} \textbf{\bibinfo{volume}{1202}}, \bibinfo{pages}{044}
  (\bibinfo{year}{2012}), \eprint{1111.4786}.

\bibitem[{\citenamefont{Cline et~al.}(2012)\citenamefont{Cline, Liu, and
  Xue}}]{Cline:2012is}
\bibinfo{author}{\bibfnamefont{J.~M.} \bibnamefont{Cline}},
  \bibinfo{author}{\bibfnamefont{Z.}~\bibnamefont{Liu}}, \bibnamefont{and}
  \bibinfo{author}{\bibfnamefont{W.}~\bibnamefont{Xue}},
  \bibinfo{journal}{Phys.Rev.} \textbf{\bibinfo{volume}{D85}},
  \bibinfo{pages}{101302} (\bibinfo{year}{2012}), \eprint{1201.4858}.

\bibitem[{\citenamefont{Boehm et~al.}(2001)\citenamefont{Boehm, Fayet, and
  Schaeffer}}]{Boehm:2000gq}
\bibinfo{author}{\bibfnamefont{C.}~\bibnamefont{Boehm}},
  \bibinfo{author}{\bibfnamefont{P.}~\bibnamefont{Fayet}}, \bibnamefont{and}
  \bibinfo{author}{\bibfnamefont{R.}~\bibnamefont{Schaeffer}},
  \bibinfo{journal}{Phys.Lett.} \textbf{\bibinfo{volume}{B518}},
  \bibinfo{pages}{8} (\bibinfo{year}{2001}), \eprint{astro-ph/0012504}.

\bibitem[{\citenamefont{Ma and Bertschinger}(1995)}]{Ma:1995ey}
\bibinfo{author}{\bibfnamefont{C.-P.} \bibnamefont{Ma}} \bibnamefont{and}
  \bibinfo{author}{\bibfnamefont{E.}~\bibnamefont{Bertschinger}},
  \bibinfo{journal}{Astrophys.J.} \textbf{\bibinfo{volume}{455}},
  \bibinfo{pages}{7} (\bibinfo{year}{1995}), \eprint{astro-ph/9506072}.

\bibitem[{\citenamefont{Lewis et~al.}(2000)\citenamefont{Lewis, Challinor, and
  Lasenby}}]{Lewis:1999bs}
\bibinfo{author}{\bibfnamefont{A.}~\bibnamefont{Lewis}},
  \bibinfo{author}{\bibfnamefont{A.}~\bibnamefont{Challinor}},
  \bibnamefont{and} \bibinfo{author}{\bibfnamefont{A.}~\bibnamefont{Lasenby}},
  \bibinfo{journal}{Astrophys.J.} \textbf{\bibinfo{volume}{538}},
  \bibinfo{pages}{473} (\bibinfo{year}{2000}), \eprint{astro-ph/9911177}.

\bibitem[{\citenamefont{{Cyr-Racine} and
  {Sigurdson}}(2011)}]{2011PhRvD..83j3521C}
\bibinfo{author}{\bibfnamefont{F.-Y.} \bibnamefont{{Cyr-Racine}}}
  \bibnamefont{and}
  \bibinfo{author}{\bibfnamefont{K.}~\bibnamefont{{Sigurdson}}},
  \bibinfo{journal}{\prd} \textbf{\bibinfo{volume}{83}}, \bibinfo{eid}{103521}
  (\bibinfo{year}{2011}), \eprint{1012.0569}.

\bibitem[{\citenamefont{Pitrou}(2011)}]{Pitrou:2010ai}
\bibinfo{author}{\bibfnamefont{C.}~\bibnamefont{Pitrou}},
  \bibinfo{journal}{Phys.Lett.} \textbf{\bibinfo{volume}{B698}},
  \bibinfo{pages}{1} (\bibinfo{year}{2011}), \eprint{1012.0546}.

\bibitem[{\citenamefont{Samushia et~al.}(2013)\citenamefont{Samushia, Reid,
  White, Percival, Cuesta, Lombriser, Manera, Nichol, Schneider, Bizyaev
  et~al.}}]{Samushia:2013vn}
\bibinfo{author}{\bibfnamefont{L.}~\bibnamefont{Samushia}},
  \bibinfo{author}{\bibfnamefont{B.~A.} \bibnamefont{Reid}},
  \bibinfo{author}{\bibfnamefont{M.}~\bibnamefont{White}},
  \bibinfo{author}{\bibfnamefont{W.~J.} \bibnamefont{Percival}},
  \bibinfo{author}{\bibfnamefont{A.~J.} \bibnamefont{Cuesta}},
  \bibinfo{author}{\bibfnamefont{L.}~\bibnamefont{Lombriser}},
  \bibinfo{author}{\bibfnamefont{M.}~\bibnamefont{Manera}},
  \bibinfo{author}{\bibfnamefont{R.~C.} \bibnamefont{Nichol}},
  \bibinfo{author}{\bibfnamefont{D.~P.} \bibnamefont{Schneider}},
  \bibinfo{author}{\bibfnamefont{D.}~\bibnamefont{Bizyaev}},
  \bibnamefont{et~al.}, \bibinfo{journal}{Monthly Notices of the Royal
  Astronomical Society, Volume 429, Issue} \textbf{\bibinfo{volume}{2}},
  \bibinfo{pages}{p.1514} (\bibinfo{year}{2013}), \eprint{1206.5309}.

\bibitem[{\citenamefont{{Raccanelli}
  et~al.}(2013{\natexlab{a}})\citenamefont{{Raccanelli}, {Bertacca},
  {Pietrobon}, {Schmidt}, {Samushia}, {Bartolo}, {Dor{\'e}}, {Matarrese}, and
  {Percival}}}]{Raccanelli:2012fk}
\bibinfo{author}{\bibfnamefont{A.}~\bibnamefont{{Raccanelli}}},
  \bibinfo{author}{\bibfnamefont{D.}~\bibnamefont{{Bertacca}}},
  \bibinfo{author}{\bibfnamefont{D.}~\bibnamefont{{Pietrobon}}},
  \bibinfo{author}{\bibfnamefont{F.}~\bibnamefont{{Schmidt}}},
  \bibinfo{author}{\bibfnamefont{L.}~\bibnamefont{{Samushia}}},
  \bibinfo{author}{\bibfnamefont{N.}~\bibnamefont{{Bartolo}}},
  \bibinfo{author}{\bibfnamefont{O.}~\bibnamefont{{Dor{\'e}}}},
  \bibinfo{author}{\bibfnamefont{S.}~\bibnamefont{{Matarrese}}},
  \bibnamefont{and} \bibinfo{author}{\bibfnamefont{W.~J.}
  \bibnamefont{{Percival}}}, \bibinfo{journal}{\mnras}
  (\bibinfo{year}{2013}{\natexlab{a}}), \eprint{1207.0500}.

\bibitem[{\citenamefont{de~Putter et~al.}(2012)\citenamefont{de~Putter, Mena,
  Giusarma, Ho, Cuesta, Seo, Ross, White, Bizyaev, Brewington
  et~al.}}]{Putter:2012kx}
\bibinfo{author}{\bibfnamefont{R.}~\bibnamefont{de~Putter}},
  \bibinfo{author}{\bibfnamefont{O.}~\bibnamefont{Mena}},
  \bibinfo{author}{\bibfnamefont{E.}~\bibnamefont{Giusarma}},
  \bibinfo{author}{\bibfnamefont{S.}~\bibnamefont{Ho}},
  \bibinfo{author}{\bibfnamefont{A.}~\bibnamefont{Cuesta}},
  \bibinfo{author}{\bibfnamefont{H.-J.} \bibnamefont{Seo}},
  \bibinfo{author}{\bibfnamefont{A.}~\bibnamefont{Ross}},
  \bibinfo{author}{\bibfnamefont{M.}~\bibnamefont{White}},
  \bibinfo{author}{\bibfnamefont{D.}~\bibnamefont{Bizyaev}},
  \bibinfo{author}{\bibfnamefont{H.}~\bibnamefont{Brewington}},
  \bibnamefont{et~al.} (\bibinfo{year}{2012}), \eprint{1201.1909}.

\bibitem[{\citenamefont{Zhao et~al.}(2012)\citenamefont{Zhao, Saito, Percival,
  Ross, Montesano, Viel, Schneider, Ernst, Manera, Miralda-Escude
  et~al.}}]{Zhao:2012ly}
\bibinfo{author}{\bibfnamefont{G.-B.} \bibnamefont{Zhao}},
  \bibinfo{author}{\bibfnamefont{S.}~\bibnamefont{Saito}},
  \bibinfo{author}{\bibfnamefont{W.~J.} \bibnamefont{Percival}},
  \bibinfo{author}{\bibfnamefont{A.~J.} \bibnamefont{Ross}},
  \bibinfo{author}{\bibfnamefont{F.}~\bibnamefont{Montesano}},
  \bibinfo{author}{\bibfnamefont{M.}~\bibnamefont{Viel}},
  \bibinfo{author}{\bibfnamefont{D.~P.} \bibnamefont{Schneider}},
  \bibinfo{author}{\bibfnamefont{D.~J.} \bibnamefont{Ernst}},
  \bibinfo{author}{\bibfnamefont{M.}~\bibnamefont{Manera}},
  \bibinfo{author}{\bibfnamefont{J.}~\bibnamefont{Miralda-Escude}},
  \bibnamefont{et~al.} (\bibinfo{year}{2012}), \eprint{1211.3741}.

\bibitem[{\citenamefont{{Samushia} et~al.}(2012)\citenamefont{{Samushia},
  {Percival}, and {Raccanelli}}}]{Samushia:2011uq}
\bibinfo{author}{\bibfnamefont{L.}~\bibnamefont{{Samushia}}},
  \bibinfo{author}{\bibfnamefont{W.~J.} \bibnamefont{{Percival}}},
  \bibnamefont{and}
  \bibinfo{author}{\bibfnamefont{A.}~\bibnamefont{{Raccanelli}}},
  \bibinfo{journal}{\mnras} \textbf{\bibinfo{volume}{420}},
  \bibinfo{pages}{2102} (\bibinfo{year}{2012}), \eprint{1102.1014}.

\bibitem[{\citenamefont{Reid et~al.}(2012)\citenamefont{Reid, Samushia, White,
  Percival, Manera, Padmanabhan, Ross, S{\'a}nchez, Bailey, Bizyaev
  et~al.}}]{Reid:2012zr}
\bibinfo{author}{\bibfnamefont{B.~A.} \bibnamefont{Reid}},
  \bibinfo{author}{\bibfnamefont{L.}~\bibnamefont{Samushia}},
  \bibinfo{author}{\bibfnamefont{M.}~\bibnamefont{White}},
  \bibinfo{author}{\bibfnamefont{W.~J.} \bibnamefont{Percival}},
  \bibinfo{author}{\bibfnamefont{M.}~\bibnamefont{Manera}},
  \bibinfo{author}{\bibfnamefont{N.}~\bibnamefont{Padmanabhan}},
  \bibinfo{author}{\bibfnamefont{A.~J.} \bibnamefont{Ross}},
  \bibinfo{author}{\bibfnamefont{A.~G.} \bibnamefont{S{\'a}nchez}},
  \bibinfo{author}{\bibfnamefont{S.}~\bibnamefont{Bailey}},
  \bibinfo{author}{\bibfnamefont{D.}~\bibnamefont{Bizyaev}},
  \bibnamefont{et~al.} (\bibinfo{year}{2012}), \eprint{1203.6641}.

\bibitem[{\citenamefont{{Ross} et~al.}(2013)\citenamefont{{Ross}, {Percival},
  {Carnero}, {Zhao}, {Manera}, {Raccanelli}, {Aubourg}, {Bizyaev},
  {Brewington}, {Brinkmann} et~al.}}]{Ross:2012ys3}
\bibinfo{author}{\bibfnamefont{A.~J.} \bibnamefont{{Ross}}},
  \bibinfo{author}{\bibfnamefont{W.~J.} \bibnamefont{{Percival}}},
  \bibinfo{author}{\bibfnamefont{A.}~\bibnamefont{{Carnero}}},
  \bibinfo{author}{\bibfnamefont{G.-b.} \bibnamefont{{Zhao}}},
  \bibinfo{author}{\bibfnamefont{M.}~\bibnamefont{{Manera}}},
  \bibinfo{author}{\bibfnamefont{A.}~\bibnamefont{{Raccanelli}}},
  \bibinfo{author}{\bibfnamefont{E.}~\bibnamefont{{Aubourg}}},
  \bibinfo{author}{\bibfnamefont{D.}~\bibnamefont{{Bizyaev}}},
  \bibinfo{author}{\bibfnamefont{H.}~\bibnamefont{{Brewington}}},
  \bibinfo{author}{\bibfnamefont{J.}~\bibnamefont{{Brinkmann}}},
  \bibnamefont{et~al.}, \bibinfo{journal}{\mnras}
  \textbf{\bibinfo{volume}{428}}, \bibinfo{pages}{1116} (\bibinfo{year}{2013}),
  \eprint{1208.1491}.

\bibitem[{\citenamefont{{Anderson} et~al.}(2012)\citenamefont{{Anderson},
  {Aubourg}, {Bailey}, {Bizyaev}, {Blanton}, {Bolton}, {Brinkmann},
  {Brownstein}, {Burden}, {Cuesta} et~al.}}]{Anderson:2012ve}
\bibinfo{author}{\bibfnamefont{L.}~\bibnamefont{{Anderson}}},
  \bibinfo{author}{\bibfnamefont{E.}~\bibnamefont{{Aubourg}}},
  \bibinfo{author}{\bibfnamefont{S.}~\bibnamefont{{Bailey}}},
  \bibinfo{author}{\bibfnamefont{D.}~\bibnamefont{{Bizyaev}}},
  \bibinfo{author}{\bibfnamefont{M.}~\bibnamefont{{Blanton}}},
  \bibinfo{author}{\bibfnamefont{A.~S.} \bibnamefont{{Bolton}}},
  \bibinfo{author}{\bibfnamefont{J.}~\bibnamefont{{Brinkmann}}},
  \bibinfo{author}{\bibfnamefont{J.~R.} \bibnamefont{{Brownstein}}},
  \bibinfo{author}{\bibfnamefont{A.}~\bibnamefont{{Burden}}},
  \bibinfo{author}{\bibfnamefont{A.~J.} \bibnamefont{{Cuesta}}},
  \bibnamefont{et~al.}, \bibinfo{journal}{\mnras}
  \textbf{\bibinfo{volume}{427}}, \bibinfo{pages}{3435} (\bibinfo{year}{2012}),
  \eprint{1203.6594}.

\bibitem[{\citenamefont{Sanchez et~al.}(2013)\citenamefont{Sanchez, Kazin,
  Beutler, Chuang, Cuesta, Eisenstein, Manera, Montesano, Nichol, Padmanabhan
  et~al.}}]{Sanchez:2013qf}
\bibinfo{author}{\bibfnamefont{A.~G.} \bibnamefont{Sanchez}},
  \bibinfo{author}{\bibfnamefont{E.~A.} \bibnamefont{Kazin}},
  \bibinfo{author}{\bibfnamefont{F.}~\bibnamefont{Beutler}},
  \bibinfo{author}{\bibfnamefont{C.-H.} \bibnamefont{Chuang}},
  \bibinfo{author}{\bibfnamefont{A.~J.} \bibnamefont{Cuesta}},
  \bibinfo{author}{\bibfnamefont{D.~J.} \bibnamefont{Eisenstein}},
  \bibinfo{author}{\bibfnamefont{M.}~\bibnamefont{Manera}},
  \bibinfo{author}{\bibfnamefont{F.}~\bibnamefont{Montesano}},
  \bibinfo{author}{\bibfnamefont{B.}~\bibnamefont{Nichol}},
  \bibinfo{author}{\bibfnamefont{N.}~\bibnamefont{Padmanabhan}},
  \bibnamefont{et~al.} (\bibinfo{year}{2013}), \eprint{1303.4396}.

\bibitem[{\citenamefont{Smith et~al.}(2003)}]{Smith:2002dz}
\bibinfo{author}{\bibfnamefont{R.}~\bibnamefont{Smith}} \bibnamefont{et~al.}
  (\bibinfo{collaboration}{Virgo Consortium}),
  \bibinfo{journal}{Mon.Not.Roy.Astron.Soc.} \textbf{\bibinfo{volume}{341}},
  \bibinfo{pages}{1311} (\bibinfo{year}{2003}), \eprint{astro-ph/0207664}.

\bibitem[{\citenamefont{Xu et~al.}(2012)\citenamefont{Xu, Padmanabhan,
  Eisenstein, Mehta, and Cuesta}}]{Xu:2012uq}
\bibinfo{author}{\bibfnamefont{X.}~\bibnamefont{Xu}},
  \bibinfo{author}{\bibfnamefont{N.}~\bibnamefont{Padmanabhan}},
  \bibinfo{author}{\bibfnamefont{D.~J.} \bibnamefont{Eisenstein}},
  \bibinfo{author}{\bibfnamefont{K.~T.} \bibnamefont{Mehta}}, \bibnamefont{and}
  \bibinfo{author}{\bibfnamefont{A.~J.} \bibnamefont{Cuesta}}
  (\bibinfo{year}{2012}), \eprint{1202.0091}.

\bibitem[{\citenamefont{Taruya et~al.}(2009)\citenamefont{Taruya, Nishimichi,
  Saito, and Hiramatsu}}]{Taruya:2009bs}
\bibinfo{author}{\bibfnamefont{A.}~\bibnamefont{Taruya}},
  \bibinfo{author}{\bibfnamefont{T.}~\bibnamefont{Nishimichi}},
  \bibinfo{author}{\bibfnamefont{S.}~\bibnamefont{Saito}}, \bibnamefont{and}
  \bibinfo{author}{\bibfnamefont{T.}~\bibnamefont{Hiramatsu}}
  (\bibinfo{year}{2009}), \eprint{0906.0507}.

\bibitem[{\citenamefont{Taruya et~al.}(2010)\citenamefont{Taruya, Nishimichi,
  and Saito}}]{Taruya:2010ij}
\bibinfo{author}{\bibfnamefont{A.}~\bibnamefont{Taruya}},
  \bibinfo{author}{\bibfnamefont{T.}~\bibnamefont{Nishimichi}},
  \bibnamefont{and} \bibinfo{author}{\bibfnamefont{S.}~\bibnamefont{Saito}}
  (\bibinfo{year}{2010}), \eprint{1006.0699}.

\bibitem[{\citenamefont{Kwan et~al.}(2011)\citenamefont{Kwan, Lewis, and
  Linder}}]{Kwan:2011fu}
\bibinfo{author}{\bibfnamefont{J.}~\bibnamefont{Kwan}},
  \bibinfo{author}{\bibfnamefont{G.~F.} \bibnamefont{Lewis}}, \bibnamefont{and}
  \bibinfo{author}{\bibfnamefont{E.~V.} \bibnamefont{Linder}},
  \bibinfo{journal}{ApJ, 2012, 748, 78}  (\bibinfo{year}{2011}),
  \eprint{1105.1194}.

\bibitem[{\citenamefont{Jennings}(2012)}]{Jennings:2012fv}
\bibinfo{author}{\bibfnamefont{E.}~\bibnamefont{Jennings}},
  \bibinfo{journal}{2012MNRAS.427L..25J}  (\bibinfo{year}{2012}),
  \eprint{1207.1439}.

\bibitem[{\citenamefont{de~la Torre and Guzzo}(2012)}]{Torre:2012dz}
\bibinfo{author}{\bibfnamefont{S.}~\bibnamefont{de~la Torre}} \bibnamefont{and}
  \bibinfo{author}{\bibfnamefont{L.}~\bibnamefont{Guzzo}}
  (\bibinfo{year}{2012}), \eprint{1202.5559}.

\bibitem[{\citenamefont{{Raccanelli} et~al.}(2010)\citenamefont{{Raccanelli},
  {Samushia}, and {Percival}}}]{Raccanelli:2010fk}
\bibinfo{author}{\bibfnamefont{A.}~\bibnamefont{{Raccanelli}}},
  \bibinfo{author}{\bibfnamefont{L.}~\bibnamefont{{Samushia}}},
  \bibnamefont{and} \bibinfo{author}{\bibfnamefont{W.~J.}
  \bibnamefont{{Percival}}}, \bibinfo{journal}{\mnras}
  \textbf{\bibinfo{volume}{409}}, \bibinfo{pages}{1525} (\bibinfo{year}{2010}),
  \eprint{1006.1652}.

\bibitem[{\citenamefont{Yoo}(2010)}]{Yoo:2010cr}
\bibinfo{author}{\bibfnamefont{J.}~\bibnamefont{Yoo}},
  \bibinfo{journal}{Phys.Rev.D} \textbf{\bibinfo{volume}{82}},
  \bibinfo{pages}{083508} (\bibinfo{year}{2010}), \eprint{1009.3021}.

\bibitem[{\citenamefont{Challinor and Lewis}(2011)}]{Challinor:2011nx}
\bibinfo{author}{\bibfnamefont{A.}~\bibnamefont{Challinor}} \bibnamefont{and}
  \bibinfo{author}{\bibfnamefont{A.}~\bibnamefont{Lewis}},
  \bibinfo{journal}{Phys.Rev.D} \textbf{\bibinfo{volume}{84}},
  \bibinfo{pages}{043516} (\bibinfo{year}{2011}), \eprint{1105.5292}.

\bibitem[{\citenamefont{Bonvin and Durrer}(2011)}]{Bonvin:2011oq}
\bibinfo{author}{\bibfnamefont{C.}~\bibnamefont{Bonvin}} \bibnamefont{and}
  \bibinfo{author}{\bibfnamefont{R.}~\bibnamefont{Durrer}},
  \bibinfo{journal}{Phys.Rev.D} \textbf{\bibinfo{volume}{84}},
  \bibinfo{pages}{063505} (\bibinfo{year}{2011}), \eprint{1105.5280}.

\bibitem[{\citenamefont{{Bertacca} et~al.}(2012)\citenamefont{{Bertacca},
  {Maartens}, {Raccanelli}, and {Clarkson}}}]{Bertacca:2012hc}
\bibinfo{author}{\bibfnamefont{D.}~\bibnamefont{{Bertacca}}},
  \bibinfo{author}{\bibfnamefont{R.}~\bibnamefont{{Maartens}}},
  \bibinfo{author}{\bibfnamefont{A.}~\bibnamefont{{Raccanelli}}},
  \bibnamefont{and}
  \bibinfo{author}{\bibfnamefont{C.}~\bibnamefont{{Clarkson}}},
  \bibinfo{journal}{JCAP} \textbf{\bibinfo{volume}{10}}, \bibinfo{eid}{025}
  (\bibinfo{year}{2012}), \eprint{1205.5221}.

\bibitem[{\citenamefont{Yoo et~al.}(2012{\natexlab{a}})\citenamefont{Yoo,
  Hamaus, Seljak, and Zaldarriaga}}]{Yoo:2012dq}
\bibinfo{author}{\bibfnamefont{J.}~\bibnamefont{Yoo}},
  \bibinfo{author}{\bibfnamefont{N.}~\bibnamefont{Hamaus}},
  \bibinfo{author}{\bibfnamefont{U.}~\bibnamefont{Seljak}}, \bibnamefont{and}
  \bibinfo{author}{\bibfnamefont{M.}~\bibnamefont{Zaldarriaga}},
  \bibinfo{journal}{Phys.Rev.D} \textbf{\bibinfo{volume}{86}},
  \bibinfo{pages}{063514} (\bibinfo{year}{2012}{\natexlab{a}}),
  \eprint{1109.0998}.

\bibitem[{\citenamefont{Yoo et~al.}(2012{\natexlab{b}})\citenamefont{Yoo,
  Hamaus, Seljak, and Zaldarriaga}}]{Yoo:2012bh}
\bibinfo{author}{\bibfnamefont{J.}~\bibnamefont{Yoo}},
  \bibinfo{author}{\bibfnamefont{N.}~\bibnamefont{Hamaus}},
  \bibinfo{author}{\bibfnamefont{U.}~\bibnamefont{Seljak}}, \bibnamefont{and}
  \bibinfo{author}{\bibfnamefont{M.}~\bibnamefont{Zaldarriaga}},
  \bibinfo{journal}{Phys.Rev.D} \textbf{\bibinfo{volume}{86}},
  \bibinfo{pages}{063514} (\bibinfo{year}{2012}{\natexlab{b}}),
  \eprint{1206.5809}.

\bibitem[{\citenamefont{Montanari and Durrer}(2012)}]{Montanari:2012kl}
\bibinfo{author}{\bibfnamefont{F.}~\bibnamefont{Montanari}} \bibnamefont{and}
  \bibinfo{author}{\bibfnamefont{R.}~\bibnamefont{Durrer}},
  \bibinfo{journal}{Phys. Rev.} \textbf{\bibinfo{volume}{D86}},
  \bibinfo{pages}{063503} (\bibinfo{year}{2012}), \eprint{1206.3545}.

\bibitem[{\citenamefont{{Raccanelli}
  et~al.}(2013{\natexlab{b}})\citenamefont{{Raccanelli}, {Bertacca}, {Dore},
  and {Maartens}}}]{Raccanelli:2013tg}
\bibinfo{author}{\bibfnamefont{A.}~\bibnamefont{{Raccanelli}}},
  \bibinfo{author}{\bibfnamefont{D.}~\bibnamefont{{Bertacca}}},
  \bibinfo{author}{\bibfnamefont{O.}~\bibnamefont{{Dore}}}, \bibnamefont{and}
  \bibinfo{author}{\bibfnamefont{R.}~\bibnamefont{{Maartens}}},
  \bibinfo{journal}{ArXiv e-prints}  (\bibinfo{year}{2013}{\natexlab{b}}),
  \eprint{1306.6646}.

\bibitem[{\citenamefont{Cyr-Racine and
  Sigurdson}(2013{\natexlab{b}})}]{Cyr-Racine:2013jua}
\bibinfo{author}{\bibfnamefont{F.-Y.} \bibnamefont{Cyr-Racine}}
  \bibnamefont{and} \bibinfo{author}{\bibfnamefont{K.}~\bibnamefont{Sigurdson}}
  (\bibinfo{year}{2013}{\natexlab{b}}), \eprint{1306.1536}.

\bibitem[{\citenamefont{Bashinsky and Seljak}(2004)}]{Bashinsky:2003tk}
\bibinfo{author}{\bibfnamefont{S.}~\bibnamefont{Bashinsky}} \bibnamefont{and}
  \bibinfo{author}{\bibfnamefont{U.}~\bibnamefont{Seljak}},
  \bibinfo{journal}{Phys.Rev.} \textbf{\bibinfo{volume}{D69}},
  \bibinfo{pages}{083002} (\bibinfo{year}{2004}), \eprint{astro-ph/0310198}.

\bibitem[{\citenamefont{Hou et~al.}(2011)\citenamefont{Hou, Keisler, Knox,
  Millea, and Reichardt}}]{Hou:2011ec}
\bibinfo{author}{\bibfnamefont{Z.}~\bibnamefont{Hou}},
  \bibinfo{author}{\bibfnamefont{R.}~\bibnamefont{Keisler}},
  \bibinfo{author}{\bibfnamefont{L.}~\bibnamefont{Knox}},
  \bibinfo{author}{\bibfnamefont{M.}~\bibnamefont{Millea}}, \bibnamefont{and}
  \bibinfo{author}{\bibfnamefont{C.}~\bibnamefont{Reichardt}}
  (\bibinfo{year}{2011}), \eprint{1104.2333}.

\bibitem[{\citenamefont{Foot}(2013)}]{Foot:2012ai}
\bibinfo{author}{\bibfnamefont{R.}~\bibnamefont{Foot}},
  \bibinfo{journal}{Phys.Lett.} \textbf{\bibinfo{volume}{B718}},
  \bibinfo{pages}{745} (\bibinfo{year}{2013}), \eprint{1208.6022}.

\bibitem[{\citenamefont{{Lewis} and {Challinor}}(2006)}]{LewChal06}
\bibinfo{author}{\bibfnamefont{A.}~\bibnamefont{{Lewis}}} \bibnamefont{and}
  \bibinfo{author}{\bibfnamefont{A.}~\bibnamefont{{Challinor}}},
  \bibinfo{journal}{Physics Reports} \textbf{\bibinfo{volume}{429}},
  \bibinfo{pages}{1} (\bibinfo{year}{2006}), \eprint{arXiv:astro-ph/0601594}.

\bibitem[{\citenamefont{{Planck Collaboration}
  et~al.}(2013)\citenamefont{{Planck Collaboration}, {Ade}, {Aghanim},
  {Armitage-Caplan}, {Arnaud}, {Ashdown}, {Atrio-Barandela}, {Aumont},
  {Baccigalupi}, {Banday} et~al.}}]{planck17_2013}
\bibinfo{author}{\bibnamefont{{Planck Collaboration}}},
  \bibinfo{author}{\bibfnamefont{P.~A.~R.} \bibnamefont{{Ade}}},
  \bibinfo{author}{\bibfnamefont{N.}~\bibnamefont{{Aghanim}}},
  \bibinfo{author}{\bibfnamefont{C.}~\bibnamefont{{Armitage-Caplan}}},
  \bibinfo{author}{\bibfnamefont{M.}~\bibnamefont{{Arnaud}}},
  \bibinfo{author}{\bibfnamefont{M.}~\bibnamefont{{Ashdown}}},
  \bibinfo{author}{\bibfnamefont{F.}~\bibnamefont{{Atrio-Barandela}}},
  \bibinfo{author}{\bibfnamefont{J.}~\bibnamefont{{Aumont}}},
  \bibinfo{author}{\bibfnamefont{C.}~\bibnamefont{{Baccigalupi}}},
  \bibinfo{author}{\bibfnamefont{A.~J.} \bibnamefont{{Banday}}},
  \bibnamefont{et~al.}, \bibinfo{journal}{ArXiv e-prints}
  (\bibinfo{year}{2013}), \eprint{1303.5077}.

\bibitem[{\citenamefont{Story et~al.}(2012)\citenamefont{Story, Reichardt, Hou,
  Keisler, Aird et~al.}}]{Story:2012wx}
\bibinfo{author}{\bibfnamefont{K.}~\bibnamefont{Story}},
  \bibinfo{author}{\bibfnamefont{C.}~\bibnamefont{Reichardt}},
  \bibinfo{author}{\bibfnamefont{Z.}~\bibnamefont{Hou}},
  \bibinfo{author}{\bibfnamefont{R.}~\bibnamefont{Keisler}},
  \bibinfo{author}{\bibfnamefont{K.}~\bibnamefont{Aird}}, \bibnamefont{et~al.}
  (\bibinfo{year}{2012}), \eprint{1210.7231}.

\bibitem[{\citenamefont{Dunkley et~al.}(2013)\citenamefont{Dunkley, Calabrese,
  Sievers, Addison, Battaglia et~al.}}]{Dunkley:2013vu}
\bibinfo{author}{\bibfnamefont{J.}~\bibnamefont{Dunkley}},
  \bibinfo{author}{\bibfnamefont{E.}~\bibnamefont{Calabrese}},
  \bibinfo{author}{\bibfnamefont{J.}~\bibnamefont{Sievers}},
  \bibinfo{author}{\bibfnamefont{G.}~\bibnamefont{Addison}},
  \bibinfo{author}{\bibfnamefont{N.}~\bibnamefont{Battaglia}},
  \bibnamefont{et~al.} (\bibinfo{year}{2013}), \eprint{1301.0776}.

\bibitem[{\citenamefont{Das et~al.}(2013)\citenamefont{Das, Louis, Nolta,
  Addison, Battistelli et~al.}}]{Das:2013zf}
\bibinfo{author}{\bibfnamefont{S.}~\bibnamefont{Das}},
  \bibinfo{author}{\bibfnamefont{T.}~\bibnamefont{Louis}},
  \bibinfo{author}{\bibfnamefont{M.~R.} \bibnamefont{Nolta}},
  \bibinfo{author}{\bibfnamefont{G.~E.} \bibnamefont{Addison}},
  \bibinfo{author}{\bibfnamefont{E.~S.} \bibnamefont{Battistelli}},
  \bibnamefont{et~al.} (\bibinfo{year}{2013}), \eprint{1301.1037}.

\bibitem[{\citenamefont{{Okamoto} and {Hu}}(2003)}]{okahu03}
\bibinfo{author}{\bibfnamefont{T.}~\bibnamefont{{Okamoto}}} \bibnamefont{and}
  \bibinfo{author}{\bibfnamefont{W.}~\bibnamefont{{Hu}}},
  \bibinfo{journal}{\prd} \textbf{\bibinfo{volume}{67}}, \bibinfo{eid}{083002}
  (\bibinfo{year}{2003}), \eprint{arXiv:astro-ph/0301031}.

\bibitem[{\citenamefont{Padmanabhan et~al.}(2012)\citenamefont{Padmanabhan, Xu,
  Eisenstein, Scalzo, Cuesta et~al.}}]{Padmanabhan:2012hf}
\bibinfo{author}{\bibfnamefont{N.}~\bibnamefont{Padmanabhan}},
  \bibinfo{author}{\bibfnamefont{X.}~\bibnamefont{Xu}},
  \bibinfo{author}{\bibfnamefont{D.~J.} \bibnamefont{Eisenstein}},
  \bibinfo{author}{\bibfnamefont{R.}~\bibnamefont{Scalzo}},
  \bibinfo{author}{\bibfnamefont{A.~J.} \bibnamefont{Cuesta}},
  \bibnamefont{et~al.} (\bibinfo{year}{2012}), \eprint{1202.0090}.

\bibitem[{\citenamefont{Beutler et~al.}(2011)\citenamefont{Beutler, Blake,
  Colless, Jones, Staveley-Smith et~al.}}]{Beutler:2011hx}
\bibinfo{author}{\bibfnamefont{F.}~\bibnamefont{Beutler}},
  \bibinfo{author}{\bibfnamefont{C.}~\bibnamefont{Blake}},
  \bibinfo{author}{\bibfnamefont{M.}~\bibnamefont{Colless}},
  \bibinfo{author}{\bibfnamefont{D.~H.} \bibnamefont{Jones}},
  \bibinfo{author}{\bibfnamefont{L.}~\bibnamefont{Staveley-Smith}},
  \bibnamefont{et~al.}, \bibinfo{journal}{Mon.Not.Roy.Astron.Soc.}
  \textbf{\bibinfo{volume}{416}}, \bibinfo{pages}{3017} (\bibinfo{year}{2011}),
  \eprint{1106.3366}.

\bibitem[{\citenamefont{{Dawson} et~al.}(2013)\citenamefont{{Dawson},
  {Schlegel}, {Ahn}, {Anderson}, {Aubourg}, {Bailey}, {Barkhouser}, {Bautista},
  {Beifiori}, {Berlind} et~al.}}]{dawsonetal13}
\bibinfo{author}{\bibfnamefont{K.~S.} \bibnamefont{{Dawson}}},
  \bibinfo{author}{\bibfnamefont{D.~J.} \bibnamefont{{Schlegel}}},
  \bibinfo{author}{\bibfnamefont{C.~P.} \bibnamefont{{Ahn}}},
  \bibinfo{author}{\bibfnamefont{S.~F.} \bibnamefont{{Anderson}}},
  \bibinfo{author}{\bibfnamefont{{\'E}.}~\bibnamefont{{Aubourg}}},
  \bibinfo{author}{\bibfnamefont{S.}~\bibnamefont{{Bailey}}},
  \bibinfo{author}{\bibfnamefont{R.~H.} \bibnamefont{{Barkhouser}}},
  \bibinfo{author}{\bibfnamefont{J.~E.} \bibnamefont{{Bautista}}},
  \bibinfo{author}{\bibfnamefont{A.}~\bibnamefont{{Beifiori}}},
  \bibinfo{author}{\bibfnamefont{A.~A.} \bibnamefont{{Berlind}}},
  \bibnamefont{et~al.}, \bibinfo{journal}{Astronomical Journal}
  \textbf{\bibinfo{volume}{145}}, \bibinfo{eid}{10} (\bibinfo{year}{2013}),
  \eprint{1208.0022}.

\bibitem[{\citenamefont{{York} et~al.}(2000)\citenamefont{{York}, {Adelman},
  {Anderson}, {Anderson}, {Annis}, {Bahcall}, {Bakken}, {Barkhouser},
  {Bastian}, {Berman} et~al.}}]{yorketal00}
\bibinfo{author}{\bibfnamefont{D.~G.} \bibnamefont{{York}}},
  \bibinfo{author}{\bibfnamefont{J.}~\bibnamefont{{Adelman}}},
  \bibinfo{author}{\bibfnamefont{J.~E.} \bibnamefont{{Anderson}},
  \bibfnamefont{Jr.}}, \bibinfo{author}{\bibfnamefont{S.~F.}
  \bibnamefont{{Anderson}}},
  \bibinfo{author}{\bibfnamefont{J.}~\bibnamefont{{Annis}}},
  \bibinfo{author}{\bibfnamefont{N.~A.} \bibnamefont{{Bahcall}}},
  \bibinfo{author}{\bibfnamefont{J.~A.} \bibnamefont{{Bakken}}},
  \bibinfo{author}{\bibfnamefont{R.}~\bibnamefont{{Barkhouser}}},
  \bibinfo{author}{\bibfnamefont{S.}~\bibnamefont{{Bastian}}},
  \bibinfo{author}{\bibfnamefont{E.}~\bibnamefont{{Berman}}},
  \bibnamefont{et~al.}, \bibinfo{journal}{Astronomical Journal}
  \textbf{\bibinfo{volume}{120}}, \bibinfo{pages}{1579} (\bibinfo{year}{2000}),
  \eprint{arXiv:astro-ph/0006396}.

\bibitem[{\citenamefont{{Eisenstein} et~al.}(2011)\citenamefont{{Eisenstein},
  {Weinberg}, {Agol}, {Aihara}, {Allende Prieto}, {Anderson}, {Arns},
  {Aubourg}, {Bailey}, {Balbinot} et~al.}}]{sdss3}
\bibinfo{author}{\bibfnamefont{D.~J.} \bibnamefont{{Eisenstein}}},
  \bibinfo{author}{\bibfnamefont{D.~H.} \bibnamefont{{Weinberg}}},
  \bibinfo{author}{\bibfnamefont{E.}~\bibnamefont{{Agol}}},
  \bibinfo{author}{\bibfnamefont{H.}~\bibnamefont{{Aihara}}},
  \bibinfo{author}{\bibfnamefont{C.}~\bibnamefont{{Allende Prieto}}},
  \bibinfo{author}{\bibfnamefont{S.~F.} \bibnamefont{{Anderson}}},
  \bibinfo{author}{\bibfnamefont{J.~A.} \bibnamefont{{Arns}}},
  \bibinfo{author}{\bibfnamefont{{\'E}.}~\bibnamefont{{Aubourg}}},
  \bibinfo{author}{\bibfnamefont{S.}~\bibnamefont{{Bailey}}},
  \bibinfo{author}{\bibfnamefont{E.}~\bibnamefont{{Balbinot}}},
  \bibnamefont{et~al.}, \bibinfo{journal}{\aj} \textbf{\bibinfo{volume}{142}},
  \bibinfo{eid}{72} (\bibinfo{year}{2011}), \eprint{1101.1529}.

\bibitem[{\citenamefont{{White} et~al.}(2011)\citenamefont{{White}, {Blanton},
  {Bolton}, {Schlegel}, {Tinker}, {Berlind}, {da Costa}, {Kazin}, {Lin}, {Maia}
  et~al.}}]{whiteetal11}
\bibinfo{author}{\bibfnamefont{M.}~\bibnamefont{{White}}},
  \bibinfo{author}{\bibfnamefont{M.}~\bibnamefont{{Blanton}}},
  \bibinfo{author}{\bibfnamefont{A.}~\bibnamefont{{Bolton}}},
  \bibinfo{author}{\bibfnamefont{D.}~\bibnamefont{{Schlegel}}},
  \bibinfo{author}{\bibfnamefont{J.}~\bibnamefont{{Tinker}}},
  \bibinfo{author}{\bibfnamefont{A.}~\bibnamefont{{Berlind}}},
  \bibinfo{author}{\bibfnamefont{L.}~\bibnamefont{{da Costa}}},
  \bibinfo{author}{\bibfnamefont{E.}~\bibnamefont{{Kazin}}},
  \bibinfo{author}{\bibfnamefont{Y.-T.} \bibnamefont{{Lin}}},
  \bibinfo{author}{\bibfnamefont{M.}~\bibnamefont{{Maia}}},
  \bibnamefont{et~al.}, \bibinfo{journal}{\apj} \textbf{\bibinfo{volume}{728}},
  \bibinfo{eid}{126} (\bibinfo{year}{2011}), \eprint{1010.4915}.

\bibitem[{\citenamefont{{Ahn} et~al.}(2012)\citenamefont{{Ahn}, {Alexandroff},
  {Allende Prieto}, {Anderson}, {Anderton}, {Andrews}, {Aubourg}, {Bailey},
  {Balbinot}, {Barnes} et~al.}}]{DR9}
\bibinfo{author}{\bibfnamefont{C.~P.} \bibnamefont{{Ahn}}},
  \bibinfo{author}{\bibfnamefont{R.}~\bibnamefont{{Alexandroff}}},
  \bibinfo{author}{\bibfnamefont{C.}~\bibnamefont{{Allende Prieto}}},
  \bibinfo{author}{\bibfnamefont{S.~F.} \bibnamefont{{Anderson}}},
  \bibinfo{author}{\bibfnamefont{T.}~\bibnamefont{{Anderton}}},
  \bibinfo{author}{\bibfnamefont{B.~H.} \bibnamefont{{Andrews}}},
  \bibinfo{author}{\bibfnamefont{{\'E}.}~\bibnamefont{{Aubourg}}},
  \bibinfo{author}{\bibfnamefont{S.}~\bibnamefont{{Bailey}}},
  \bibinfo{author}{\bibfnamefont{E.}~\bibnamefont{{Balbinot}}},
  \bibinfo{author}{\bibfnamefont{R.}~\bibnamefont{{Barnes}}},
  \bibnamefont{et~al.}, \bibinfo{journal}{\apjs}
  \textbf{\bibinfo{volume}{203}}, \bibinfo{eid}{21} (\bibinfo{year}{2012}),
  \eprint{1207.7137}.

\bibitem[{\citenamefont{{Feldman} et~al.}(1994)\citenamefont{{Feldman},
  {Kaiser}, and {Peacock}}}]{FKP}
\bibinfo{author}{\bibfnamefont{H.~A.} \bibnamefont{{Feldman}}},
  \bibinfo{author}{\bibfnamefont{N.}~\bibnamefont{{Kaiser}}}, \bibnamefont{and}
  \bibinfo{author}{\bibfnamefont{J.~A.} \bibnamefont{{Peacock}}},
  \bibinfo{journal}{\apj} \textbf{\bibinfo{volume}{426}}, \bibinfo{pages}{23}
  (\bibinfo{year}{1994}), \eprint{arXiv:astro-ph/9304022}.

\bibitem[{\citenamefont{{Reid} et~al.}(2010)\citenamefont{{Reid}, {Percival},
  {Eisenstein}, {Verde}, {Spergel}, {Skibba}, {Bahcall}, {Budavari}, {Frieman},
  {Fukugita} et~al.}}]{reidetal10}
\bibinfo{author}{\bibfnamefont{B.~A.} \bibnamefont{{Reid}}},
  \bibinfo{author}{\bibfnamefont{W.~J.} \bibnamefont{{Percival}}},
  \bibinfo{author}{\bibfnamefont{D.~J.} \bibnamefont{{Eisenstein}}},
  \bibinfo{author}{\bibfnamefont{L.}~\bibnamefont{{Verde}}},
  \bibinfo{author}{\bibfnamefont{D.~N.} \bibnamefont{{Spergel}}},
  \bibinfo{author}{\bibfnamefont{R.~A.} \bibnamefont{{Skibba}}},
  \bibinfo{author}{\bibfnamefont{N.~A.} \bibnamefont{{Bahcall}}},
  \bibinfo{author}{\bibfnamefont{T.}~\bibnamefont{{Budavari}}},
  \bibinfo{author}{\bibfnamefont{J.~A.} \bibnamefont{{Frieman}}},
  \bibinfo{author}{\bibfnamefont{M.}~\bibnamefont{{Fukugita}}},
  \bibnamefont{et~al.}, \bibinfo{journal}{\mnras}
  \textbf{\bibinfo{volume}{404}}, \bibinfo{pages}{60} (\bibinfo{year}{2010}),
  \eprint{0907.1659}.

\bibitem[{\citenamefont{{Ross} et~al.}(2012)\citenamefont{{Ross}, {Percival},
  {S{\'a}nchez}, {Samushia}, {Ho}, {Kazin}, {Manera}, {Reid}, {White},
  {Tojeiro} et~al.}}]{arossetal12}
\bibinfo{author}{\bibfnamefont{A.~J.} \bibnamefont{{Ross}}},
  \bibinfo{author}{\bibfnamefont{W.~J.} \bibnamefont{{Percival}}},
  \bibinfo{author}{\bibfnamefont{A.~G.} \bibnamefont{{S{\'a}nchez}}},
  \bibinfo{author}{\bibfnamefont{L.}~\bibnamefont{{Samushia}}},
  \bibinfo{author}{\bibfnamefont{S.}~\bibnamefont{{Ho}}},
  \bibinfo{author}{\bibfnamefont{E.}~\bibnamefont{{Kazin}}},
  \bibinfo{author}{\bibfnamefont{M.}~\bibnamefont{{Manera}}},
  \bibinfo{author}{\bibfnamefont{B.}~\bibnamefont{{Reid}}},
  \bibinfo{author}{\bibfnamefont{M.}~\bibnamefont{{White}}},
  \bibinfo{author}{\bibfnamefont{R.}~\bibnamefont{{Tojeiro}}},
  \bibnamefont{et~al.}, \bibinfo{journal}{\mnras}
  \textbf{\bibinfo{volume}{424}}, \bibinfo{pages}{564} (\bibinfo{year}{2012}),
  \eprint{1203.6499}.

\bibitem[{\citenamefont{{Seljak}}(2000)}]{seljak2000}
\bibinfo{author}{\bibfnamefont{U.}~\bibnamefont{{Seljak}}},
  \bibinfo{journal}{\mnras} \textbf{\bibinfo{volume}{318}},
  \bibinfo{pages}{203} (\bibinfo{year}{2000}), \eprint{arXiv:astro-ph/0001493}.

\bibitem[{\citenamefont{{Seljak}}(2001)}]{seljak01}
\bibinfo{author}{\bibfnamefont{U.}~\bibnamefont{{Seljak}}},
  \bibinfo{journal}{\mnras} \textbf{\bibinfo{volume}{325}},
  \bibinfo{pages}{1359} (\bibinfo{year}{2001}),
  \eprint{arXiv:astro-ph/0009016}.

\bibitem[{\citenamefont{{Schulz} and {White}}(2006)}]{schulzwhite06}
\bibinfo{author}{\bibfnamefont{A.~E.} \bibnamefont{{Schulz}}} \bibnamefont{and}
  \bibinfo{author}{\bibfnamefont{M.}~\bibnamefont{{White}}},
  \bibinfo{journal}{Astroparticle Physics} \textbf{\bibinfo{volume}{25}},
  \bibinfo{pages}{172} (\bibinfo{year}{2006}), \eprint{arXiv:astro-ph/0510100}.

\bibitem[{\citenamefont{{Guzik} et~al.}(2007)\citenamefont{{Guzik},
  {Bernstein}, and {Smith}}}]{guziketal07}
\bibinfo{author}{\bibfnamefont{J.}~\bibnamefont{{Guzik}}},
  \bibinfo{author}{\bibfnamefont{G.}~\bibnamefont{{Bernstein}}},
  \bibnamefont{and} \bibinfo{author}{\bibfnamefont{R.~E.}
  \bibnamefont{{Smith}}}, \bibinfo{journal}{\mnras}
  \textbf{\bibinfo{volume}{375}}, \bibinfo{pages}{1329} (\bibinfo{year}{2007}),
  \eprint{arXiv:astro-ph/0605594}.

\bibitem[{\citenamefont{{Scherrer} and {Weinberg}}(1998)}]{scherrwein98}
\bibinfo{author}{\bibfnamefont{R.~J.} \bibnamefont{{Scherrer}}}
  \bibnamefont{and} \bibinfo{author}{\bibfnamefont{D.~H.}
  \bibnamefont{{Weinberg}}}, \bibinfo{journal}{\apj}
  \textbf{\bibinfo{volume}{504}}, \bibinfo{pages}{607} (\bibinfo{year}{1998}),
  \eprint{arXiv:astro-ph/9712192}.

\bibitem[{\citenamefont{{Coles} et~al.}(1999)\citenamefont{{Coles}, {Melott},
  and {Munshi}}}]{colesetal99}
\bibinfo{author}{\bibfnamefont{P.}~\bibnamefont{{Coles}}},
  \bibinfo{author}{\bibfnamefont{A.~L.} \bibnamefont{{Melott}}},
  \bibnamefont{and} \bibinfo{author}{\bibfnamefont{D.}~\bibnamefont{{Munshi}}},
  \bibinfo{journal}{\apjl} \textbf{\bibinfo{volume}{521}}, \bibinfo{pages}{L5}
  (\bibinfo{year}{1999}), \eprint{arXiv:astro-ph/9904253}.

\bibitem[{\citenamefont{{Saito} et~al.}(2009)\citenamefont{{Saito}, {Takada},
  and {Taruya}}}]{saitoetal09}
\bibinfo{author}{\bibfnamefont{S.}~\bibnamefont{{Saito}}},
  \bibinfo{author}{\bibfnamefont{M.}~\bibnamefont{{Takada}}}, \bibnamefont{and}
  \bibinfo{author}{\bibfnamefont{A.}~\bibnamefont{{Taruya}}},
  \bibinfo{journal}{\prd} \textbf{\bibinfo{volume}{80}}, \bibinfo{eid}{083528}
  (\bibinfo{year}{2009}), \eprint{0907.2922}.

\bibitem[{\citenamefont{{Tegmark} et~al.}(2006)\citenamefont{{Tegmark},
  {Eisenstein}, {Strauss}, {Weinberg}, {Blanton}, {Frieman}, {Fukugita},
  {Gunn}, {Hamilton}, {Knapp} et~al.}}]{Tegetal2006}
\bibinfo{author}{\bibfnamefont{M.}~\bibnamefont{{Tegmark}}},
  \bibinfo{author}{\bibfnamefont{D.~J.} \bibnamefont{{Eisenstein}}},
  \bibinfo{author}{\bibfnamefont{M.~A.} \bibnamefont{{Strauss}}},
  \bibinfo{author}{\bibfnamefont{D.~H.} \bibnamefont{{Weinberg}}},
  \bibinfo{author}{\bibfnamefont{M.~R.} \bibnamefont{{Blanton}}},
  \bibinfo{author}{\bibfnamefont{J.~A.} \bibnamefont{{Frieman}}},
  \bibinfo{author}{\bibfnamefont{M.}~\bibnamefont{{Fukugita}}},
  \bibinfo{author}{\bibfnamefont{J.~E.} \bibnamefont{{Gunn}}},
  \bibinfo{author}{\bibfnamefont{A.~J.~S.} \bibnamefont{{Hamilton}}},
  \bibinfo{author}{\bibfnamefont{G.~R.} \bibnamefont{{Knapp}}},
  \bibnamefont{et~al.}, \bibinfo{journal}{Phys.Rev.D}
  \textbf{\bibinfo{volume}{74}}, \bibinfo{pages}{123507}
  (\bibinfo{year}{2006}), \eprint{arXiv:astro-ph/0608632}.

\bibitem[{\citenamefont{{Percival} et~al.}(2007)\citenamefont{{Percival},
  {Cole}, {Eisenstein}, {Nichol}, {Peacock}, {Pope}, and
  {Szalay}}}]{percetal07}
\bibinfo{author}{\bibfnamefont{W.~J.} \bibnamefont{{Percival}}},
  \bibinfo{author}{\bibfnamefont{S.}~\bibnamefont{{Cole}}},
  \bibinfo{author}{\bibfnamefont{D.~J.} \bibnamefont{{Eisenstein}}},
  \bibinfo{author}{\bibfnamefont{R.~C.} \bibnamefont{{Nichol}}},
  \bibinfo{author}{\bibfnamefont{J.~A.} \bibnamefont{{Peacock}}},
  \bibinfo{author}{\bibfnamefont{A.~C.} \bibnamefont{{Pope}}},
  \bibnamefont{and} \bibinfo{author}{\bibfnamefont{A.~S.}
  \bibnamefont{{Szalay}}}, \bibinfo{journal}{\mnras}
  \textbf{\bibinfo{volume}{381}}, \bibinfo{pages}{1053} (\bibinfo{year}{2007}),
  \eprint{0705.3323}.

\bibitem[{\citenamefont{{Manera} et~al.}(2013)\citenamefont{{Manera},
  {Scoccimarro}, {Percival}, {Samushia}, {McBride}, {Ross}, {Sheth}, {White},
  {Reid}, {S{\'a}nchez} et~al.}}]{maneraetal13}
\bibinfo{author}{\bibfnamefont{M.}~\bibnamefont{{Manera}}},
  \bibinfo{author}{\bibfnamefont{R.}~\bibnamefont{{Scoccimarro}}},
  \bibinfo{author}{\bibfnamefont{W.~J.} \bibnamefont{{Percival}}},
  \bibinfo{author}{\bibfnamefont{L.}~\bibnamefont{{Samushia}}},
  \bibinfo{author}{\bibfnamefont{C.~K.} \bibnamefont{{McBride}}},
  \bibinfo{author}{\bibfnamefont{A.~J.} \bibnamefont{{Ross}}},
  \bibinfo{author}{\bibfnamefont{R.~K.} \bibnamefont{{Sheth}}},
  \bibinfo{author}{\bibfnamefont{M.}~\bibnamefont{{White}}},
  \bibinfo{author}{\bibfnamefont{B.~A.} \bibnamefont{{Reid}}},
  \bibinfo{author}{\bibfnamefont{A.~G.} \bibnamefont{{S{\'a}nchez}}},
  \bibnamefont{et~al.}, \bibinfo{journal}{\mnras}
  \textbf{\bibinfo{volume}{428}}, \bibinfo{pages}{1036} (\bibinfo{year}{2013}),
  \eprint{1203.6609}.

\bibitem[{\citenamefont{Lewis and Bridle}(2002)}]{Lewis:2002ah}
\bibinfo{author}{\bibfnamefont{A.}~\bibnamefont{Lewis}} \bibnamefont{and}
  \bibinfo{author}{\bibfnamefont{S.}~\bibnamefont{Bridle}},
  \bibinfo{journal}{Phys.Rev.} \textbf{\bibinfo{volume}{D66}},
  \bibinfo{pages}{103511} (\bibinfo{year}{2002}), \eprint{astro-ph/0205436}.

\bibitem[{\citenamefont{Hanson et~al.}(2013)}]{Hanson:2013hsb}
\bibinfo{author}{\bibfnamefont{D.}~\bibnamefont{Hanson}} \bibnamefont{et~al.}
  (\bibinfo{collaboration}{SPTpol Collaboration}) (\bibinfo{year}{2013}),
  \eprint{1307.5830}.

\bibitem[{\citenamefont{Austermann et~al.}(2012)\citenamefont{Austermann, Aird,
  Beall, Becker, Bender et~al.}}]{Austermann:2012ga}
\bibinfo{author}{\bibfnamefont{J.}~\bibnamefont{Austermann}},
  \bibinfo{author}{\bibfnamefont{K.}~\bibnamefont{Aird}},
  \bibinfo{author}{\bibfnamefont{J.}~\bibnamefont{Beall}},
  \bibinfo{author}{\bibfnamefont{D.}~\bibnamefont{Becker}},
  \bibinfo{author}{\bibfnamefont{A.}~\bibnamefont{Bender}},
  \bibnamefont{et~al.}, \bibinfo{journal}{Proc.SPIE Int.Soc.Opt.Eng.}
  \textbf{\bibinfo{volume}{8452}}, \bibinfo{pages}{84520E}
  (\bibinfo{year}{2012}), \eprint{1210.4970}.

\bibitem[{\citenamefont{Pilbratt et~al.}(2010)\citenamefont{Pilbratt,
  Riedinger, Passvogel, Crone, Doyle et~al.}}]{Pilbratt:2010mv}
\bibinfo{author}{\bibfnamefont{G.}~\bibnamefont{Pilbratt}},
  \bibinfo{author}{\bibfnamefont{J.}~\bibnamefont{Riedinger}},
  \bibinfo{author}{\bibfnamefont{T.}~\bibnamefont{Passvogel}},
  \bibinfo{author}{\bibfnamefont{G.}~\bibnamefont{Crone}},
  \bibinfo{author}{\bibfnamefont{D.}~\bibnamefont{Doyle}},
  \bibnamefont{et~al.}, \bibinfo{journal}{Astron.Astrophys.}
  \textbf{\bibinfo{volume}{518}}, \bibinfo{pages}{L1} (\bibinfo{year}{2010}),
  \eprint{1005.5331}.

\bibitem[{\citenamefont{Smith et~al.}(2006)\citenamefont{Smith, Hu, and
  Kaplinghat}}]{Smith:2006nk}
\bibinfo{author}{\bibfnamefont{K.~M.} \bibnamefont{Smith}},
  \bibinfo{author}{\bibfnamefont{W.}~\bibnamefont{Hu}}, \bibnamefont{and}
  \bibinfo{author}{\bibfnamefont{M.}~\bibnamefont{Kaplinghat}},
  \bibinfo{journal}{Phys.Rev.} \textbf{\bibinfo{volume}{D74}},
  \bibinfo{pages}{123002} (\bibinfo{year}{2006}), \eprint{astro-ph/0607315}.

\bibitem[{\citenamefont{Ade et~al.}(2013{\natexlab{b}})}]{Ade:2013lmv}
\bibinfo{author}{\bibfnamefont{P.}~\bibnamefont{Ade}} \bibnamefont{et~al.}
  (\bibinfo{collaboration}{Planck Collaboration})
  (\bibinfo{year}{2013}{\natexlab{b}}), \eprint{1303.5080}.

\end{thebibliography}
%%%%%%%%%

\end{document}